\pdfoutput=1

\def\showtableofcontents{1}
\def\showauthnotes{0}

\documentclass[11pt]{article}

\newcommand{\accept}{p}

\newcommand{\compress}{\mathsf{compress}}
\newcommand{\decup}{\mathsf{Decup}}
\newcommand{\distinguish}{\Delta}
\newcommand{\game}{\mathsf{Game}}
\newcommand{\prs}{R}
\newcommand{\bprs}{\bR}
\newcommand{\operator}{\mathrm{op}}
\newcommand{\oracle}{\mathcal{O}}
\newcommand{\queryreg}{\calH_{\mathsf{query}}}
\newcommand{\rescale}{D}
\newcommand{\spectral}{\mathsf{Spectral}}
\newcommand{\trunc}{\mathrm{trunc}}
\newcommand{\weights}{\mathsf{wt}}
\newcommand{\width}{\mathsf{width}}
\newcommand{\wt}{\mathsf{wt}}

\ifnum\showauthnotes=1
\newcommand{\jnote}[1]{\textcolor{red}{\small {\textbf{(John:} #1\textbf{) }}}}
\newcommand{\fermi}[1]{\textcolor{blue}{\small {\textbf{(Fermi:} #1\textbf{) }}}}
\newcommand{\alex}[1]{\textcolor{green}{\small {\textbf{(Alex:} #1\textbf{) }}}}

\else 
\newcommand{\jnote}[1]{}
\newcommand{\fermi}[1]{}
\newcommand{\alex}[1]{}
\fi

\usepackage{tikz}
\usetikzlibrary{quantikz2}
\usepackage{style}

\title{A one-query lower bound for unitary synthesis \\ and breaking quantum cryptography}

\author{Alex Lombardi\thanks{Princeton. Email: \texttt{alex.lombardi@princeton.edu}. Research was done in part while the author was a Simons-Berkeley Postdoctoral Fellow.} \and Fermi Ma\thanks{Simons Institute $\&$ UC Berkeley. Email: \texttt{fermima1@gmail.com}.} \and John Wright\thanks{UC Berkeley. Email: \texttt{jswright@berkeley.edu}.}}

\date{}

\begin{document}

\maketitle
\thispagestyle{empty}

\begin{abstract}
    The Unitary Synthesis Problem (Aaronson-Kuperberg 2007) asks whether any $n$-qubit unitary $U$ can be implemented by an efficient quantum algorithm $A$ augmented with an oracle that \emph{computes an arbitrary Boolean function} $f$. In other words, can the task of implementing any unitary be efficiently reduced to the task of implementing any Boolean function? 
    
    In this work, we prove a one-query lower bound for unitary synthesis. We show that there exist unitaries $U$ such that no quantum polynomial-time oracle algorithm $A^f$ can implement $U$, even approximately, if it only makes one (quantum) query to $f$. Our approach also has implications for quantum cryptography: we prove (relative to a random oracle) the existence of quantum cryptographic primitives that remain secure against all one-query adversaries $A^{f}$. Since such one-query algorithms can decide any language, solve any classical search problem, and even prepare any quantum state, our result suggests that implementing random unitaries and breaking quantum cryptography may be harder than all of these tasks. 

    To prove this result, we formulate unitary synthesis as an efficient challenger-adversary game, which enables proving lower bounds by analyzing the maximum success probability of an adversary $A^f$. Our main technical insight is to identify a natural spectral relaxation of the one-query optimization problem, which we bound using tools from random matrix theory. 

    We view our framework as a potential avenue to rule out polynomial-query unitary synthesis, and we state conjectures in this direction. 
    
\end{abstract}

\newpage
\ifnum\showtableofcontents=1
{\newpage
\hypersetup{linktocpage}
\tableofcontents
\thispagestyle{empty}
}

\newpage
\setcounter{page}{1}

\section{Introduction}


This paper is about \emph{unitary synthesis}, the task of implementing a given $n$-qubit unitary transformation $U$ as a quantum circuit. Unitary synthesis is ubiquitous throughout quantum computing, since virtually any quantum computational task --- be it preparing a state, performing a measurement, or transforming one state into another --- can be done by implementing \emph{some} unitary. Of course, not every unitary can be implemented efficiently. As a special case, consider the classical task of evaluating an $(n-1)$-bit Boolean function $f:\{0, 1\}^{n-1} \rightarrow \{0, 1\}$. This can be solved by implementing an $n$-qubit unitary transformation, namely the unitary $U: \ket{x,b} \mapsto \ket{x,b \oplus f(x)}$,
and so Shannon's classic counting argument~\cite{Sha49} implies that even these unitaries require $\Omega(2^n/n)$ gates to implement. But are worst-case unitaries hard to compute only because they can solve hard classical problems? Or is it possible that unitaries could still be hard even if it were easy to solve all classical problems?

This question was first posed in $2006$ in an influential work by Aaronson and Kuperberg~\cite{AK07},
and it was later dubbed ``the Unitary Synthesis Problem'' by Aaronson in his 2016 Barbados lectures~\cite{Aar16}. Formally, they considered $\poly(n)$-size quantum oracle circuits $A^{(\cdot)}$ that have the ability to make quantum queries to an arbitrary Boolean function $f: \{0,1\}^\ell \rightarrow \{0,1\}$ on $\ell = \poly(n)$ bits.
This gives these circuits the power to instantaneously compute any Boolean function of their choice,
and Aaronson and Kuperberg asked if this power enables them to efficiently implement any unitary transformation as well. More concretely, they asked the following question.

\begin{quote}
    \textbf{The Unitary Synthesis Problem \cite{AK07,Aar16}:} Is there a universal efficient oracle circuit $A^{(\cdot)}$ such that for any unitary $U$, there is a corresponding Boolean function $f$ for which $A^{f}$ implements $U$?
\end{quote}


In other words, the Unitary Synthesis Problem asks whether the task of implementing an arbitrary unitary can be efficiently reduced to computing Boolean functions. Notably, if the answer turns out to be negative, this would give strong evidence (in the form of a black-box separation) that the hardest quantum problems are harder than the hardest classical problems. 

Since it was first posed,
the Unitary Synthesis Problem has become arguably \emph{the} central open problem in the rapidly growing field of unitary complexity, which we will discuss in more detail in \Cref{sec:background,sec:related-work} below. To date, there is no clear consensus on what the true complexity of unitary synthesis should be: for all we knew, it might require as little as one query to the oracle, or as many as $2^{\Omega(n)}$. 

One reason this question is subtle is that algorithms that make just one query to an arbitrary Boolean function are already quite powerful. For example, it turns out that such algorithms can solve the \emph{state synthesis problem}, in which the goal is to produce an arbitrary quantum state $\ket{\psi}$ \cite{Aar16,INN+22,Ros23} (see \cref{sec:related-work} for discussion). The state synthesis and unitary synthesis problems share a number of similarities, and there has been some speculation that extending state synthesis techniques could lead to positive results on unitary synthesis (for example,  see \cite[Section 7.2]{INN+22}). An excellent treatment of these and related problems can be found in Aaronson's Barbados notes~\cite{Aar16}, Rosenthal's Ph.D. thesis~\cite{Ros23b},
as well as in recent course notes of Yuen~\cite{Yue22,Yue22b}. 

\paragraph{How hard is unitary synthesis?} There are several inefficient algorithms for the Unitary Synthesis Problem,
the most basic of which queries an oracle $\widetilde{O}(2^{2n})$ times to learn a classical description of $U$
and then implements it using $\widetilde{O}(2^{2n})$ gates.
As noted by Yuen~\cite{Yue22}, this basic algorithm can be implemented with a single quantum query to $f$ using the Bernstein-Vazirani algorithm~\cite{BV97}, at the expense of making the query extremely large: in particular, it requires a quantum query to a Boolean function $f:\{0, 1\}^{\ell} \rightarrow \{0, 1\}$ on inputs of length $\ell = \widetilde{O}(2^{2n})$. If we restrict to algorithms that only make efficient queries to $f$, i.e., queries that only evaluate $f$ on $\ell = \poly(n)$-length inputs, the best known query complexity is $O(2^{n/2})$, achieved by a Grover-style algorithm due to Rosenthal~\cite{Ros22}.

On the other hand, prior to this work, no general query lower bound for the Unitary Synthesis Problem was known. There \emph{is} a well-known lower bound due to Aaronson and Kuperberg \cite{AK07} that rules out a certain class of one-query algorithms $A^f$, namely those that exactly implement a unitary operation on their first $n$ qubits for all choices of $f$. More recently, Rosenthal~\cite{Ros22} proved a lower bound ruling out a different specialized class of many-query algorithms. We discuss both of these lower bounds further in \Cref{sec:related-work}.
~However, the problem of ruling out (or constructing) even \emph{one-query} unitary synthesis algorithms has remained open since Aaronson and Kuperberg first posed it nearly two decades ago
(cf.\ Open Problem 4 and footnote 13 in~\cite{AK07}).

In this work, we resolve this open question and prove the first one-query lower bound for the Unitary Synthesis Problem.

\begin{theorem}[informal, see \cref{thm:main-at-end-of-proof}]\label{thm:main-intro}
There is no efficient oracle circuit $A^{(\cdot)}$ that approximately implements an arbitrary $n$-qubit unitary $U$ by making one quantum query to a $U$-dependent Boolean function $f$.
\end{theorem}

Our lower bound applies even if the oracle circuit is allowed to use an unbounded number of non-oracle gates and ancilla qubits.
In fact, it even applies to circuits that are allowed to query $f$ on inputs which are extremely long, but not \emph{too} extremely long;
technically, we require that $A^{(\cdot)}$ can only query Boolean functions $f:\{0, 1\}^{\ell} \rightarrow \{0, 1\}$ on $\ell = o(2^n)$ bits.
~Note that some restriction on the size of the queries is necessary, due to the one-query Bernstein-Vazirani-style algorithm mentioned above, which queries a Boolean function on $\ell = \widetilde{O}(2^{2n})$ bits. Finally, our lower bound also extends to circuits that query arbitrary functions $f$ with $\poly(n)$ bits of output, and to circuits that make $\poly(n)$-many \emph{non-adaptive} queries, i.e., queries of the form
\begin{align*}
    \ket{x_1}\ket{b_1}\hdots \ket{x_t}\ket{b_t}\mapsto \ket{x_1}\ket{b_1 \oplus f_1(x_1)}\hdots \ket{x_t}\ket{b_t \oplus f_t(x_t)}.
\end{align*}
This is because such queries can be simulated using a single query to a more complex function, via another Bernstein-Vazirani trick (see \cref{rem:parallel-queries}). 

We prove \cref{thm:main-intro} by leveraging a connection between the unitary synthesis problem and \emph{quantum cryptography}, as we discuss next.


\subsection{Unitary synthesis and quantum cryptography}\label{sec:background}

\paragraph{Background and motivation.} The past few years have seen a surge of interest in so-called \emph{inherently quantum problems}, which are computational tasks in which either the input is a quantum state, the output is a quantum state, or both. These include many of the most important tasks in quantum computing, such as breaking computationally-secure quantum bit commitments, performing quantum state tomography, preparing the ground state of a local Hamiltonian, and decoding black hole radiation. The central goal of this area is to classify these problems according to the computational resources needed to solve them. Normally, we would do so using the language of computational complexity theory. However, after initial classification attempts, a mysterious, recurring phenomenon has emerged: computational complexity theory appears to be completely unable to classify many of these problems at all.

As just one example of this phenomenon, let us look to the field of quantum cryptography, where some of the most exciting work involving inherently quantum problems is being done today. This is due to the remarkable discovery that certain quantum cryptographic primitives --- such as pseudorandom states and quantum bit commitments — are sufficient for a wide array of cryptographic applications, and yet appear to be weaker than traditional ``minimal’’ cryptographic assumptions such as one-way functions or pseudorandom generators (PRGs).
\paragraph{Pseudorandom states.} Of these quantum primitives, we focus on single-copy pseudorandom states (PRSes), introduced by Ji, Liu, and Song~\cite{JLS18},
which can be seen as a quantum analogue of PRGs.\footnote{\cite{JLS18} actually required a PRS to satisfy a stronger ``many-copy'' security notion, and subsequent works studied the weaker notion of single-copy security (e.g.,~\cite{MY22}). We will not consider many-copy security in this work, but briefly mention that any polynomial-copy PRS can be broken given a \emph{single query} to a $\mathsf{PSPACE}$ oracle~\cite{Kre21}. In contrast, our connection to the unitary synthesis problem makes essential use of the single-copy security notion.} Classically, a PRG is a set of $K \ll N \coloneqq 2^n$ efficiently computable $n$-bit strings $\{x_k\}_{k \in [K]}$
in which a string $x_{\bk}$ drawn uniformly at random from the set is computationally indistinguishable from a truly random $n$-bit string.
Quantumly, a (single-copy) PRS is a set of $K \ll N$ efficiently computable $n$-qubit quantum states $\{\ket{\psi_k}\}_{k \in [K]}$
in which a state $\ket{\psi_{\bk}}$ drawn uniformly at random from the set is computationally indistinguishable from a Haar random $n$-qubit state.
Single-copy PRSes are known to imply the existence of quantum bit commitments \cite{Yan22,MY22,BCQ23}, which are a key ingredient in many cryptographic protocols, ranging from zero-knowledge proof systems~\cite{BG22,GJMZ23} to secure multiparty computation~\cite{GLSV21,BCKM21,AQY22}.

With these definitions in mind, what can we say about the \emph{computational complexity} of breaking cryptographic pseudorandomness? Classically, it is easy to see that secure PRGs do not exist if $\PTIME = \NP$.
~In fact, there is a polynomial-time black-box (Turing, or even Karp) reduction $A^{(\cdot)}$ which can break PRGs given oracle access to a function $f:\{0, 1\}^* \rightarrow \{0, 1\}$ that decides an $\NP$-complete language.
~This explains why proving the existence of unconditionally secure PRGs has so far been unsuccessful,
as doing so would imply the breakthrough complexity theoretic lower bound $\PTIME \neq \NP$.

In the quantum setting, what can we say about the computational complexity of breaking a PRS?
Is there a complexity assumption that we can make,
such as $\BQP = \QMA$,
which would imply that PRSes can be broken in polynomial-time? The answer to this question is currently unknown, and the difficulty stems from the fact that the computational task associated with breaking a PRS is an inherently quantum problem. In particular, the adversary’s goal is to distinguish between a pseudorandom state and a Haar random state, given one of the two at random---a quantum-input, classical-output task.
On the other hand, traditional complexity classes such as $\PTIME$ and $\PSPACE$, and even quantum complexity classes such as $\QMA$, only capture problems with classical inputs. For example, even though the witness for a $\QMA$ statement is a quantum state, the \emph{input} to the problem is always a classical string, such as the description of a local Hamiltonian. 

\paragraph{How hard is it to break quantum cryptography?} As a result of this mismatch between classical-input and quantum-input problems, it is not at all clear how breaking a PRS is related to traditional complexity assumptions. For example, a recent work of Kretschmer, Qian, Sinha, and Tal~\cite{KQST23} has shown that the existence of PRSes is independent of the $\PTIME$-versus-$\NP$ question, at least in the oracle setting, by constructing an oracle relative to which PRSes exist but $\PTIME = \NP$.
~However, \cite{KQST23} derives security of their candidate PRS from the hardness of an oracle problem, $\mathsf{OR}\circ \mathsf{FORRELATION}$, which is easily solvable in $\mathsf{PSPACE}$. Despite this, it is not clear whether a $\mathsf{PSPACE}$ oracle should be powerful enough to break every PRS — in fact, it is not even clear that an oracle for the \emph{halting problem} would suffice. 

This raises a tantalizing question: what if the existence of PRSes is independent of traditional complexity altogether? Could we show that breaking a PRS does not black-box reduce to deciding any language?
~Let us now relate this back to the Unitary Synthesis Problem. Given a PRS, there always exists a unitary $U$ which one could use to break the PRS if one could implement it efficiently,
namely any unitary which maps $\mathrm{span}\{\ket{\psi_1}, \ldots, \ket{\psi_K}\}$ to $\mathrm{span}\{\ket{1}, \ldots, \ket{K}\}$. 
~If an efficient quantum oracle circuit $A^{(\cdot)}$ can synthesize such a $U$ given oracle access to some Boolean function $f$, then the PRS can be efficiently broken relative to $f$. 

Our second main result is to rule out any single-query algorithm for this task, relative to a random oracle.

\begin{theorem}[informal, see \cref{thm:main-prs-body}]\label{thm:main-prs-intro}
 Relative to a random oracle, there exists a PRS (and a quantum bit commitment scheme) secure against all one-query oracle algorithms $A^{f}$ for \emph{every Boolean function} $f$. 
\end{theorem}
\noindent \cref{thm:main-prs-intro} offers the strongest evidence to date that the security of PRSes
might be independent of \emph{all} of traditional computational complexity. Our two results, when taken together, demonstrate the close connection between the Unitary Synthesis Problem
and the security of PRSes;
as we will see below, we essentially prove these two results simultaneously,
because in constructing a PRS which cannot be broken with one query,
we are implicitly constructing a unitary which cannot be synthesized with one query.

\subsection{Our approach}

We prove \cref{thm:main-intro,thm:main-prs-intro} by analyzing an oracle version of the single-copy PRS security game, which we call the ``Oracle State Distinguishing Game'' (see \cref{sec:game-def}).
~To state this task, let us define two pieces of relevant notation.
First, given a Boolean function $h:\{0,1\}^n\rightarrow \{\pm 1\}$, we define the corresponding \emph{binary phase state} as
\begin{equation*}
        \ket{\psi_h} \coloneqq \frac{1}{\sqrt{N}} \cdot \sum_{x\in \{0,1\}^n} h(x) \cdot \ket{x}.
    \end{equation*}
Next, a \emph{function family} is a function
    $\prs:[K] \times \{0, 1\}^n \rightarrow \{\pm 1\}$.
    We think of $\prs$ as defining a family of $K$ Boolean functions as follows:
    for each $1 \leq k \leq K$, we let $R_k:\{0, 1\}^n \rightarrow \{\pm 1\}$ be the function $R_k(\cdot) \coloneqq R(k, \cdot)$. In general, we require $K \ll N$; a typical setting will be $K = N/2$.
    
\begin{definition}[Oracle State Distinguishing Game]
    Let $\bprs: [K] \times \{0, 1\}^n \rightarrow \{\pm 1\}$ be a uniformly random function family.
    The \emph{Oracle State Distinguishing Game} involves two parties, a challenger and an adversary.
    The adversary is modeled as an oracle circuit $A^{(\cdot)}$ which is allowed to query an arbitrary Boolean function $f$ depending on $\bprs$.
    The game is played as follows.
    \begin{enumerate}
    \item The challenger samples a random bit $\bb \in \{0, 1\}$. \item The challenger generates a random $n$-qubit state $\ket{\bpsi}$ in one of two ways:
    \begin{itemize}
        \item[$\circ$] If $\bb = 0$, the challenger samples a uniformly random $\bk \sim [K]$ and generates $\ket{\bpsi} \coloneqq \ket{\psi_{R_{\bk}}}$,
        the binary phase state corresponding to the Boolean function $\bprs_{\bk}$.
        \item[$\circ$] If $\bb = 1$, the challenger samples a uniformly random Boolean function $\bh:\{0,1\}^n \rightarrow \{\pm 1\}$ and sets $\ket{\bpsi} \coloneqq \ket{\psi_{\bh}}$, the binary phase state corresponding to $\bh$.
    \end{itemize}
    \item The challenger sends $\ket{\bpsi}$ to the adversary. 
    \item The adversary runs the oracle circuit $A^f$ on $\ket{\bpsi}$ and outputs a bit $\bb' \in \{0, 1\}$.
    \item If $\bb' = \bb$, then the adversary wins. Otherwise, they lose.
    \end{enumerate}
\end{definition}

Intuitively, the function family $\bprs$
specifies a family of pseudorandom states $\{\ket{\psi_{\bprs_k}}\}_{k \in [K]}$,
and the adversary's goal is to distinguish a randomly chosen state from this from a uniformly random binary phase state $\ket{\psi_{\bh}}$. As discussed above, an algorithm for the Unitary Synthesis Problem
yields a successful adversary for the Oracle State Distinguishing Game,
and so a query lower bound for the Oracle State Distinguishing Game
implies a query lower bound for the Unitary Synthesis Problem.
We show the following lower bound for the Oracle State Distinguishing Game.

\begin{theorem}\label{thm:main-distinguishing-intro}
    Suppose that $A^{f}$ is a one-query oracle circuit that achieves advantage $\epsilon$ in the Oracle State Distinguishing Game. Then, $A^f$ must make a query of size at least $\ell = \Omega(K\epsilon^2)$ bits.
\end{theorem}

This lower bound implies that for typical settings of $K$ (such as $K = N/2$), to achieve a non-negligible distinguishing probability, the adversary's query must have length exponential in~$n$; in particular, a superpolynomial-length query is required whenever $\epsilon \geq n^{\omega(1)} /\sqrt{K}$. This dependence on $K$ is optimal, as there are polynomial-time 1-query algorithms which do achieve distinguishing advantage $\Omega(1/\sqrt{K})$ (see~\cref{sec:appendix-attack}).

As discussed above, \Cref{thm:main-distinguishing-intro} immediately implies \cref{thm:main-intro}, our one-query lower bound for the Unitary Synthesis Problem.
In fact, since we show that the adversary's distinguishing advantage is negligible, this gives a unitary $U_{\bprs}$ which is hard to synthesize even in an extremely weak sense: no efficient one-query algorithm $A^{f}$ can correctly implement any unitary that even remotely approximates the behavior of $U_{\bprs}$.
In addition, since the Oracle State Distinguishing Game is an oracle analogue of the security game for a single-copy PRS family, standard techniques (see \cref{sec:prs}) allow us to transform \Cref{thm:main-distinguishing-intro} into a proof that, relative to a random oracle, there exist PRS families and quantum bit commitment schemes secure against all one-query adversaries.
This gives \cref{thm:main-prs-intro}.

These results demonstrate the usefulness of the Oracle State Distinguishing Game as a means for studying the Unitary Synthesis Problem,
and we believe that it is also a useful avenue for proving stronger lower bounds against algorithms which use more than one query.
To this end, we make the following conjecture.

\begin{conjecture}[Strong Non-Synthesis Conjecture]\label{conjecture:main}
    For all  $K \geq n^{\omega(1)}$, any polynomial-query oracle algorithm $A^{f}$ wins the Oracle State Distinguishing Game with advantage at most $\negl(n)$. 
\end{conjecture}
\noindent
A proof of \Cref{conjecture:main} would imply a negative resolution to the Unitary Synthesis Problem.
In addition, it would imply the existence of single-copy PRSes (and thus, quantum bit commitments) secure against all efficient polynomial-query adversaries, relative to a random oracle. In other words, computationally secure quantum cryptography would not black-box imply the existence of any hard language. We note that the lower bound $K \geq n^{\omega(1)}$ in \Cref{conjecture:main} is necessary; as discussed above, if $K = \poly(n)$, then there is a simple attack that achieves $1/\sqrt{K} = 1/\poly(n)$ advantage.\footnote{In fact, there is another attack that achieves advantage close to $1$ in this regime, based on the LMR algorithm \cite{LMR14,Yue22b}. The adversary can make a single call to its $\bprs$-dependent oracle $f$ to generate $m = \poly(n)$ copies of each state $\ket{\psi_{\bprs_k}}$. Then for each $1 \leq k \leq K$, the adversary can test if the challenge state $\ket{\bpsi}$ is equal to $\ket{\psi_{\bprs_k}}$ by measuring $\ket{\bpsi} \otimes \ket{\psi_{\bprs_k}}^{\otimes m}$ with $\{\Pi_{\mathrm{sym}}, \Id - \Pi_{\mathrm{sym}}\}$, where $\Pi_{\mathrm{sym}}$ is the projector onto the symmetric subspace.
If they are \emph{not} equal, doing so will only perturb the state $\ket{\bpsi}$ slightly, allowing the adversary to reuse it for further tests.} 

In~\cref{subsec:future}, we state weaker conjectures which correspond to simpler cases of~\cref{conjecture:main}. In particular, in \cref{conj:simple}, we give a self-contained mathematical conjecture which corresponds to the simplest class of oracle adversaries that we do not know how to rule out.


\paragraph{Additional remarks.} We make two final observations about the Oracle State Distinguishing Game.
First, note that the adversary's task is to perform a measurement $\{M_0, M_1\}$ which distinguishes between the two cases of the game.
In particular, writing $U_{\bprs}$ for the unitary written above,
the adversary would like to carry out the measurement specified by the two projectors
\begin{equation*}
    M_0 := U_{\bprs}^\dagger \cdot (\ketbra{1} + \cdots + \ketbra {K}) \cdot U_{\bprs}
    \quad \text{and} \quad
    M_1 := \Id - M_0.
\end{equation*}
This is an example of a \emph{measurement synthesis} task,
an inherently quantum problem
in which the input is quantum but the output is classical.
Measurement synthesis has been discussed
much less than state synthesis and unitary synthesis
in the literature
(the only work we are aware of that discusses it is~\cite{BEM+23}).
However, our results suggest that it is measurement synthesis
that is the hard problem at the core of unitary synthesis.
Combined with the fact that state synthesis has efficient one-query algorithms~\cite{Ros23},
this suggests that the crucial distinction between classical problems
and inherently quantum problems is whether the input, and not necessarily the output, is classical or quantum.

Second, we note that the Oracle State Distinguishing Game is fairly robust to the precise distribution of states used to specify it.
For example, rather than specifying the game in terms of random binary phase states, we could have specified it using Haar random states.
In this version of the game, $K$ independent Haar random states $\ket{\bpsi_1}, \ldots, \ket{\bpsi_K}$ are sampled in advance.
Then the adversary is given either ($b=0$) one of these $K$ states sampled uniformly at random, or ($b=1$) a new Haar random state $\ket{\bpsi}$, and asked to distinguish between these two cases.
Though we do not prove it here,
our lower bound in \Cref{thm:main-distinguishing-intro} also holds for this variant of the Oracle State Distinguishing Game.
One nice property of this distribution is that hardness of the Oracle State Distinguishing Game for this distribution directly implies hardness of Unitary Synthesis for a Haar-random unitary $U$. We refer the reader to~\cref{sec:unitary-synthesis} for further discussion.

\subsection{Related Work}\label{sec:related-work}
In this section, we elaborate on some works related to the Unitary Synthesis Problem and our results. We discuss (1) prior lower bounds, (2) positive results on the closely-related state synthesis problem, and (3) related work in unitary complexity theory.

\subsubsection{Lower bounds for unitary synthesis}

The best known prior lower bound for the Unitary Synthesis Problem
comes from the original paper on this topic by Aaronson and Kuperberg~\cite{AK07}.
To understand their lower bound, let us first make more explicit the computational model we are assuming for our oracle circuit $A^{(\cdot)}$.
A general oracle circuit $A^{(\cdot)}$ may wish to make use of additional \emph{ancilla qubits},
in which case it will be structured as follows:
it will have an $n$-qubit input register and an input ancilla register initialized to $\ket{0^a}$,
as well as an $n$-qubit output register and an $a$-qubit output ``junk'' register.
Indeed, if $A^{(\cdot)}$ does \emph{not} have ancillas,
then it is unable to query any oracle $f$ on inputs of length greater than $n$,
which turns out to make $A^{(\cdot)}$ quite weak.
This is because for such an $A^{(\cdot)}$,
the number of possible unitaries you can synthesize when ranging over all functions $f$ is bounded by $2^{2^n}$, which is simply not enough to ``cover all unitaries'' by a counting argument.
(See \cref{sec:appendix-counting} for a simple lower bound along these lines.)

Now we can state the Aaronson and Kuperberg~\cite{AK07} lower bound.
They showed a one-query lower bound against any oracle circuit $A^{(\cdot)}$ which has the following property: for every choice of oracle $f$, the oracle circuit $A^{f}$ is required to \emph{exactly implement} an $n$-qubit unitary on its first $n$ qubits. Mathematically, this means that for any $n$-qubit state $\ket{\psi}$, we must have that
\begin{equation}
    A^f \cdot \ket{\psi} \otimes \ket{0^a}
    = (U_f \cdot \ket{\psi}) \otimes \ket{\mathsf{junk}_f}, \label{eq:AK-model}
\end{equation}
where $U_f$ is some $n$-qubit unitary which depends on $f$,
and $\ket{\mathsf{junk}_f}$ is some $a$-qubit junk state which depends on $f$.
This defines a class of oracle algorithms that turns out to be highly restrictive, for several reasons. We list two.
\begin{enumerate}
    \item
    The class excludes algorithms $A^f$ such that \Cref{eq:AK-model} only holds approximately, even with inverse-exponential precision. 
    \item The model requires that the circuit $A^f$ implements a unitary for \emph{every} oracle $f$.
    On the other hand, there are many examples of oracle circuits not belonging to this class which expect the oracle $f$ to be ``properly formatted'', and do not synthesize any unitary if $f$ is not properly formatted.

\end{enumerate}
    To elaborate on (2), consider the following simple attack: the oracle circuit $A^f$ queries $f$ to learn an $\ell$-bit classical string $s$ on an ancilla space, and then applies an $n$-qubit unitary $U_s$ that depends on $s$. By the Bernstein-Vazirani trick, $A^f$ can learn an $\ell$-bit string $s$ in a single query by first preparing the uniform superposition on $\ell$ qubits, then querying the Boolean function $f_{s}(x) \coloneqq s \cdot x$, and finally applying a Hadamard transform. Even though this oracle circuit $A^f$ always implements a unitary on the first $n$ qubits when $f$ computes an inner-product function, this is not guaranteed in general: for arbitrary $f$, the oracle circuit may obtain a superposition over different $s$, in which case the operation on the first $n$-qubits is not guaranteed to be unitary.

Indeed, Aaronson and Kuperberg are able to prove their lower bound against this class by a \emph{counting argument}: they prove that the number of distinct unitaries that a one-query oracle circuit $A^{(\cdot)}$ in this class can synthesize, ranging over all oracles $f$, is at most $4^{2^n}$~\cite[Theorem 6.7]{AK07}, irrespective of the number of ancilla qubits $a$. Unfortunately, as we discuss in \cref{sec:tech-adversary,sec:appendix-counting}, these types of counting arguments are insufficient to prove a general query lower bound.


A more recent lower bound, due to Rosenthal~\cite{Ros22}, shows that unitary synthesis is hard relative to a \emph{state synthesis} oracle. Roughly speaking, this lower bound states that synthesizing a unitary $U$ requires roughly $2^{n/2}$ queries to an oracle that, on any classical input $\ket{x}$, for $x \in \{0, 1\}^n$, outputs the state $\ket{x} \tensor U\ket{x}$. This shows that the power to produce any state of the form $U\ket{x}$ is insufficient to implement $U$ efficiently. However, the technique says little about the problem of synthesizing $U$ relative to an \emph{arbitrary} function oracle $f$.

\subsubsection{Relationship to state synthesis.}

Let us contrast our one-query lower bound for the Unitary Synthesis Problem
with the state of affairs for a related problem known as \emph{state synthesis}.
State synthesis is the task of implementing a quantum circuit that outputs a specified $n$-qubit quantum state $\ket{\psi}$ when run on the all-0's input. Alternatively, one can view state synthesis as an easier version of unitary synthesis, where the goal is merely to implement the unitary correctly on the all $0$'s input, rather than on \emph{all} possible inputs.

Like unitary synthesis, state synthesis requires large quantum circuits:
it can be shown via counting arguments
that there exist worst-case states on $n$ qubits that require circuits of size $\Omega(2^n/n)$ to compute approximately
(see the excellent discussion of this in~\cite[Section 1.3.4]{Ros23b}).

It turns out, however, that state synthesis becomes easy if Boolean functions are easy~\cite{Aar16,INN+22,Ros23}.
In particular, Rosenthal's state-of-the-art result~\cite{Ros23} gives a quantum-polynomial time oracle algorithm $A^{(\cdot)}$ such that for any $n$-qubit pure state $\ket{\psi}$,
there exists a Boolean function $f:\{0,1\}^m \rightarrow \{\pm 1\}$ such that $A^{f}(1^n)$ makes \emph{one} quantum query to $f$
and outputs $\ket{\psi}$ up to inverse exponential precision. For some intuition behind this result,
observe that \emph{binary phase states} $\frac{1}{\sqrt{2^n}} \cdot \sum_{x \in \{0, 1\}^n} f(x)\cdot\ket{x}$ are trivial to synthesize with one query: simply prepare the uniform superposition $\frac 1 {\sqrt{2^n}} \cdot \sum_{x \in \{0, 1\}^n} \ket{x}$, and make one query to the phase oracle $\calO_f: \ket{x} \rightarrow f(x) \cdot \ket{x}$.
It turns out that worst-case states can then be synthesized via a careful reduction to the binary phase state case. This can be viewed as a one-query reduction from the task of state synthesis to the problem of computing an arbitrary Boolean function. In contrast, our main result shows that no such reduction is possible for unitary synthesis.

\subsubsection{Quantum cryptography and unitary complexity}

A connection between (plain model) quantum cryptography and the Unitary Synthesis Problem was recently discovered by Kretschmer \cite{Kre23},
who showed that if the Unitary Synthesis Problem is resolved in the positive,
then showing the existence of a secure PRS implies that $\BPP \neq \NEXP$.
This result says that traditional complexity theory \emph{does} have something to say about the existence of PRSes,
but only if unitaries are easy to synthesize. 

Beyond ``traditional complexity theory,'' a very recent and intriguing line of work has introduced a complexity theory of inherently quantum problems,
with complexity classes corresponding to both state synthesis problems and unitary synthesis problems~\cite{RY22,INN+22,Ros23,MY23,BEM+23,DGLM23}.
As above, this line of work argues that traditional complexity theory is ill-equipped to address the complexity of inherently quantum problems,
as traditional complexity theory is only about classical-input, classical-output problems, i.e.,\ functions $f:\{0, 1\}^* \rightarrow \{0, 1\}$.
In this new theory of unitary complexity,
the existence of secure PRSes \emph{does} have complexity theoretic implications
(in particular, it implies the separation $\mathsf{unitaryBQP} \neq \mathsf{unitaryPSPACE}$).

An important open direction is to study the relationship between these new inherently quantum complexity theories and the traditional ``classical'' complexity theory.
Interestingly, Kretschmer's result above \cite{Kre23} suggests that these seemingly different complexity theories might be closer than they first appear, if the Unitary Synthesis Problem is resolved in the positive.
In particular, his result, stated more broadly, is the following: suppose the Unitary Synthesis Problem is resolved in the positive. Then $\mathsf{unitaryBQP} \neq \mathsf{unitaryPSPACE}$ implies that $\BPP \neq \NEXP$.
In this light, our \cref{thm:main-intro}, providing negative evidence for the Unitary Synthesis Problem, can also be interpreted as providing positive evidence that these complexity theories are in fact distinct.

\subsection{Organization}

The remainder of this paper is organized as follows.
\Cref{sec:tech-overview} gives a technical overview of our proofs.
\Cref{sec:notation} includes preliminary details about oracle circuits,
building towards a simple normal form for these circuits that we will use in our proofs.
In \Cref{sec:lower-bound}, we give the proof of our main result,
the one-query lower bound for the Oracle State Distinguishing Game,
which we then use in \Cref{sec:prs} to show the existence of secure PRSes and quantum bit commitments relative to a random oracle.
\Cref{sec:appendix-matrix-chernoff} includes a second proof of our main result with slightly worse parameters.
In \Cref{sec:appendix-counting}, we show a counting lower bound against even many-query oracle circuits which can only compute a small number of distinct unitaries, generalizing the one-query lower bound of Aaronson and Kuperberg~\cite{AK07}.\ Finally, in \cref{sec:appendix-attack}, we give a one-query algorithm to match our main lower bound (\cref{thm:main-at-end-of-proof}) in its dependence on $K$. 

\subsection{Acknowledgements}

We thank Scott Aaronson, Prabhanjan Ananth, Zvika Brakerski, Ran Canetti, Lijie Chen, Shafi Goldwasser, Louis Golowich, Tarun Kathuria, William Kretschmer, Tony Metger, Sidhanth Mohanty, Anand Natarajan, Ryan O'Donnell, Luowen Qian, Prasad Raghavendra, Greg Rosenthal, Nick Spooner, Nikhil Srivastava, Vinod Vaikuntanathan, Ramon van Handel, Umesh Vazirani, Thomas Vidick, Henry Yuen, and Mark Zhandry for helpful discussions.

We thank Scott Aaronson and Greg Rosenthal for comments on an earlier draft of this paper.

\newpage

\section{Technical overview}\label{sec:tech-overview}

We will sketch the proof of \cref{thm:main-distinguishing-intro}, beginning by describing our mathematical model for single-query adversaries in \Cref{sec:tech-adversary}.
Following this, we will develop our proof strategy in the context of three different and increasingly complicated types of adversaries.
     First, in \Cref{sec:tech-advice}, we will look at adversaries which use their one query to prepare a quantum advice state.
     Next, in \Cref{sec:tech-no-isometry}, we will look at adversaries which have no ancilla qubits and do not apply any gates prior to their oracle query.
     Finally, in \Cref{sec:tech-general}, we will look at general single-query adversaries.
     
\subsection{Modeling the adversary}\label{sec:tech-adversary}

A single-query adversary can be modeled as a quantum circuit with an input register of $n$ qubits and an ancilla register of $a$ qubits, for a size of $m = n + a$ total qubits.
Given an $n$-qubit input state $\ket{\psi}$, the adversary acts as follows.
\begin{enumerate}
    \item The adversary will initialize its ancilla qubits to $\ket{0^a}$. Then, it applies a unitary $U$ to $\ket{\psi} \ket{0^a}$. Equivalently, it applies the \emph{isometry} $V \coloneqq U \cdot (\Id \otimes \ket{0^a})$ to $\ket{\psi}$.
    
    \item It then queries its oracle $f:\{0, 1\}^m \rightarrow \{\pm 1\}$. This applies the unitary $\oracle_f$ to its state, where $\oracle_f$ is the unitary defined as
    \begin{align*}
        \calO_f\cdot \ket{z} = f(z)\cdot \ket{z}, \ \ \text{ for all } z \in \{0,1\}^m.
    \end{align*}
    \item Finally, the adversary performs a binary projective measurement $\{\meas, \Id-\meas\}$ on its state. This produces a measurement outcome $\bb' \in \{0, 1\}$,
    which it outputs as its guess.
\end{enumerate}
After the oracle, the adversary's state is $\oracle_f \cdot V \cdot \ket{\psi}$.
Thus, the probability it outputs $\bb' = 0$ is
\begin{equation}\label{eq:prob-of-zero}
    \Vert \Pi \cdot \oracle_f \cdot V \cdot \ket{\psi} \Vert^2
    = \bra{\psi} \cdot V^\dagger \cdot \oracle_f \cdot \Pi \cdot \oracle_f \cdot V \cdot \ket{\psi}.
\end{equation}
Intuitively, one should think of the size $m$ as ``small'', say $m = \poly(n)$.
This is because $m$ is also the length of the adversary's oracle query, and it is necessary for us to assume a bound on the query length so that the problem remains nontrivial.
Otherwise, there is a simple attack based on the Bernstein-Vazirani algorithm~\cite{BV97} which solves the problem using a single extremely large query of length $K \cdot N$, which we describe below.

\begin{example}[A one-query attack with exponential query size]
     Define $f_R: \{0,1\}^{KN}\rightarrow \{\pm 1\}$ so that $f_R(z) = (-1)^{z \cdot r}$, where $r \in \{0,1\}^{KN}$ is a binary vector representation of $\prs$. Then if the adversary queries $f_R$ on the uniform superposition over all $z \in \{0,1\}^{KN}$, it obtains the state
    \begin{align*}
    \frac{1}{\sqrt{KN}} \cdot \sum_{z \in \{0,1\}^{KN}} f_R(z)\cdot  \ket{z} = \frac{1}{\sqrt{KN}} \cdot \sum_{z \in \{0,1\}^{KN}} (-1)^{z \cdot r}\cdot \ket{z},
    \end{align*}
    The state on the right-hand side is simply the Hadamard transform of $\ket{r}$, and thus the adversary can obtain the entire truth table of $\prs$.
\end{example}

As it turns out, once we assume our adversary has ``small'' query length, it can be converted to one with ``small'' size $m$ as well (see \Cref{sec:adversary-space}). Hence, we may assume that the adversary's oracle is applied to all $m$ qubits.
We will now carry out the following change in notation that will be applied throughout the paper:
to simplify notation,
we will set $N := 2^n$ and $M:= 2^m$
and associate the set $\{1, \ldots, N\}$ with $\{0, 1\}^n$ and $\{1, \ldots, M\}$ with $\{0, 1\}^m$.
As a result, a ``Boolean'' function is now formatted as $h:[N]\rightarrow \{\pm 1\}$ and is associated with the phase state
\begin{equation*}
    \ket{\psi_h} = \frac{1}{\sqrt{N}} \cdot \sum_{x=1}^N h(x) \cdot\ket{x},
\end{equation*}
a function family is formatted $R:[K]\times[N] \rightarrow \{\pm 1\}$,
and an oracle function is formatted $f:[M]\rightarrow \{\pm 1\}$.
With this change in notation, the input state space becomes $\C^N$ and the adversary's state space becomes $\C^M$,
so that (i) the isometry $V$ maps $\mathbb{C}^N$ to $\mathbb{C}^{\qdim}$, and (ii) the oracle unitary $\calO_f$ has dimension $\qdim \times \qdim$.
Thus, for any particular function family $R: [K] \times [N] \rightarrow \{\pm 1\}$, and any one-query adversary $A^{(\cdot)}$, the maximum achievable distinguishing advantage is equal to
\begin{align*}
    \max_{f : [\qdim] \rightarrow \{\pm 1\}} \Big|\E_{\bk \sim [K]} \Big[ \Pr[A^f( \ket{\psi_{R_{\bk}}}) \text{ outputs } ``0"]\Big] - \E_{\bh} \Big[ \Pr[A^f( \ket{\psi_{\bh}}) \text{ outputs } ``0"]\Big]\Big|,
\end{align*} 
where here and throughout this section
we are writing $\bh$ for a uniformly random Boolean function $\bh : [N] \rightarrow \{\pm 1\}$.
Substituting in \Cref{eq:prob-of-zero}, this is equal to

\begin{align}\label{eq:used-to-have-a-box}
    \max_{f}\Big|\E_{\bk \sim [K]} \Big[ \bra{\psi_{R_{\bk}}} \cdot V^\dagger \cdot \calO_f \cdot \meas \cdot \calO_f \cdot V \cdot \ket{\psi_{R_{\bk}}}\Big] - \E_{\bh} \Big[ \bra{\psi_{\bh}} \cdot V^\dagger \cdot \calO_f \cdot \meas \cdot \calO_f \cdot V \cdot \ket{\psi_{\bh}} \Big] \Big|.
\end{align}
Our goal is to prove \cref{thm:main-distinguishing-intro}, which can be phrased more formally as follows.

\begin{theorem}[\cref{thm:main-distinguishing-intro}, rephrased]
    Let $A^{(\cdot)}$ be a single-query adversary for the Oracle State Distinguishing Game that acts on an $M$-dimensional Hilbert space.
    Then with high probability over the choice of $\bR: [K]\times [N]\rightarrow \{\pm 1\}$, $A^{(\cdot)}$ achieves maximum distinguishing advantage at most 
        $O\Big(\sqrt{\frac{\log \qdim}{K}}\Big)$.
\end{theorem}

Now let us now briefly discuss one potential approach for proving \Cref{thm:main-distinguishing-intro}: counting arguments.
These are based on the simple observation is that the distinguishing advantage is easily upper-bounded for any \emph{fixed} oracle $f$, which corresponds to an adversary that does not depend on $\bprs$. This can be argued using standard concentration of measure tools from probability theory, and the resulting concentration bound one can show is \emph{extremely good}: in particular, the probability that a fixed $A^f$ has distinguishing advantage at least $\epsilon$ is at most $2^{-\Omega(\epsilon^2 KN)}$.
Given this degree of concentration, it is tempting to simply union bound over all choices of $f$ to upper-bound the maximum distinguishing advantage; this is known as a counting argument. Unfortunately, this approach quickly begins to fail as the adversary's space grows: the number of possible functions $f$ is $2^{\qdim}$, where $\qdim$ is potentially much larger than $KN$. Recall that $KN \leq N^2 = 2^{2n}$, while $\qdim$ could be (at least) $2^{\poly(n)}$, for an arbitrary $\poly(n)$. Thus, this type of counting argument cannot give a general one-query lower bound.
That said, it \emph{can} rule out some interesting special cases of adversaries, which we discuss in \cref{sec:appendix-counting}. 
Finally, we note that there is a more powerful version of counting arguments known as \emph{chaining} (cf.\ \cite[Chapter 8]{Ver18}),
but we were unable to successfully apply chaining arguments to this problem.

In the next few subsections,
we will describe an alternative approach for bounding the maximum distinguishing advantage across all choices of $f$ simultaneously via matrix concentration inequalities.

\subsection{Adversaries with quantum advice}\label{sec:tech-advice}

We begin with the simple but conceptually useful special case of one-query adversaries, namely those that use the query to $f$ to synthesize an $f$-dependent \emph{advice state}. In other words, the adversary acts as follows.
\begin{enumerate}
    \item First, it applies an isometry $V$ that acts by appending a fixed $m$-qubit state $\ket{\phi}$.
    Thus, the $n$-qubit input state $\ket{\psi}$ is mapped to the $(n+m)$-qubit state $\ket{\psi}\tensor \ket{\phi}$.
    (We are abusing notation in this subsection by writing $m$ only for the qubits in the advice state, rather than for all of the qubits.
    We will return to the normal definition of $m$ in \Cref{sec:tech-no-isometry,sec:tech-general} below.)
    \item Next, it makes an oracle query $\calO_f$ that acts as the identity on the input state $\ket{\psi}$ and only modifies $\ket{\phi}$. Then the adversary's state becomes $\ket{\psi} \otimes \ket{\phi_f}$, where $\ket{\phi_f}$ is some $f$-dependent state.
\end{enumerate}
In total, for such an adversary, $\calO_f \cdot V \cdot \ket{\psi} = \ket{\psi}\otimes \ket{\phi_f}$.
Attacks of this form can synthesize many kinds of states: for example, if $\ket{\phi}$ is a uniform superposition, then $\ket{\phi_f}$ can be any binary phase state. 
(We remark that there are techniques in the cryptography literature for proving lower bounds against quantum advice \cite{HXY19,CLQ20,CGLQ20,Liu23}. However, the techniques seem to be highly tailored to the advice setting and are not related to our approach.)

Supposing the adversary works in this manner,
we can compute its maximum distinguishing advantage
on a uniformly random $\bprs:[K]\times[N]\rightarrow \{\pm 1\}$ as
\begin{align*}
    \max_{f:[M]\rightarrow \{\pm 1\}}\left|\E_{\bk \sim [K]} \Big[ \bra{\psi_{\bprs_{\bk}}} \bra{\phi_f}\cdot \meas \cdot  \ket{\psi_{\bprs_{\bk}}} \ket{\phi_f} \Big] - \E_{\bh} \Big[ \bra{\psi_{\bh}}\bra{\phi_f}   \cdot \meas  \cdot \ket{\psi_{\bh}} \ket{\phi_f}  \Big] \right|,
\end{align*}
by \Cref{eq:used-to-have-a-box}.
The benefit of focusing on advice states is that we can factor out the $f$-dependent term $\ket{\phi_f}$ from each expectation. To do so, for any Boolean function $h:[N] \rightarrow \{\pm 1\}$, let $\meas_h$ denote the $\qdim \times \qdim$-dimensional matrix 
\begin{align*}
    \meas_{h} \coloneqq (\bra{\psi_{h}} \otimes \Id) \cdot \meas \cdot (\ket{\psi_{h}} \otimes \Id).
\end{align*}
Note that $0 \leq \meas_h \leq \Id$, since $\ket{\psi_h}$ is a unit vector and $\meas$ is a projection.
Then we can rewrite the distinguishing advantage as
\begin{align}
    & \hspace{.5cm} \max_{f:[M]\rightarrow \{\pm 1\}}\Big|\E_{\bk \sim [K]} \Big[ \bra{\phi_f} \cdot \Pi_{\bprs_{\bk}} \cdot \ket{\phi_f} \Big] - \E_{\bh} \Big[ \bra{\phi_f} \cdot \Pi_{\bh} \cdot \ket{\phi_f} \Big] \Big| \nonumber\\
    &= \max_{f:[M]\rightarrow \{\pm 1\}} \Big| \bra{\phi_f} \cdot \Big( \E_{\bk \sim [K]} \big[ \Pi_{\bprs_{\bk}} \big] - \E_{\bh} \big[  \Pi_{\bh}\big] \Big) \cdot \ket{\phi_f} \Big| \label{eq:factored-advice}.
\end{align}
Since $\ket{\phi_f}$ is a unit vector, we can upper bound this by a maximum over all unit vectors, i.e.
\begin{align*}
    (\ref{eq:factored-advice}) &\leq \max_{ \norm{\ket{v}} = 1} \Big| \bra{v} \cdot \Big( \E_{\bk \sim [K]} \big[ \Pi_{\bprs_{\bk}} \big] - \E_{\bh} \big[  \Pi_{\bh}\big] \Big) \cdot \ket{v} \Big| \\
    &= \Big\Vert\E_{\bk \sim [K]} \big[ \Pi_{\bprs_{\bk}} \big] - \E_{\bh} \big[\Pi_{\bh}\big]\Big\Vert_{\mathrm{op}}\\
    &= \Big\Vert\E_{\bk \sim [K]} \big[Z_{\bprs_{\bk}} \big]\Big\Vert_{\mathrm{op}}, \ \ \text{for} \ \ Z_{h} \coloneqq \Pi_{h} - \E_{\bh} \big[ \Pi_{\bh} \big].
\end{align*}
Here, we are writing $\Vert\cdot\Vert_{\mathrm{op}}$ for the operator norm. Thus, we have reduced our problem to bounding the operator norm of the average of $K$ random matrices $Z_{\bR_1}, \ldots, Z_{\bR_K}$. 

We will bound this operator norm using the technique of \emph{matrix concentration}, which generalizes scalar concentration bounds (such as Chernoff-Hoeffding bounds) to the random matrix setting.~
Specifically, the matrix Hoeffding inequality (roughly) says the following (see~\cite{Tro12}, Theorem 1.3 or~\cref{thm:matrix-hoeffding} for the precise statement).

\begin{theorem}[Matrix Hoeffding (informal)] If $K$ independent and identically distributed mean-zero random $D \times D$ Hermitian matrices $\bZ_1,\dots,\bZ_K$ always have bounded operator norm, then with high probability, 
    \begin{align*}
        \Big\Vert\E_{\bk \sim [K]} \bZ_{\bk}\Big\Vert_\operator \leq O\Big(\sqrt{\frac{\log(D\cdot K)}{K}}\Big).
    \end{align*} 
\end{theorem}
\noindent (Note that the scalar Hoeffding bound can be recovered by taking $D = 1$ above.)
To apply the matrix Hoeffding inequality to our problem, we need to verify that when $\bprs:[K]\times[N]\rightarrow \{\pm 1\}$ is uniformly random, our matrices $Z_{\bR_1},\dots,\allowbreak Z_{\bR_K}$ satisfy these properties. Indeed:
\begin{itemize}
    \item[$\circ$] $Z_{\bR_1},\dots,\allowbreak Z_{\bR_K}$ are independent and identically distributed
    since each $\bR_k$ is an independent, uniformly random Boolean function $\bR_k:[N] \rightarrow \{\pm 1\}$.
    \item[$\circ$] For each $1\leq k\leq K$, $Z_{\bR_k}$ has expectation zero: 
    \begin{align*}
        \E_{\bR} \big[ Z_{\bR_k} \big] = \E_{\bR} \big[\Pi_{\bR_k}\big] - \E_{\bh} \big[ \Pi_{\bh} \big] = 0.
    \end{align*}
    \item[$\circ$] For each $1 \leq k \leq K$, the operator norm $\norm{Z_{\bR_k}}_{\mathrm{op}}$ is always bounded by $2$, since 
    \begin{align*}
    \norm{Z_{\bR_k}}_{\mathrm{op}} = \Big\Vert \Pi_{\bR_k} - \E_{\bh} \big[ \Pi_{\bh} \big]\Big\Vert_{\mathrm{op}} \leq \norm{\Pi_{\bR_k}}_{\mathrm{op}} + \Big\Vert\E_{\bh} \big[ \Pi_{\bh} \big]\Big\Vert_{\mathrm{op}} \leq 2.
    \end{align*}
\end{itemize} 
As a result, since in our setting $D = \qdim$, an $\epsilon$-distinguisher requires $\log(\qdim) = \Omega(K\epsilon^2)$, as claimed. In other words, the adversary needs a \emph{huge} advice state to win the distinguishing game.

In summary, our strategy involved identifying a well-behaved quantity that governs the advantage of $A^f$ across all choices of $f$ simultaneously. As we have seen, the operator norm is an example of such a quantity: although bounding the quadratic form $\bra{v} \cdot (\E_{\bk \sim [K]}Z_{\bR_{\bk}} ) \cdot \ket{v}$ for \emph{all} vectors $\ket{v}$ would naively require the concentration of $O(1/\epsilon)^{\qdim}$ different scalars (corresponding to an $\epsilon$-net over $\mathbb C^{\qdim}$), matrix concentration shows that the operator norm behaves as if it has $\qdim$, rather than $2^{\qdim}$, ``independent degrees of freedom''. 

\subsection{Adversaries with a trivial isometry}
\label{sec:tech-no-isometry}

Let us recall \Cref{eq:used-to-have-a-box}, our expression for the adversary's maximum distinguishing advantage:
\begin{align*}
    \max_{f:[\qdim]\rightarrow \{\pm 1\}}\Big|\E_{\bk \sim [K]} \Big[ \bra{\psi_{R_{\bk}}}\cdot V^\dagger \cdot \calO_f \cdot \meas \cdot \calO_f \cdot V \cdot\ket{\psi_{R_{\bk}}}\Big] - \E_{\bh} \Big[ \bra{\psi_{\bh}}\cdot V^\dagger \cdot \calO_f \cdot \meas \cdot \calO_f \cdot V \cdot\ket{\psi_{\bh}} \Big] \Big|.
\end{align*} 
The advice state case above suggests the following approach to bounding this expression:

\begin{enumerate}
    \item Factor the dependence on the oracle $\mathcal O_f$ to the ``outside'' of the expression, and
    \item Rely on a matrix concentration inequality to bound the advantage for all $f$ simultaneously.
\end{enumerate}
Unfortunately, the advice state case does not tell us whether this approach is possible, or how to carry it out, in general. To gain some intuition, we will analyze another simple special case, the case where $V = \Id$, in which the adversary does not use any ancilla qubits and only applies the identity unitary.
In this case, $M = N$, and we will allow the adversary to query an arbitrary oracle $f:[N] \rightarrow \{\pm 1\}$.
Then the because $V = \Id$, the adversary's maximum distinguishing advantage on a uniformly random $\bprs:[K] \times [N] \rightarrow \{\pm 1\}$ is given by
\begin{align}\label{eq:trivial-dist}
    \max_{f : \{0,1\}^n \rightarrow \{\pm 1\}}\Big|\E_{\bk \sim [K]} \Big[ \bra{\psi_{\bprs_{\bk}}} \cdot\calO_f \cdot \meas \cdot \calO_f \cdot \ket{\psi_{\bprs_{\bk}}}\Big] - \E_{\bh} \Big[ \bra{\psi_{\bh}} \cdot\calO_f \cdot \meas \cdot \calO_f \cdot\ket{\psi_{\bh}} \Big] \Big|.
\end{align} 
Towards ``factoring out'' the $\calO_f$ dependence to the outside of the expression, we make use of the fact that any binary phase state $\ket{\psi_h}$ can be written as the product of a diagonal $\{\pm 1\}$-matrix and the uniform superposition state:
\begin{align*}
    \ket{\psi_h} = \frac{1}{\sqrt{N}} \cdot \sum_{x=1}^N h(x) \cdot \ket{x} =   \underbrace{\Bigg(\sum_{x=1}^N h(x)\cdot  \ketbra{x} \Bigg)}_{D_h} \cdot \underbrace{\Bigg(\frac{1}{\sqrt{N}} \cdot \sum_{x=1}^N \ket{x}\Bigg)}_{\ket{+_N}} = D_h \cdot \ket{+_N}.
\end{align*}
The key benefit of this ``diagonal decomposition'' is that the diagonal matrices $D_h$ and $\mathcal O_f$ \emph{commute}, which allows us to rewrite the state $\calO_f \cdot \ket{\psi_h}$ as follows:
\begin{align*}
    \calO_f \cdot \ket{\psi_h}= \calO_f  \cdot D_{h} \cdot\ket{+_N} = D_{h}\cdot \calO_f\cdot \ket{+_N} = D_h \cdot \ket{\phi_f},
\end{align*}
where $\ket{\phi_f}\coloneqq \calO_f\cdot \ket{+_N}$ is the binary phase state corresponding to $f$. Plugging this back into our expression for the maximum distinguishing advantage, we can again employ a spectral relaxation:
\begin{align*}
    \eqref{eq:trivial-dist} ={}&\max_{f : \{0,1\}^n \rightarrow \{\pm 1\}}\Big|\E_{\bk \sim [K]} \Big[ \bra{\phi_f} \cdot D_{\bprs_{\bk}} \cdot \meas \cdot D_{\bprs_{\bk}} \cdot \ket{\phi_f} \Big] - \E_{\bh} \Big[ \bra{\phi_f} \cdot D_{\bh} \cdot \meas \cdot D_{\bh} \cdot\ket{\phi_f}  \Big] \Big|\\
    ={}& \max_{f : \{0,1\}^n \rightarrow \{\pm 1\}}\Big|\bra{\phi_f}\cdot \Big( \E_{\bk \sim [K]} \Big[ D_{\bprs_{\bk}} \cdot \meas \cdot D_{\bprs_{\bk}} \Big] - \E_{\bh} \Big[ D_{\bh} \cdot \meas \cdot D_{\bh} \Big] \Big) \cdot\ket{\phi_f}\Big|\\
    \leq{}& \Big\Vert \E_{\bk \sim [K]} \Big[ D_{\bprs_{\bk}} \cdot \meas \cdot D_{\bprs_{\bk}} \Big] - \E_{\bh} \Big[ D_{\bh} \cdot \meas \cdot D_{\bh} \Big] \Big\Vert_\operator\\
    ={}& \Big\Vert\E_{\bk \sim [K]} Z_{\bprs_{\bk}}\Big\Vert_\operator, \ \ \text{for} \ \ Z_{\bprs_k} \coloneqq D_{\bprs_{k}} \cdot \meas \cdot D_{\bprs_{k}} - \E_{\bh} \Big[ D_{\bh} \cdot \meas \cdot D_{\bh} \Big].
\end{align*}
As in the advice state case, our problem has again reduced to bounding the operator norm of $\E_{\bk \sim [K]} Z_{\bR_{\bk}}$ for a uniformly random $\bprs$. And just like before, the matrices $Z_{\bR_1},\dots,Z_{\bR_K}$ are mean-zero, independent and identically distributed, and their norm is bounded by $2$, since
\begin{align*}
    \norm{Z_{\bR_k}}_\operator \leq \norm{ D_{\bprs_{\bk}} \cdot \meas \cdot D_{\bprs_{\bk}}}_\operator + \Big\Vert\E_{\bh} \big[ D_{\bh} \cdot \meas \cdot D_{\bh} \big]\Big\Vert_\operator \leq 2,
\end{align*}
where the second inequality uses the fact that the $D_h$ is a unitary matrix for any Boolean function $h:[N]\rightarrow \{\pm 1\}$, so $\norm{D_h \cdot M \cdot D_h}_{\operator} \leq 1$. Thus, we can apply the matrix Hoeffding inequality as before. Since the $Z_{\bR_{1}}, \ldots, Z_{\bR_{K}}$ are $N \times N$ matrices, matrices, we obtain a bound on the maximum distinguishing advantage of
\begin{equation*}
    O\left(\sqrt{\frac{\log(N)}K}\right).
\end{equation*}
To summarize, the key new idea in this special case was to introduce a diagonal decomposition which holds for arbitrary phase states $\ket{\psi_{h}}$.

\subsection{The general one-query bound}\label{sec:tech-general}

Now we consider the case of a general adversary.
Let us recall one last time \Cref{eq:used-to-have-a-box}, our expression for the adversary's maximum distinguishing advantage:
\begin{align*}
    \max_{f : [\qdim] \rightarrow \{\pm 1\}}\Big|\E_{\bk \sim [K]} \left[ \bra{\psi_{R_{\bk}}} \cdot V^\dagger \cdot \calO_f \cdot \meas \cdot \calO_f \cdot V \cdot \ket{\psi_{R_{\bk}}}\right] - \E_{\bh} \Big[ \bra{\psi_{\bh}}\cdot  V^\dagger \cdot \calO_f \cdot \meas \cdot \calO_f \cdot V \cdot \ket{\psi_{\bh}} \Big] \Big|.
\end{align*}
The previous special case suggests the following strategy for bounding this expression:
\begin{enumerate}
    \item First, for any Boolean function $h:[N] \rightarrow \{\pm 1\}$, find a ``diagonal decomposition'' of the state $V \cdot \ket{\psi_{h}}$ of the form
\begin{align*}
    V \cdot \ket{\psi_{h}} = \big(h\text{-dependent diagonal matrix}\big) \cdot \ket{\text{fixed state}}, 
\end{align*}
    \item Next, use this decomposition to re-express $\calO_f \cdot V \cdot \ket{\psi_{h}}$ as
\begin{align*}
    \calO_f \cdot V \cdot \ket{\psi_{h}} &= \calO_f \cdot \big(h\text{-dependent diagonal matrix}\big) \cdot \ket{\text{fixed state}} \\
    &= \big(h\text{-dependent diagonal matrix}\big) \cdot \calO_f \cdot \ket{\text{fixed state}}\\
    &= \big(h\text{-dependent diagonal matrix}\big) \cdot \ket{f\text{-dependent state}}.
\end{align*}
\end{enumerate}
Importantly, the diagonal decomposition should satisfy the following two properties:

\begin{enumerate}
    \item the fixed state should have \emph{unit norm}, so that we can perform a spectral relaxation, and
    \item the $h$-dependent diagonal matrix should have \emph{bounded operator norm}, so that we can apply the matrix Hoeffding inequality.
\end{enumerate}
Unfortunately, it turns out that for a general isometry $V$, a diagonal decomposition satisfying the above requirements does not exist. Consider the following example.
\begin{example}[No nice diagonal decomposition]
\label{ex:overview-bad-decomposition}
    Let $\qdim = N$ and let $V$ be the $N \times N$ Hadamard transform $V = H^{\otimes n}$. For all Boolean vectors $r \in \{0,1\}^n$, let $h_r: \{0,1\}^n \rightarrow \{0,1\}$ denote the inner product function $h_r(x) = r \cdot x \pmod{2}$. Then 
    \begin{align*}
        \ket{\psi_{h_r}} = \frac{1}{\sqrt{N}} \cdot \sum_{x \in \{0,1\}^n} (-1)^{r \cdot x}\cdot \ket{x} = H^{\otimes n} \cdot\ket{r},
    \end{align*}
    and thus $V \cdot \ket{\psi_{h_r}} = H^{\otimes n} \cdot \ket{\psi_{h_r}} = \ket{r}$. Suppose that we try to write each $\ket{\psi_{h_r}}$ as the product of an $h_r$-dependent diagonal matrix and the uniform superposition state $\ket{+_N} \coloneqq (1/\sqrt{N})\cdot \sum_{x \in \{0, 1\}^n} \ket{x}$. That is, for each $r \in \{0,1\}^n$:
    \begin{align*}
        V \cdot \ket{\psi_{h_r}} = \ket{r} = D_r \cdot \Big(\frac{1}{\sqrt{N}} \cdot \sum_{x \in \{0, 1\}^n} \ket{x}\Big).
    \end{align*}
    Then the only choice of $D_r$ satisfying the above is $D_r = \sqrt{N}\cdot \ketbra{r}$, which has \emph{exponentially large} operator norm. Moreover, there is nothing special about the uniform superposition --- no matter what fixed state we use, there will exist $h_r$ such that $D_r$ has exponentially large operator norm. 
\end{example}

Thus,~\Cref{ex:overview-bad-decomposition} shows that we cannot hope for a diagonal decomposition that satisfies our desired conditions for \emph{all} binary phase states $\ket{\psi_h}$. 
Nevertheless, we will show that a meaningful diagonal decomposition is still possible in the general case. The key
~insight, which we will show next, is that for any isometry $V$, there exists a diagonal decomposition of $V \cdot \ket{\psi_{\bh}}$ in which the $\bh$-dependent diagonal matrix has bounded operator norm with extremely high probability over the choice of $\bh:[N]\rightarrow \{\pm 1\}$.

\subsubsection{The weight vector decomposition}
\label{subsubsec:overview-weight-vector}
Our goal is to find a diagonal decomposition of the form
\[ V \cdot \ket{\psi_{\bh}} = D_{V, \bh} \cdot \ket{\phi},
\]
in which $D_{V, \bh}$ has a ``small'' operator norm, with high probability over a uniformly random $\bh$.
Let us consider what would be implied if such a decomposition were to exist,
and then work backwards to construct the decomposition.

If such a decomposition exists, then for each $1 \leq i \leq M$,
let us consider the $i$-th coordinates of the left-hand and right-hand sides, which are given by
\begin{equation}\label{eq:weight-equality}
    \bra{i} \cdot V \cdot \ket{\psi_{\bh}} = \bra{i} \cdot D_{V, \bh} \cdot \ket{\phi}
    = (D_{V, \bh})_{i, i} \cdot \phi_i.
\end{equation}
Because $D_{V, \bh}$ has ``small'' operator norm for a ``typical'' $\bh$,
this means that $(D_{V, \bh})_{i, i}$ is ``small'' for a ``typical'' $\bh$.
Hence, for such an $\bh$,
the right-hand side of \Cref{eq:weight-equality}
must be roughly equal to $\phi_i$, the magnitude of the $i$-th coordinate in $\ket{\phi}$.
(More correctly, it must be not too much \emph{larger} than $\phi_i$.)
This, in turn, implies that the left-hand side of \Cref{eq:weight-equality} must be roughly equal to $\phi_i$ as well, at least for a ``typical''~$\bh$.
This motivates studying the magnitude of the $i$-th coordinate of $V \cdot \ket{\psi_{\bh}}$ for a ``typical'' Boolean function $\bh$.
We can do so by looking at its average squared magnitude
\begin{equation}\label{eq:mag-sq}
    p_i \coloneqq \underset{\bh}{\E}[\left|\bra{i} \cdot V \cdot \ket{\psi_{\bh}}\right|^2 ].
\end{equation}
In other words, 
$p_i$ denotes the probability that measuring the state $V\ket{\psi_{\bh}}$ in the standard basis results in an outcome of $i$.
Then we expect  the $i$-th coordinate of $V \cdot \ket{\psi_{\bh}}$ to have magnitude roughly $\sqrt{p_i}$, and that suggests the following choice for our fixed state in the diagonal decomposition:
\[\ket{\wt_V} \coloneqq \sum_{i =1}^{\qdim} \sqrt{p_i}\cdot \ket{i},
\]
which we refer to as the \emph{weight vector} for $V$. We observe that $\ket{\wt_V}$ is indeed a unit vector because
\[ \braket{\wt_V} = \sum_{i=1}^{\qdim} p_i = \underset{\bh}{\E}\Big[ \sum_{i=1}^{\qdim} \left|\bra{i} \cdot V \cdot \ket{\psi_{\bh}}\right|^2 \Big] = \underset{\bh}{\E}\Big[ \norm{V \cdot \ket{\psi_{\bh}}}^2 \Big] = 1,
\]
where the last equality holds because $V$ is an isometry. Intuitively, the state $\ket{\wt_V}$ encodes how much \emph{weight} the isometry $V$ places on each individual coordinate $1 \leq i \leq \qdim$. 

To compute the full diagonal decomposition, we write the isometry $V: \mathbb{C}^N \rightarrow \mathbb{C}^{\qdim}$ explicitly as 
\begin{align*}
    V = \sum_{i=1}^M \sum_{x=1}^N v_{i,x}\cdot \ketbra{i}{x} = \sum_{i=1}^M\ket{i}\cdot \Big( \sum_{x=1}^N v_{i,x}\bra{x}\Big) = \sum_{i=1}^{\qdim} \ketbra{i}{v_i},
\end{align*}
so that the $i$th coordinate of $V\cdot \ket{\psi_{\bh}}$ is given by
\[ \alpha_{\bh, i}
\coloneqq \bra{i} \cdot V\cdot \ket{\psi_{\bh}}
= \bra{i} \cdot \Big(\sum_{i=1}^{\qdim} \ketbra{i}{v_i}\Big) \cdot \ket{\psi_{\bh}}
= \braket{v_i}{\psi_{\bh}},
\]
yielding the decomposition
\[ V \cdot \ket{\psi_{\bh}} = \sum_{i=1}^{\qdim} \alpha_{\bh,i} \cdot \ket{i}
= \underbrace{\left(\sum_{i=1}^{\qdim} \frac{\alpha_{\bh,i}}{\sqrt{p_i}} \cdot \ketbra{i} \right)}_{D_{V,\bh}} \cdot \Big(\sum_{i =1}^{\qdim} \sqrt{p_i}\cdot \ket{i}\Big) = D_{V, \bh} \cdot \ket{\wt_V}. 
\]
\subsubsection{Bounding the operator norm of $D_{V, \bh}$.}
Our next step is to determine whether the random matrix $D_{V, \bh}$ actually \emph{has} bounded operator norm with high probability. Its operator norm is given by
\[ \norm{D_{V,\bh}}_{\operator} = \max_{1 \leq i \leq M} \frac{\left|\alpha_{\bh, i}\right|}{{\sqrt{p_i}}},
\]
and we know from \Cref{eq:mag-sq} that for every $1\leq i \leq M$,
\begin{equation*}
    \E_{\bh}[|\alpha_{\bh,i}|^2] = p_i.
\end{equation*}
Therefore, if each coordinate $\alpha_{\bh, i}$ has good enough (scalar) concentration, we can bound $\norm{D_{\bh}}_{\operator}$ with high probability.

Thus, we have reduced the problem to understanding the concentration of the random variables
\[ \frac{\alpha_{\bh, i}}{\sqrt{p_i}} = \frac 1 {\sqrt{p_i}} \cdot \braket{v_i}{\psi_{\bh}} = \sum_{x=1}^N \frac{v_{i,x}}{\sqrt{p_i}}\cdot \bh(x).
\]
To see that this expression is small with high probability, we observe that it is a weighted linear combination of the $N$ i.i.d.\ $\{\pm 1\}$ random variables $\{\bh(x)\}_{x\in [N]}$, has mean zero, and has variance
\[
\E_{\bh}\Big[\Big|\frac{\alpha_{\bh,i}}{\sqrt{p_i}}\Big|^2\Big]
= \frac{\E_{\bh}[|\alpha_{\bh,i}|^2]}{p_i}
= \frac{p_i}{p_i}
= 1.
\]
Therefore, standard (scalar) concentration tools (see \Cref{thm:complex-concentration}) tell us that this random variable exhibits ``sub-Gaussian concentration,'' implying (in this case) that it is larger than any $t$ with probability at most $2\cdot \exp(-t^2/2)$. Union bounding over all $\qdim$ coordinates, we conclude that $\norm{D_{V, \bh}}_{\operator} > t$ with probability at most $2M \cdot \exp(-t^2/2)$, and so it is, for example, unlikely to be much larger than $O(\sqrt{\log \qdim})$.

\subsubsection{Putting everything together}
With the weight-vector decomposition in hand,
we can proceed to bounding the adversary's maximum distinguishing advantage along similar lines as in \Cref{sec:tech-no-isometry}.
To begin, we can rewrite the state $\calO_f \cdot V \cdot \ket{\psi_h}$ as follows:
\begin{align*}
    \calO_f \cdot V \cdot \ket{\psi_h} = \calO_f \cdot D_{V,h}\cdot \ket{\wt_V} = D_{V,h} \cdot \calO_f \cdot \ket{\wt_V},
\end{align*}
for every function $h$. In other words, the choice of $\bR$-dependent function $f$ here is accounted for as the vector $\calO_f \cdot \ket{\wt_V}$, which is a unit vector for all functions $f$. 

By an argument similar to the one in \Cref{sec:tech-no-isometry}, we can then bound the adversary's maximum distinguishing advantage by the operator norm 
\begin{align*}
    \norm{ \E_{\bk \sim [K]} Z_{\bprs_{\bk}} }_{\operator}, \ \ \text{where} \ Z_{h} \coloneqq D_{V,h}^\dagger \cdot \meas \cdot D_{V,h} - \E_{\bh} \Big[ D_{V,\bh}^\dagger \cdot \meas \cdot D_{V,\bh} \Big].
\end{align*}
Thus, we have again reduced our problem to bounding the operator norm of an average of $K$ independent and identically distributed 
matrices $Z_{\bprs_k}$
whose operator norms are bounded with high probability.
In particular, since the operator norm of $\Vert D_{V, \bh} \Vert$ is usually no more than $O(\sqrt{\log M})$, over a uniformly random $\bh$, the operator norm of $D_{V,\bh}^\dagger \cdot \meas \cdot D_{V,\bh}$ is usually no more than $O(\log M)$.
We would like to then conclude that $\norm{ \E_{\bk \sim [K]} Z_{\bprs_{\bk}} }_{\operator}$ is at most \begin{equation*}
    O\Big(\sqrt{\frac{\log \qdim}{K}}\Big)
\end{equation*}
with high probability, which would imply our claimed result. 

Unfortunately, we cannot quite apply the matrix Hoeffding inequality directly, which requires that the matrices have bounded operator norm with probability $1$, not just with high probability. Getting around this issue requires some additional technical ideas, and we give two ways of handling it in the main body of the paper.
\begin{enumerate}
    \item Our first approach is to truncate the diagonal matrices $D_{V, \bR_k}$ so that any entries whose magnitude exceeds some number $B$ are scaled down so that their magnitude is equal to~$B$.
    The result is that all matrices now have bounded operator norm, which means we are in fact able to apply the matrix Hoeffding inequality.
    Ultimately, this results in a bound of $O(1/K^{1/4})$ on the adversary's distinguishing advantage for reasonably small values of $M$ (say, $M \leq \exp(K^{1/8}/4)$), which is more than enough to prove the one-query lower bound for the Unitary Synthesis Problem.
    That said, this bound is not quite strong enough to prove the bound claimed in \cref{thm:main-distinguishing-intro}.
    \item To prove the precise bound claimed in \cref{thm:main-distinguishing-intro}
    and thereby achieve the correct asymptotic dependence on $K$, we give a somewhat different analysis.
    \begin{enumerate}
        \item First, we show that it suffices to bound the \emph{expected} distinguishing advantage on a random $\bro$, rather than proving a bound with high probability. To show this, we show that the maximum distinguishing advantage concentrates extremely well around its expectation (see \cref{lem:adv-tail}).
        \item To bound the expected distinguishing advantage, we use a different technique called ``decoupling'', which is common in the random matrix theory literature \cite{Ver11,Van17}. At a high level, the technique (when combined with the ideas from this technical overview) allows us to reduce to bounding the expected operator norm of the random matrix
        \[\E_{\bk \sim [K]}[D_{V, \bprs_{\bk}}^\dagger \cdot \meas \cdot D_{V, \bprs_{\bk}'}],
        \]
        where $\bro$ and $\bro'$ are independent and uniformly random function families. This is easier to give a sharp bound on because the dependence on each of $\bro$ and $\bro'$ is \emph{linear} rather than quadratic, allowing us to prove an optimal bound on the expected value by applying a different matrix concentration inequality for matrix Rademacher series (\cref{thm:khintchine}).
        Unlike the matrix Hoeffding inequality, this matrix concentration inequality does not require the matrices to have bounded operator norm with probability 1 (though it does require the matrices to be random Rademacher matrices), which is why it gives stronger bounds than we achieve using our first approach.
    \end{enumerate}    
\end{enumerate}
The second proof, which achieves the optimal dependence on $K$, is presented in \cref{sec:lower-bound}. 
The first proof is presented in \cref{sec:appendix-matrix-chernoff}. 
We believe that theoretical computer scientists might find the first proof more straightforward to follow.

\subsection{Future directions: beyond one query}
\label{subsec:future}

\cref{thm:main-distinguishing-intro} proves that efficient one-query oracle algorithms achieve at most negligible advantage in the Oracle State Distinguishing Game (and thus cannot synthesize arbitrary unitaries). We conjecture (see~\cref{conjecture:main}) that efficient oracle circuits making $\poly(n)$-many sequential queries cannot win our distinguishing game. Towards resolving the full conjecture, we believe it may be useful to focus on the special case of \emph{two-query} adversaries. In this subsection, we present several conjectures --- all weaker than~\cref{conjecture:main} --- that capture the simplest unresolved special cases of two-query attacks. 

First, we will need the following observation about the power of \emph{classical} oracle queries. Let us fix a function family $\prs: [K] \times [N] \rightarrow \{\pm 1\}$ and a projective measurement $\calP = \{\meas_i\}_{i \in [M]}$ with $M$ outcomes. Suppose the adversary, upon receiving $\ket{\bpsi}$, applies an isometry $V: \C^N \rightarrow \C^M$ followed by an oracle query $\oracle_f$. Next, it performs the measurement $\calP$, obtaining some outcome in $\{1,\dots,D\}$. Depending on whether the adversary's input state is sampled from the ``pseudorandom'' or ``random'' distribution, the outcome of measuring $\calP$ is distributed as either:
\begin{itemize}
    \item $\mathsf{Dist}_0$: The result of applying $\calP$ to $\oracle_f \cdot V \cdot \ket{\psi_{\ro_{\bk}}}$, for a random $\bk \sim [K]$.
    \item $\mathsf{Dist}_1$: The result of applying $\calP$ to $\oracle_f \cdot V \cdot \ket{\psi_{\bh}}$, for a random $\bh: [N] \rightarrow \{\pm 1\}$.
\end{itemize}
We observe that if the total variation distance (or statistical distance) between $\mathsf{Dist}_0$ and $\mathsf{Dist}_1$ is $\epsilon$, then, by making one classical oracle query, the adversary can distinguish the two cases with advantage $\epsilon$. This is because the second query can be made to the Boolean function $g: [M] \rightarrow \{\pm 1\}$, defined as $g(i) = \mathsf{Sign}(\Pr[i \sim \mathsf{Dist}_0] - \Pr[i \sim \mathsf{Dist}_1])$. If the output of $g$ is $+ 1$, the adversary guesses that it was in the pseudorandom case, and if the output of $g$ is $-1$, the adversary guesses that it was in the Haar random case, and attains distinguishing advantage $\varepsilon$.

\paragraph{The ``1.5-query'' conjecture.} We conjecture that adversaries that make one quantum query followed by one classical query cannot win our distinguishing game.

\begin{conjecture}[The 1.5-query conjecture.]
    Fix an isometry $V: \C^N \rightarrow \C^M$ and an $M$-outcome projective measurement $\calP = \{ \meas_i \}_{i \in [M]}$ acting on $\C^M$. For any subset $S \subseteq [M]$, let $\meas_S \coloneqq \sum_{i \in S} \meas_i$. With high probability over a uniformly random $\bR: [K] \times [N] \rightarrow \{\pm 1\}$, the ``1.5-query'' adversary's distinguishing advantage,
    \begin{align*}
     \max_{S \subseteq [M]} \max_{f: [M] \rightarrow \{\pm 1\}} \E_{\bk \sim [K]} \Big[ \bra{\psi_{\bprs_{\bk}}} V^\dagger \cdot \oracle_f \cdot \meas_S \cdot \calO_f \cdot V \ket{\psi_{\bprs_{\bk}}}\Big] - \E_{\bh} \Big[ \bra{\psi_{\bh}} V^\dagger \cdot \calO_f \cdot \meas_S \cdot \oracle_f \cdot V \ket{\psi_{\bh}} \Big],
    \end{align*}
    is at most $\negl(n)$.
\end{conjecture}

One potential approach towards bounding this expression is to observe that if we fix a subset $S$, the remaining expression has the same form as the maximum distinguishing advantage for a one-query adversary. We can therefore apply the same weight-vector decomposition described in~\cref{subsubsec:overview-weight-vector} and invoke a spectral relaxation. The result is that the following is an upper bound for the adversary's distinguishing advantage: 
\begin{align}
    \max_{S \subseteq [M]} \Big\Vert \underset{\bk}{\E}\Big[D_{V, \bprs_{\bk}}^\dagger  \cdot \meas_S \cdot  D_{V, \bprs_{\bk}}\Big] - \underset{\bh}{\E} \Big[ D_{V, \bh}^\dagger  \cdot \meas_S \cdot  D_{V, \bh}\Big] \Big\Vert_{\operator}, \label{1.5-query-adv-3}
\end{align}
where the definitions of $D_{V,\bh}$ and $D_{V,\bR_{\bk}}$ are the same as in~\cref{subsubsec:overview-weight-vector}.

The central difficulty we face is that matrix concentration inequalities are not sufficient to bound~\cref{1.5-query-adv-3}. Indeed, they can be applied for any \emph{fixed} choice of $S$, but it is unclear how to bound the operator norm of $2^M$ matrices simultaneously, one for each $S$. Nevertheless, we believe that the expression (\ref{1.5-query-adv-3}) is in fact negligible with high probability over $\bro$. 

\paragraph{The ``$(1+\epsilon)$-query'' conjecture.} Finally, we highlight a sub-class of $1.5$-query adversaries that we do not know how to rule out, which we refer to as $(1+\epsilon)$-query adversaries. Instead of making an arbitrary first query to the oracle, these adversaries use their first query to synthesize an advice state $\ket{v} \in \C^{M/N}$ (similar to the adversaries we considered in~\cref{sec:tech-advice}); note that while $\ket{v}$ is technically restricted to states of a certain type, treating it as an arbitrary unit vector is essentially without loss of generality.\footnote{For example, the adversary can implement Rosenthal’s state synthesis algorithm \cite{Ros23} to prepare an arbitrary quantum advice state.}

\begin{conjecture}[The $(1+\epsilon)$-query conjecture.]
    Fix any $M$-outcome projective measurement $\calP = \{ \meas_i \}_{i \in [M]}$ acting on $\C^M$. With high probability over a uniformly random $\bR: [K] \times [N] \rightarrow \{\pm 1\}$, the adversary's distinguishing advantage
    \begin{align}
    \max_{S \subseteq [M]} \Big\Vert \E_{\bk \sim [K]} \Big[ (\bra{\psi_{\bprs_{\bk}}} \otimes \Id) \cdot \meas_S \cdot (\ket{\psi_{\bprs_{\bk}}} \otimes \Id) \Big] - \E_{\bh} \Big[ (\bra{\psi_{\bh}} \otimes \Id) \cdot \meas_S \cdot (\ket{\psi_{\bh}} \otimes \Id) \Big] \Big\Vert_\operator \label{1.5-query-2}
    \end{align} 
    is at most $\negl(n)$.
\end{conjecture}
Again, the difficulty we face in bounding~\cref{1.5-query-2} is that matrix concentration inequalities only seem to apply when the subset $S$ is fixed, and not when we maximize over all $S$.

\paragraph{A simple mathematical conjecture.} Finally, in order to state the simplest mathematical conjecture that captures this ``simultaneous matrix concentration'' problem, we give a slightly different version of the above $(1+\varepsilon)$-query conjecture (which corresponds to the case where $\ket{\psi_{\bR_k}}$ and $\ket{\psi_{\bh}}$ are Haar random).

\begin{conjecture}
\label{conj:simple}
    Let $c > 0$ be any constant. Set parameters $N = 2^n$, $K = N/2$ and $M = 2^n \cdot 2^{n^c}$. Let $\{\meas_i\}_{i \in [M]}$ be projectors acting on $\C^M$ such that $\sum_{i \in [M]} \meas_i = \Id_M$. Sample $K$ Haar-random unit vectors $\ket*{\bpsi_1},\dots,\ket*{\bpsi_K} \in \mathbb C^N$. Then with probability $1-\negl(n)$,
    \begin{align*}
        \max_{S \subseteq [M]} \Big\Vert \E_{\bk \sim [K]} \Big[ (\bra{\bpsi_{\bk}} \otimes \Id) \cdot \meas_S \cdot (\ket{\bpsi_{\bk}} \otimes \Id) \Big] - \E_{\ket{\bpsi}} \Big[ (\bra{\bpsi} \otimes \Id) \cdot \meas_S \cdot (\ket{\bpsi} \otimes \Id) \Big] \Big\Vert_\operator = \negl(n).
    \end{align*}
    where each $\Id$ is $M/N \times M/N$-dimensional and $\ket{\bpsi}$ is Haar-random.
\end{conjecture}

We note that it would be extremely surprising to us if \cref{conj:simple} turns out to be false, since that would imply that a two-query algorithm can win (the Haar-random state version of) the Oracle State Distinguishing Game.

\newpage

\section{The Oracle State Distinguishing Game}\label{sec:game-def}

The purpose of this section is to define the Oracle State Distinguishing Game and prove several fundamental properties about it. We note that our main proof in \cref{sec:lower-bound} can be understood without reading \cref{sec:unitary-synthesis,sec:game-concentration,sec:adversary-space}.

The section is organized as follows. In \cref{sec:notation}, we introduce some notation and formalism for oracle algorithms. 
In \Cref{subsec:game-def}, we define the Oracle State Distinguishing Game. In \Cref{sec:unitary-synthesis} we show that hardness of the Oracle State Distinguishing Game for $T$-query adversaries implies hardness of $T$-query unitary synthesis for any parameter $T$.

In \cref{sec:game-concentration}, we appeal to concentration of measure to give (for any oracle adversary $A$) an upper tail inequality on the optimal distinguishing advantage in the oracle state distinguishing game, which implies that it suffices to bound the adversary's expected distinguishing advantage over the choice of $\bro$.  

In \Cref{sec:adversary-space}, we show that two complexity measures of an oracle algorithm --- \emph{query length} and \emph{space complexity} --- are tightly related in the oracle state distinguishing game. The assumption that our adversaries are space-efficient as well as query-efficient will be crucial in both proofs of the one-query lower bound. 

Finally, in \cref{sec:one-query-model}, we give an explicit ``normal form'' for one-query adversaries (using \cref{sec:adversary-space}), setting up simplified notation that suffices for \cref{sec:lower-bound}. 

\subsection{Preliminary notation}\label{sec:notation}

We will use \textbf{boldface} to denote random variables.
We will write $\ln(\cdot)$ for the natural logarithm and $\log_2(\cdot)$ for the base-2 logarithm.

\begin{notation}[Register size versus dimension]\label{not:size-versus-dim}
A quantum register consisting of $m$ qubits has dimension $M = 2^m$.
Viewing it as a space of $m$ qubits, it is natural to index the basis by binary strings $x \in \{0, 1\}^m$. On the other hand, viewing it as a space of dimension $M$, it is natural to index the basis by integers $1 \leq i \leq M$. 
We can associate these two indexing schemes by associating the number $i$ with the string $x$ that is the $m$-bit binary representation of $i - 1$.
We will typically prefer the second indexing scheme, and will therefore typically represent $m$-qubit states as
\begin{equation*}
    \ket{\psi} = \sum_{i=1}^M \psi_i \cdot \ket{i}.
\end{equation*}
Throughout this work, we will consider algorithms which take as input a quantum state. We will typically reserve $n$ for the length of the input register in qubits and $N \coloneqq 2^n$ for the dimension of this register.
\end{notation}

Most of our quantum state inputs will come in the form of binary phase states.

\begin{definition}[Binary phase state]
    A \emph{Boolean function} is a function $h:\{0, 1\}^n \rightarrow \{\pm 1\}$.
    Due to the association between $\{0, 1\}^n$ and $[N]$ given in \Cref{not:size-versus-dim}, we will typically prefer to write such a function as $h: [N]\rightarrow \{\pm 1\}$, and we will elect to still refer to such a function as a ``Boolean function''.
    The corresponding \emph{binary phase state} is
    \begin{equation*}
        \ket{\psi_h} \coloneqq \frac{1}{\sqrt{N}} \cdot \sum_{x=1}^N h(x) \cdot \ket{x}.
    \end{equation*}
\end{definition}

\begin{definition}[Phase oracle]\label{def:phase}
    Let $f:[L]\rightarrow \{\pm 1\}$ be a Boolean function.
    Then the corresponding \emph{phase oracle} is the $L \times L$ diagonal unitary matrix $\oracle_f$ given by
    \begin{equation*}
        \oracle_f = \sum_{i=1}^{L} f(i) \cdot \ketbra{i}.
    \end{equation*}
    Operationally, for any $1 \leq i \leq L$,
    the oracle acts as $\oracle_f \cdot \ket{i} = f(i) \cdot \ket{i}$.
    If $L = 2^{\ell}$ for some integer~$\ell$,
    then we refer to $\ell$ as the \emph{input length} of the phase oracle
    and $L$ as the \emph{dimension} of the phase oracle.
\end{definition}

Phase oracles can be contrasted with \emph{bit flip oracles}.
A bit flip oracle $\oracle^{\mathsf{flip}}_g$ is specified by a function $g:[\qdim] \rightarrow \{0, 1\}$
and is defined as follows:
for each $1 \leq i \leq \qdim$ and $b \in \{0, 1\}$,
it acts as $\oracle^{\mathsf{flip}}_g \cdot \ket{i, b} = \ket{i, b \oplus g(i)}$. 
In general, a bit flip oracle can always be used to implement a phase oracle,
but the reverse is only partially true: implementing a bit flip oracle requires a \emph{controlled} phase oracle.
However, we will see below that the class of phase oracles we consider are actually powerful enough to implement controlled phase oracles,
and hence can be converted to bit flip oracles if desired.

Throughout this work, we will consider a class of circuits
which take as input a quantum state
and are allowed to perform several queries to a phase oracle.
We define these formally as follows.

\begin{definition}[Oracle circuit]
\label{def:oracle-algorithm}
    A $t$-query \emph{oracle circuit} $A^{(\cdot)}$
    begins with an input register of $n$ qubits and an ancilla register of $a$ qubits, each initialized to $\ket{0}$,
    for a total of $m = n + a$ qubits.
    It then performs the $m$-qubit unitaries $U_1, \ldots, U_{t+1}$. In addition, between each pair of unitaries,
    it performs a query to a phase oracle of input length $\ell$,
    which acts on the first $\ell$ qubits.
    We write
    $A^{(\cdot)}=(n, m, \ell, U_1, \ldots, U_{t+1})$ in order to specify these parameters.

    The precise execution of the oracle circuit depends on which Boolean function it is given query access to.
    Given a Boolean function $f:\{0, 1\}^\ell \rightarrow \{\pm 1\}$,
    we write $A^f$ for the oracle circuit given access to $f$.
    On input an $n$-qubit state, it computes the state
    \begin{equation*}
        U_{t+1} \cdot \calO_f \cdot U_t \cdots \calO_f \cdot U_2 \cdot \calO_f \cdot U_1 \cdot \ket{\psi} \otimes \ket{0^a}.
    \end{equation*}
    This is illustrated in \Cref{fig:oracle-algorithm}.
\end{definition}

\begin{figure}[h]
\centering
\begin{quantikz}
\lstick{$\ket{\psi}$} & \gate[3]{U_1} & \gate[2]{\oracle_f} & \gate[3]{U_2} & \gate[2]{\oracle_f} &  \ \ldots\ & \gate[3]{U_t} & \gate[2]{\oracle_f} & \gate[3]{U_{t+1}}&  \\
\lstick[2]{$\ket{0^a}$} & && &&  \ \ldots\ &&&&\\
& & & &&  \ \ldots\ &&&&
\end{quantikz}
\caption{The execution of $A^f$ on input $\ket{\psi}$.}
\label{fig:oracle-algorithm}
\end{figure}

\begin{remark}[Querying multiple functions]\label{rem:multi-query}
    Note that we have defined our oracle circuits
    so that every application of the oracle gate
    queries the same Boolean function $f$.
    One can consider an alternative model of $t$-query oracle circuits
    which are instead allowed to query a different Boolean function $f_i$ for each oracle call $1 \leq i \leq t$.
    However, one can simulate access to these $t$ Boolean functions using a single Boolean function $f$ defined as $f(\mathrm{bin}(i), x) \coloneqq f_i(x)$, where $\mathrm{bin}(i)$ is the $a = \lceil \log_2(t) \rceil$-bit binary encoding of $i$.
    Hence, an adversary which queries $t$ different Boolean functions
    can be simulated by an adversary which queries one Boolean function and has a small $a$-qubit overhead.
    Thus, it is essentially without loss of generality to focus on adversaries which query a single function, as we do.
    We note that this transformation is standard and appears,
    for example, at the top of page 5 in Rosenthal's Ph.D. thesis~\cite{Ros23b}.
\end{remark}
 
\begin{remark}[Querying many-bit functions]\label{rem:parallel-queries}
    Yet another model of $t$-query oracle circuits allows for making bit-flip queries to $d$-bit output functions of the form $\ket{x} \ket{y} \mapsto \ket{x} \ket{y\oplus f(x)}$ for $x\in [M], y\in \{0,1\}^d$. As pointed out in \cite{Ros23b} Section 2.1, such queries can be simulated by a single quantum query to a $1$-bit function:
    \[ \ket{x} \ket{r}\mapsto (-1)^{r\cdot f(x)} \ket{x} \ket{r}. 
    \]
    \noindent The new function $g(x, r) = r\cdot f(x)$ has domain $[M\cdot 2^d]$. This also allows us to simulate \emph{parallel queries} of the form
    \[  \ket{x_1}\ket{b_1}\hdots \ket{x_t}\ket{b_t}\mapsto \ket{x_1}\ket{b_1\oplus f(x_1)}\hdots \ket{x_t} \ket{b_t \oplus f(x_t)}
    \]
    by defining $x = (x_1, \hdots, x_t)$. Therefore, our one-query lower bounds imply lower bounds against a bounded (e.g., polynomial or sub-exponential) number of parallel queries. 
\end{remark}

\subsection{Defining the Oracle State Distinguishing Game}\label{subsec:game-def}

In this section, we define the Oracle State Distinguishing Game.
To begin, every such game is parameterized by a particular family of functions, which is defined as follows.

\begin{definition}[Function families]
    Let $K$ and $N$ be integers.
    A \emph{function family} is a function
    $R:[K] \times [N] \rightarrow \{\pm 1\}$.
    We think of $\prs$ as defining a family of $K$ Boolean functions as follows:
    for each $1 \leq k \leq K$, we let $R_k:[N] \rightarrow \{\pm 1\}$ be the function $R_k(\cdot) \coloneqq R(k, \cdot)$.
\end{definition}

We have chosen the letter ``$\prs$'' for function families as shorthand for the word ``$\prs$andom'',
    as our function families will often (though not always) be random variables.

\begin{definition}[Oracle State Distinguishing Game]
    Let $\prs:[K]\times[N]\rightarrow \{\pm 1\}$ be a function family.
    The \emph{Oracle State Distinguishing Game on $\prs$}, denoted $\game^{\prs}$, involves two parties, a challenger and an adversary.
    It is played as follows.
    \begin{enumerate}
    \item The challenger samples a random bit $\bb \in \{0, 1\}$. \item The challenger generates a random $n$-qubit state $\ket{\bpsi}$ in one of two ways:
    \begin{itemize}
        \item[$\circ$] If $\bb = 0$, the challenger samples a uniformly random $\bk \sim [K]$ and generates $\ket{\bpsi} \coloneqq \ket{\psi_{R_{\bk}}}$.
        \item[$\circ$] If $\bb = 1$, the challenger samples a uniformly random $\bx \sim [N]$ and sets $\ket{\bpsi} \coloneqq \ket{\bx}$.
    \end{itemize}
    \item The challenger sends $\ket{\bpsi}$ to the adversary. 
    \item The adversary outputs a bit $\bb' \in \{0, 1\}$.
    \item If $\bb' = \bb$, then the adversary wins. Otherwise, they lose.
    \end{enumerate}
    The \emph{Oracle State Distinguishing Game}, denoted $\game_{N, K}$ is played as follows.
    A uniformly random function family $\bprs:[K]\times [N] \rightarrow \{\pm 1\}$ is sampled,
    and then the challenger plays $\game^{\bprs}$ with the adversary.
\end{definition}

\begin{remark}[Computational complexity of the challenger]\label{rem:challenger-complexity}
In the ``$\bb=1$ case'',
the view of the adversary is that it receives a maximally mixed state $\Id_N/N$.
Hence, we can equivalently view the challenger as sampling a random state from \emph{any} distribution,
so long as an average state drawn from this distribution is maximally mixed.
For example,
we can equivalently view the challenger as sampling a uniformly random Boolean function $\bh: [N] \rightarrow \{ \pm 1\}$ and setting $\ket{\bpsi} \coloneqq \ket{\psi_{\bh}}$, or sampling $\ket{\bpsi}$ as an $N$-dimensional Haar-random state.
We will typically prefer the first of these points of view throughout this work.

We have chosen to have the challenger sample a random basis state $\ket{\bx}$ in this case to emphasize that the challenger is computationally efficient in our construction.
Note that they can also efficiently construct the state $\ket{\psi_{\prs_{\bk}}}$ in the ``$\bb = 0$ case'' given oracle access to $\prs$.
In particular, they need only query the oracle $\prs(\bk, \cdot)$ on the uniform superposition state $\ket{+_N} \coloneqq \frac{1}{\sqrt{N}}\cdot  \sum_{x = 1}^N \ket{x}$.
\end{remark}

We will model our adversary as an oracle circuit $A^{(\cdot)}$ with an $N$-dimensional input register. 
Intuitively, the adversary will be allowed to select its own preferred oracle $f$ to give it the best chance of winning the Oracle State Distinguishing Game on $\prs$.
When the game is played on a uniformly random choice of the function family $\bprs:[K] \times [N] \rightarrow \{\pm 1\}$, the adversary will be allowed to select an oracle $f_{\bprs}$ which depends on $\bprs$.
\begin{definition}[Adversary]
\label{def:adversary}
    An \emph{adversary} is specified by an oracle circuit $A^{(\cdot)}$.
    Let $L$ be the dimension of the queries the oracle circuit makes.
    Given oracle access to a Boolean function $f:[L] \rightarrow \{\pm 1\}$,
    the adversary acts as follows.
    On input the quantum state $\ket{\psi}$,
    it applies $A^f$,
    and then it measures the first qubit in the standard basis.
    It outputs the measurement outcome $\bb'\in \{0, 1\}$.
\end{definition}

Now we introduce several pieces of notation which will help us describe the adversary's winning probability in the Oracle State Distinguishing Game.

\begin{notation}[Adversary's acceptance probability]
    Let $A^{(\cdot)}$ be an adversary
    which has an $N$-dimensional input register
    and makes $L$-dimensional queries.
    Let $h:[N] \rightarrow \{\pm 1\}$ be a Boolean function,
    and let $f:[L]\rightarrow \{\pm 1\}$ be another Boolean function.
    We will use the notation 
    \begin{equation*}
        \accept_{A}(h \mid f)
        \coloneqq \Pr[\text{$A^f$ outputs ``$0$'' on $\ket{\psi_h}$}].
    \end{equation*}
\end{notation}

Let $\prs:[K]\times[N]\rightarrow \{\pm 1\}$ be a function family.
Then in the ``$\bb = 0$ case'', the probability the adversary wins $\game^{\prs}$ can be expressed in this notation as
$\E_{\bk \sim [K]}[\accept_{A}(\prs_{\bk} \mid f)].$
As for the ``$\bb=1$ case''
let us follow \Cref{rem:challenger-complexity}
and view the challenger as sampling a uniformly random Boolean function $\bh:[N] \rightarrow \{\pm 1\}$ and setting $\ket{\bpsi} \coloneqq \ket{\psi_{\bh}}$;
given this state, the adversary wins with probability $1-\accept_A(\bh \mid f)$.
Putting these two together, the probability the adversary wins $\game^{\prs}$ is
\begin{equation}\label{eq:winning-prob}
    \frac{1}{2} \cdot \E_{\bk \sim [K]}[\accept_{A}(\prs \mid f)]
    + \frac{1}{2} \cdot \E_{\bh}[1 - \accept_{A}(\bh \mid f)].
\end{equation}
Note that the adversary can trivially win with probability $1/2$ by always outputting $\bb' = 0$.
Thus, we care about the amount by which the adversary's acceptance probability differs from $1/2$, which is known as its \emph{advantage}.

\begin{definition}[Distinguishing advantage]\label{def:advantage}
Let $\prs:[K]\times[N]\rightarrow \{\pm 1\}$ be a function family.
Let $A^{(\cdot)}$ be an adversary with an $N$-dimensional input register.
Let $L$ be the dimension of $A^{(\cdot)}$'s queries,
and let $f:[L] \rightarrow \{\pm 1\}$ be a Boolean function.
Then the \emph{distinguishing advantage of $A^f$ on $\game^{\prs}$} is defined as
\begin{equation*}
    \distinguish_{A}(\prs \mid f)
    \coloneqq 2 \cdot \Big|\Pr[\text{$A^f$ wins on $\game^{\prs}$}] - \frac{1}{2}\Big|.
\end{equation*}
The factor of 2 in front was chosen so that the distinguishing advantage is a number between $0$ and~$1$ and is equal to~$1$ if the adversary always wins (or loses).
If we plug in \Cref{eq:winning-prob} for the adversary's winning probability, we can rewrite the distinguishing advantage as
\begin{equation*}
    \distinguish_{A}(\prs \mid f)
    = \Big|\E_{\bk \sim [K]}[\accept_{A}(\prs_{\bk} \mid f)]
    - \E_{\bh}[\accept_{A}(\bh \mid f)]\Big|,
\end{equation*}
where $\bh:[N] \rightarrow \{\pm 1\}$ is a uniformly random Boolean function.
This equation is the form that we will most typically express the distinguishing advantage in,
and it explains why we refer to this as the \emph{distinguishing} advantage, which is because it expresses how well the adversary's output can be used to distinguish between the two cases.
\end{definition}

The adversary's goal is to maximize the distinguishing advantage,
and it can do so by picking the best possible function $f:[L] \rightarrow \{\pm 1\}$ to perform oracle queries to.
This motivates the following quantity,
which is the main quantity we will be studying throughout this paper.

\begin{definition}[Maximum distinguishing advantage]
\label{def:max-dist}
Let $\prs:[K]\times[N]\rightarrow \{\pm 1\}$ be a function family.
Let $A^{(\cdot)}$ be an adversary
    which has an $N$-dimensional input register
    and makes $L$-dimensional queries.
The \emph{maximum distinguishing advantage of $A^{(\cdot)}$ on $\game^{\prs}$} is defined as
\begin{equation*}
    \distinguish_{A}(\prs)
    \coloneqq \max_{f:[L] \rightarrow \{\pm 1\}}\{\distinguish_{A}(\prs\mid f)\}.
\end{equation*}
Finally, the \emph{maximum distinguishing advantage of $A^{(\cdot)}$ on $\game_{K, N}$} is equal to
\begin{equation*}
    \distinguish_{A}^{\avg} \coloneqq \E_{\bprs}[\distinguish_{A}(\bprs)],
\end{equation*}
where $\bprs:[K]\times[N]\rightarrow \{\pm 1\}$ is a uniformly random function family.
\end{definition}

The goal of this work is to show that $\distinguish_{A}^{\avg}$ is small for any adversary $A^{(\cdot)}$ which makes a single query of length $\ell = o(K)$. Moreover, we will prove that in the same parameter regime, with high probability over $\bro$, $\distinguish_A(\bro)$ is small.

\subsection{Relationship to the Unitary Synthesis Problem}\label{sec:unitary-synthesis}

In this section, we formalize the Unitary Synthesis Problem
and its relationship to the Oracle State Distinguishing Game.
Or, rather, we will suggest one possible way of formalizing the Unitary Synthesis Problem,
as there seems to be no generally agreed upon precise formulation of the problem.
For example, the task is to approximate a general $n$-qubit unitary $U$, but there are many different ways of defining what it means to approximate a unitary.
This was addressed by Scott Aaronson in a comment on the Shtetl-Optimized blog~\cite{Aar21}, in which he said the following.
\begin{quote}
\textit{
``The unitary synthesis problem is interesting for any reasonable notion of approximating $U$. In other words, we lack a positive result even for the loosest notions of approximation you mentioned, or a negative result even for the most stringent ones! Once we have some results, then we can start worrying about these distinctions.''}
\end{quote}
The last few years have seen  increasing interest in fundamentally quantum tasks, and as a result we now do have some results on problems related to unitary synthesis~\cite{RY22,Ros22,BEM+23},
and these have given several ways of precisely formalizing unitary synthesis.

Let us first recall several standard notions from quantum information theory.
Given two $n$-qubit density matrices $\rho_1$ and $\rho_2$, their \emph{trace distance} is given by
\begin{equation*}
    \mathrm{D}_{\mathrm{tr}}(\rho_1, \rho_2) \coloneqq \frac{1}{2} \cdot \Vert \rho_1 - \rho_2 \Vert_1,
\end{equation*}
where $\Vert \cdot \Vert_1$ is the trace norm.
Given two quantum channels $\Phi_1, \Phi_2$, both with $n$-qubit inputs and outputs, their \emph{diamond distance} is given by
\begin{equation*}
    \mathrm{D}_{\diamond}(\Phi_1, \Phi_2) \coloneqq 
    \max_{\rho}\{\mathrm{D}_{\mathrm{tr}}((\Phi_1 \otimes \Id)(\rho), (\Phi_2 \otimes \Id)(\rho))\},
\end{equation*}
where the maximization is over all $2n$-qubit density matrices $\rho$, and both $\Id$ operators refer to the $n$-qubit identity channel.
For more background these distances, see~\cite[Chapter 3]{Wat18}.

Following~\cite{BEM+23}, we will define what it means to approximate a unitary in terms of the diamond distance.

\begin{definition}[Approximating a unitary]
Let $U$ be an $n$-qubit unitary,
and let $\Phi_U$ be the associated quantum channel.
Let $\Phi_{\mathrm{approx}}$ be a quantum channel with $n$-qubit input and output registers.
Then $\Phi_{\mathrm{approx}}$ is said to \emph{$\epsilon$-approximate $U$} if
$\mathrm{D}_{\diamond}(\Phi_{\mathrm{approx}}, \Phi_U) \leq \epsilon$.
\end{definition}

We will also define the channel associated with an oracle circuit $A^{(\cdot)}$ in the natural way.

\begin{definition}[Channel implemented by an oracle circuit]
    Given a $t$-query oracle circuit $A^{(\cdot)} = (n,m,\ell,U_1,\dots,U_t,U_{t+1})$ and a Boolean function $f: \{0,1\}^\ell \rightarrow \{\pm 1\}$, the associated \emph{$n$-qubit channel} $\Phi_{A^f}$ is defined as follows:
    \begin{enumerate}
        \item Given an $n$ qubit input $\ket{\psi}$, compute the state
        \begin{align*}
            U_{t+1} \cdot \calO_f \cdot U_t \cdot \cdots \calO_f \cdot U_2 \cdot \calO_f \cdot U_1 \cdot \ket{\psi} \ket*{0^{m-n}}
        \end{align*}
        \item Return the first $n$ qubits as the output (and discard the rest).
    \end{enumerate}
\end{definition}

With these definitions in hand, we can give a formal statement of the Unitary Synthesis Problem.

\begin{definition}[The Unitary Synthesis Problem]
Fix an error parameter $\epsilon(n) = 1/2^{\Omega(n)}$. Does there exist a $\poly(n)$-query oracle circuit $A^{(\cdot)}$ computable by a $\poly(n)$-sized quantum circuit such that for all $n$-qubit unitaries $U$, there exists a Boolean function $f: \{0,1\}^* \rightarrow \{\pm 1\}$ such that $\mathrm{D}_{\diamond}(\Phi_{A^f}, \Phi_U) \leq \epsilon(n)$?
\end{definition}

As discussed in~\cref{sec:background}, a bound on the maximum distinguishing advantage in the Oracle Distinguishing Game immediately implies a lower bound for the worst-case version of the Unitary Synthesis Problem, since there always \emph{exists} an information-theoretic distinguisher that wins the corresponding 
Oracle State Distinguishing Game for $R$. In fact, if we make a slight modification to the Oracle State Distinguishing Game, then a the distinguishing advantage bound would imply a slightly stronger claim, namely that Unitary Synthesis Problem is hard for a Haar-random $U$ (we note that this is technically the version of the problem stated by Aaronson and Kuperberg~\cite{AK07}).

In more detail, one can consider a variant of the Oracle State Distinguishing Game where every $\ket{\psi_{\bR_k}}$ is sampled as a Haar random state, rather than as a binary phase state (we do not give a separate analysis for the version of Oracle State Distinguishing with Haar random states, but our proof technique can easily be adapted to handle it). Next, suppose that there exists an oracle circuit $A^{(\cdot)}$ that can synthesize an $n$-qubit Haar random unitary $U$. Then for a random $U$ and any $K < N$, there exists (with high probability) a choice of $f$ such that $A^{f}$ implements the channel corresponding to $U$. In particular, this means that for a Haar-random subspace $S = \mathrm{span} \{ U^\dagger \ket{1},\dots, U^\dagger \ket{K}\}$, there exists $f$ such that $A^f$ maps $S$ to $\mathrm{span} \{ \ket{1},\dots,\ket{K}\}$. Such an oracle circuit $A^{(\cdot)}$ can be used to win the Oracle State Distinguishing Game, since the subspace $\mathrm{span} \{ \ket{\psi_{\bR_1}},\dots,\ket{\psi_{\bR_K}}\}$ is distributed as a $K$-dimensional Haar-random subspace, and the ability to map this subspace to $\mathrm{span} \{ \ket{1},\dots,\ket{K}\}$ immediately yields a distinguisher for the game.

To summarize, we have argued that a lower bound for breaking a (single-copy) pseudorandom state family --- in an oracle setting where the $K$ pseudorandom states are distributed as Haar random states --- directly implies hardness of synthesizing the first $K$ columns of a Haar-random unitary. Thus, we have the following claim.

\begin{claim}
    If the maximum distinguishing advantage of any efficient $t$-query adversary in the oracle distinguishing game is $o(1)$, then there is no efficient $t$-query oracle algorithm for the Unitary Synthesis Problem on a Haar-random unitary $U$.
\end{claim}

\subsection{Upper tail inequality for the maximum distinguishing advantage}\label{sec:game-concentration}
Throughout this subsection,
we will write $A^{(\cdot)} = (n, m, \ell, U_1, \ldots, U_{t+1})$ for a $t$-query adversary with an $(N \coloneqq 2^n)$-dimensional input register and $(L \coloneqq 2^{\ell})$-dimensional queries which is playing $\game^R$ for function families of the form $R:[K] \times [N] \rightarrow \{\pm 1\}$.

Let $\bprs:[K] \times [N] \rightarrow \{\pm 1\}$ be a uniformly random function family.
In this section, we consider the random variable $\distinguish_A(\bprs)$ corresponding to the maximum distinguishing advantage of $A$ (\cref{def:max-dist}),
and we show that it has strong one-sided concentration around its mean $\distinguish_A$.
Our main result is as follows.

\begin{lemma}[Upper tail for $\distinguish_A(\bro)$]
\label{lem:adv-tail}
    There exists a constant $c > 0$ such that the following is true.
    Let $\bprs:[K] \times[N] \rightarrow \{\pm 1\}$ be a uniformly random function family. Then for all $\epsilon \geq 0$,
    \[ \Pr_{\bprs}[\distinguish_{A}(\bprs) \geq \distinguish_{A}^{\avg} + \epsilon ] \leq 4\cdot\exp(-c \cdot \epsilon^2 KN). 
    \]
\end{lemma}

The main technical lemma we will need to prove this is the following version of Talagrand's concentration inequality, which is stated in \cite[Theorem 5.2.16]{Ver18}.

\begin{definition}[Lipschitz]
    Let $g:[-1,1]^n\rightarrow \R$ be a function.
    It has \emph{Lipschitz constant $C$} if for all $u, v \in [-1, 1]^n$,
    \begin{equation*}
        |g(u) - g(v)| \leq C \cdot \Vert u - v \Vert_2.
    \end{equation*}
\end{definition}
\begin{theorem}[Talagrand's concentration inequality]
\label{thm:talagrand}
    There exists a constant $c > 0$ such that the following is true.
    Let $g: [-1,1]^d \rightarrow \mathbb{R}$ be a convex function with Lipschitz constant $C$. Let $\bv_1,\dots,\bv_d$ be independent random variables satisfying $\abs{\bv_i} \leq 1$ for all $1 \leq i \leq d$. Then for all $t \geq 0$,
    \begin{align*}
        \Pr \left[ \abs{g(\bv_1,\dots,\bv_d) - \E[g(\bv_1,\dots,\bv_d)]} \geq t \right] \leq 2 \cdot \mathrm{exp} \left( -\frac{c \cdot t^2}{C^2} \right) .
    \end{align*}
\end{theorem}

To derive \Cref{lem:adv-tail} using Talagrand's concentration inequality,
we will view a uniformly random function family $\bprs:[K] \times[N] \rightarrow \{\pm 1\}$ as a collection of $KN$ independent $\{\pm 1\}$ random variables. 
We would then like to apply Talagrand's concentration inequality with the ``$g$'' function set to the maximum distinguishing advantage $\distinguish_A(\cdot)$, interpreted as a function of an input $R$.
However, doing so faces two difficulties:
first,~$\distinguish_A(\cdot)$ is defined only for $\{\pm 1\}$-valued inputs,
wheres the ``$g$'' function in Talagrand's concentration inequality must be defined over $[-1, 1]$ inputs.
Second, $\distinguish_A(\cdot)$ is not convex.
The first difficulty is straightforward to address,
and we begin to do so in the following definition.

\begin{definition}[Expanding the acceptance probability to bounded inputs]\label{def:bounded-inputs}
Let us fix a Boolean function $f:[L]\rightarrow \{\pm 1\}$.
Given as input the state $\ket{\psi}$, the adversary applies the oracle circuit $A^f$ and then measures the first qubit of the resulting state.
We can therefore view the adversary as applying a POVM measurement $E^f \coloneqq \{E_0^f, E_1^f\}$ to $\ket{\psi}$, where
\begin{equation*}
    E_0^f \coloneqq (A^f)^\dagger \cdot (\ketbra{0} \otimes \Id_2^{\otimes m-1}) \cdot A^f,
\end{equation*}
and $E_1^f \coloneqq \Id_N - E_0^f$.
As a result, for any Boolean function $h:[N] \rightarrow \{\pm 1\}$, we can write
\begin{equation}\label{eq:normal-definition}
    \accept_A(h \mid f)
    = \bra{\psi_h} \cdot E_0^f \cdot \ket{\psi_h}.
\end{equation}
We will now extend this expression to functions which are $[-1, 1]$-valued rather than $\{\pm 1\}$-valued.
For a bounded function $\underline{h}:[N] \rightarrow [-1,1]$ and a bit $b \in \{0, 1\}$, we define
\begin{equation*}
    \ket{\psi_{\underline{h}}} \coloneqq \frac{1}{\sqrt{N}} \cdot \sum_{x =1}^N \underline{h}(x) \cdot \ket{x},
    \quad
    \text{and}
    \quad
    \accept_{A,b}(\underline{h} \mid f)
    \coloneqq \bra{\psi_{\underline{h}}} \cdot E_b^f \cdot \ket{\psi_{\underline{h}}}.
\end{equation*}
\end{definition}

Note that $\ket{\psi_{\underline{h}}}$ is \emph{sub-normalized},
meaning that $\braket{\psi_{\underline{h}}} \leq 1$, and so it is no longer necessarily a quantum state.
In addition, if $h:[N]\rightarrow \{\pm 1\}$ is a Boolean function,
then by \Cref{eq:normal-definition},
$\accept_{A,b}(\underline{h} \mid f)$ still recovers our traditional definition of $\accept_A(\cdot \mid f)$ when $b = 0$.
As for the $b = 1$ case, note that because $h$ is a Boolean function,
\begin{align}
    \accept_{A,1}(h \mid f)
    = \bra{\psi_{h}} \cdot E_1^f \cdot \ket{\psi_{h}}
    &= \bra{\psi_{h}} \cdot (\Id_N - E_1^f) \cdot \ket{\psi_{h}}\nonumber\\
    &= 1 - \bra{\psi_{h}} \cdot E_0^f \cdot \ket{\psi_{h}}
    = 1- \accept_{A,0}(h \mid f).\label{eq:completely-obvious-equation}
\end{align}
However, this is not necessarily true of bounded functions $\underline{h}$.

Now we address the second issue, that of $\distinguish_A(\cdot)$ not being convex. 
To do so, we will have to define two variants of $\distinguish_A(\cdot)$ called $\distinguish_{A, 0}(\cdot)$ and $\distinguish_{A,1}(\cdot)$ which we will eventually show \emph{are} convex.
This motivates the following definition,
which will only be used in this subsection.

\begin{definition}[Modifying the distinguishing advantage]
Let $f:[L]\rightarrow \{\pm 1\}$ be a Boolean function
and $\underline{\prs}:[K] \times [N] \rightarrow [-1, 1]$ be a bounded function. 
For $b \in \{0, 1\}$, we define
\begin{align*}
    \distinguish_{A,b}^{}(\underline{\prs} \mid f)
    &\coloneqq \E_{\bk \sim [K]}[\accept_{A,b}(\underline{\prs}_{\bk} \mid f)]
    - \E_{\bh}[\accept_{A,b}(\bh \mid f)],\\
    \distinguish_{A,b}^{}(\underline{\prs})
    &\coloneqq \max_{f:[L] \rightarrow \{\pm 1\}}\{\distinguish_{A,b}^{}(\underline{\prs}\mid f)\}.
\end{align*}
We note that unlike in the definition of $\distinguish_{A}(\cdot)$,
there is no absolute value in the definition of $\distinguish_{A,b}(\cdot)$. (This is needed so that we can later show that it is convex.)
Finally, we define
\begin{equation*}
    \distinguish_{A,b}^{\avg} \coloneqq \E_{\bprs}[\distinguish_{A,b}^{}(\bprs)],
\end{equation*}
where $\bprs:[K] \times [N] \rightarrow \{\pm 1\}$ is a uniformly random function family.
\end{definition}

We will now make some observations about these definitions.
Let $\prs:[K]\times[N] \rightarrow \{\pm 1\}$ be a function family. Then by \Cref{eq:completely-obvious-equation},
\begin{align*}
    \distinguish_{A,1}^{}(\prs \mid f)
    &= \E_{\bk \sim [K]}[\accept_{A,1}(\prs_{\bk} \mid f)]
    - \E_{\bh}[\accept_{A,1}(\bh \mid f)]\\
    &= \E_{\bk \sim [K]}[1 - \accept_{A,0}(\prs_{\bk} \mid f)]
    - \E_{\bh}[1- \accept_{A,0}(\bh \mid f)]\\
    &= -\E_{\bk \sim [K]}[\accept_{A,0}(\prs_{\bk} \mid f)]
    + \E_{\bh}[\accept_{A,0}(\bh \mid f)]
    = - \distinguish_{A,0}^{}(\prs \mid f).
\end{align*}
(As before, this is not necessarily true of bounded functions $\underline{\prs}$.)
Thus, we have that
\begin{align*}
    \distinguish_A(\prs \mid f)
    = | \distinguish_{A, 0}^{}(\prs \mid f) |
    &= \max \{\distinguish_{A, 0}^{}(\prs \mid f), - \distinguish_{A, 0}^{}(\prs \mid f)\}\\
    &= \max \{\distinguish_{A, 0}^{}(\prs \mid f), \distinguish_{A, 1}^{}(\prs \mid f)\}.
\end{align*}
As a result,
\begin{align}
    \distinguish_{A}(\prs)
    &= \max_{f:[L] \rightarrow \{\pm 1\}}\{\distinguish_{A}(\prs \mid f)\}\nonumber\\
    &= \max_{f:[L] \rightarrow \{\pm 1\}}\big\{\max \{\distinguish_{A, 0}^{}(R \mid f), \distinguish_{A, 1}^{}(R \mid f)\}\big\}\nonumber\\
    &= \max \Big\{\max_{f:[L] \rightarrow \{\pm 1\}} \{\distinguish_{A, 0}^{}(R \mid f)\}, \max_{f:[L] \rightarrow \{\pm 1\}} \{\distinguish_{A, 1}^{}(R \mid f)\}\Big\}\nonumber\\
    & = \max\{\distinguish_{A,0}^{}(\prs), \distinguish_{A,1}^{}(\prs)\}.\label{eq:max-of-two}
\end{align}

We will show the following concentration bound for these two variants of the maximum distinguishing advantage.

\begin{lemma}[Concentration of the modified distinguishing advantages]
\label{lem:weird-advantages}
    There exists an absolute constant $c > 0$ such that the following is true.
    Let $b \in \{0, 1\}$.
    Let $\bprs:[K] \times[N] \rightarrow \{\pm 1\}$ be a uniformly random function family. Then for all $\epsilon \geq 0$,
    \[ \Pr_{\bprs}[|\distinguish^{}_{A,b}(\bprs)-\distinguish_{A,b}^{\avg}| \geq \epsilon ] \leq 2\cdot\exp(-c \cdot \epsilon^2 KN). 
    \]
\end{lemma}

Before proving \Cref{lem:weird-advantages}, let us see how it implies the main result of this subsection, \Cref{lem:adv-tail}.

\begin{proof}[Proof of \Cref{lem:adv-tail}]
    First, we note that by \Cref{eq:max-of-two}, $\distinguish_A(\prs)
    \geq\distinguish_{A,b}^{}(\prs)$ for any function family $\prs:[K]\times[N] \rightarrow \{\pm 1\}$ and any $b \in \{0, 1\}$.
    Thus, we have that $\distinguish_A^{\avg} \geq \distinguish_{A, b}^{\avg}$.
    Next, for a uniformly random function family $\bprs:[K] \times [N] \rightarrow \{\pm 1\}$,
    \begin{align*}
        &\Pr_{\bprs}[\distinguish_{A}(\bprs) \geq \distinguish_{A}^{\avg} + \epsilon ]\\
        ={}& \Pr_{\bprs}[\max \{\distinguish_{A,0}^{}(\bprs), \distinguish_{A,1}^{}(\bprs)\} \geq \distinguish_{A}^{\avg} + \epsilon ]\tag{by \Cref{eq:max-of-two}}\\
        ={}& \Pr_{\bprs}[\distinguish_{A,0}^{}(\bprs) \geq \distinguish_{A}^{\avg} + \epsilon, \text{  or  } \distinguish_{A,1}^{}(\bprs) \geq \distinguish_{A}^{\avg} + \epsilon ]\\
        \leq{}& \Pr_{\bprs}[\distinguish_{A,0}^{}(\bprs) \geq \distinguish_{A}^{\avg} + \epsilon]
        + \Pr_{\bprs}[\distinguish_{A,1}^{}(\bprs) \geq \distinguish_{A}^{\avg} + \epsilon ] \tag{by the union bound}\\
        \leq{}& \Pr_{\bprs}[\distinguish_{A,0}^{}(\bprs) \geq \distinguish_{A,0}^{\avg} + \epsilon]
        + \Pr_{\bprs}[\distinguish_{A,1}^{}(\bprs) \geq \distinguish_{A,1}^{\avg} + \epsilon ] \tag{because $\distinguish_A^{\avg} \geq \distinguish_{A,0}^{\avg}$ and $\distinguish_{A,1}^{\avg}$}\\
        \leq{}& \Pr_{\bprs}[|\distinguish_{A,0}^{}(\bprs) - \distinguish_{A,0}^{\avg}| \geq \epsilon]
        + \Pr_{\bprs}[|\distinguish_{A,1}^{}(\bprs) - \distinguish_{A,1}^{\avg}| \geq \epsilon]\\
        \leq{}& 2\cdot\exp(-c\cdot\epsilon^2 KN) + 2\cdot\exp(-c\cdot \epsilon^2 KN) \tag{by \Cref{lem:weird-advantages}}\\
        ={}& 4\cdot\exp(-c\cdot \epsilon^2 KN).
    \end{align*}
    This completes the proof.
\end{proof}

Now we focus on proving \Cref{lem:weird-advantages}.
To do so, we would like to show that $\distinguish^{}_{A,0}(\cdot)$ and $\distinguish^{}_{A,1}(\cdot)$ are convex and Lipschitz.
Prior to doing so, however,
we will first prove this for the $p_{A,b}(\cdot \mid f)$ function.

\begin{lemma}[The $p_{A,b}$ functions are convex and Lipschitz]\label{lem:accept-lipschitz}
    Let $f:[L] \rightarrow \{\pm 1\}$ be a Boolean function.
    Let $b \in \{0, 1\}$.
    Then $p_{A,b}( \cdot \mid f)$ is convex and $(2/\sqrt{N})$-Lipschitz.
\end{lemma}

\begin{proof}
Consider two bounded functions $\underline{h}, \underline{h}':[N]\rightarrow [-1,1]$.
Let $0 \leq t \leq 1$. Then
\begin{align*}
    \accept_{A,b}(t \cdot \underline{h} + (1-t) \cdot \underline{h}' \mid f) ={}& \bra*{\psi_{t \cdot \underline{h} + (1-t) \cdot \underline{h}'}} \cdot E_b^f \cdot \ket*{\psi_{t \cdot \underline{h} + (1-t) \cdot \underline{h}'}}\\
    ={}& \Vert (E_b^f)^{1/2} \cdot \ket*{\psi_{t \cdot \underline{h} + (1-t) \cdot \underline{h}'}} \Vert^2_2\\
    ={}& \Vert (E_b^f)^{1/2} \cdot (t \cdot \ket*{\psi_{\underline{h}}} + (1-t) \cdot \ket*{\psi_{\underline{h}'}}) \Vert^2_2.
\end{align*}
Now, because $\Vert \cdot \Vert_2$ is convex and $x \mapsto x^2$ is convex, we also have that $\Vert \cdot \Vert_2^2$ is convex. Hence, by Jensen's inequality, this is at most
\begin{align*}
    &  t \cdot \Vert (E_b^f)^{1/2} \ket*{\psi_{\underline{h}}} \Vert_2^2 + (1-t) \cdot \Vert (E_b^f)^{1/2} \cdot \ket*{\psi_{\underline{h}'}} \Vert^2_2 \\
    ={}& t \cdot \bra*{\psi_{\underline{h}}} \cdot E_b^f \cdot \ket*{\psi_{\underline{h}}} + (1-t) \cdot \bra*{\psi_{\underline{h}'}} \cdot E_b^f \cdot \ket*{\psi_{\underline{h}'}}\\
    ={}& t\cdot \accept_{A,b}(\underline{h} \mid f) + (1-t)\cdot \accept_{A,b}(\underline{h}' \mid f).
\end{align*}
And so $\accept_{A,b}(\cdot \mid f)$ is convex.
Next,
\begin{align*}
    &|p_{A,b}(\underline{h} \mid f)
        - p_{A,b}(\underline{h}' \mid f)|\\
    ={}& |\bra{\psi_{\underline{h}}} \cdot E_b^f \cdot \ket{\psi_{\underline{h}}} - \bra{\psi_{\underline{h}'}} \cdot E_b^f \cdot \ket{\psi_{\underline{h}'}}|\\
    ={}& |\bra{\psi_{\underline{h}}} \cdot E_b^f \cdot \ket{\psi_{\underline{h}}} - \bra{\psi_{\underline{h}}} \cdot E_b^f \cdot \ket{\psi_{\underline{h}'}}
    + \bra{\psi_{\underline{h}}} \cdot E_b^f \cdot \ket{\psi_{\underline{h}'}}
    - \bra{\psi_{\underline{h}'}} \cdot E_b^f \cdot \ket{\psi_{\underline{h}'}}|\\
    \leq{}& |\bra{\psi_{\underline{h}}} \cdot E_b^f \cdot \ket{\psi_{\underline{h}}} - \bra{\psi_{\underline{h}}} \cdot E_b^f \cdot \ket{\psi_{\underline{h}'}}|
    + |\bra{\psi_{\underline{h}}} \cdot E_b^f \cdot \ket{\psi_{\underline{h}'}}
    - \bra{\psi_{\underline{h}'}} \cdot E_b^f \cdot \ket{\psi_{\underline{h}'}}|\\
    ={}& |\bra{\psi_{\underline{h}}} \cdot E_b^f \cdot (\ket{\psi_{\underline{h}}} - \ket{\psi_{\underline{h}'}})|
    + |(\bra{\psi_{\underline{h}}} - \bra{\psi_{\underline{h}'}}) \cdot E_b^f \cdot \ket{\psi_{\underline{h}'}}|.
\end{align*}
By Cauchy-Schwarz, we can bound the first term by
\begin{align*}
    \Vert E_b^f \cdot \ket{\psi_{\underline{h}}} \Vert_2 \cdot \Vert \ket{\psi_{\underline{h}}} - \ket{\psi_{\underline{h}'}}\Vert_2
    &\leq 
    \Vert\ket{\psi_{\underline{h}}} \Vert_2 \cdot \Vert \ket{\psi_{\underline{h}}} - \ket{\psi_{\underline{h}'}}\Vert_2\tag{because $0 \preceq E_b^f \preceq I$}\\
    &\leq 
     \Vert \ket{\psi_{\underline{h}}} - \ket{\psi_{\underline{h}'}}\Vert_2 \tag{because $\ket{\psi_{\underline{h}}}$ is sub-normalized}\\
     &=\frac{1}{\sqrt{N}} \cdot \Vert h - h'\Vert_2. \tag{by definition of $\ket{\psi_h}$ and $\ket{\psi_{h'}}$}
\end{align*}
A similar argument shows that the second term is also bounded by $\Vert h - h'\Vert_2 / \sqrt{N}$. Putting these together, this shows that $p_{A,b}( \cdot \mid f)$ is $(2/\sqrt{N})$-Lipschitz.
\end{proof}

Next, we use this lemma to show that $\distinguish_{A,b}^{}(\cdot \mid f)$ is also convex and Lipschitz.

\begin{lemma}[The $\distinguish_{A,b}^{}(\cdot \mid f)$ functions are convex and Lipschitz]\label{lem:noabs-lipschitz}
    Let $f:[L] \rightarrow \{\pm 1\}$ be a Boolean function.
    Let $b \in \{0, 1\}$.
    Then the map
    \begin{equation}\label{eq:the-map}
        \underline{R} \mapsto \E_{\bk \sim [K]}[\accept_{A,b}(\underline{R}_{\bk} \mid f)]
    \end{equation}
    is convex and $(2/\sqrt{KN})$-Lipschitz.
    In addition, $\distinguish_{A,b}^{}(\cdot \mid f)$ is also convex and $(2/\sqrt{KN})$-Lipschitz.
\end{lemma}
\begin{proof}
We first prove the lemma for the map in \Cref{eq:the-map}.
Consider two bounded functions $\underline{R}, \underline{R}':[K]\times[N]\rightarrow [-1,1]$.
Let $0 \leq t \leq 1$. Then by \Cref{lem:accept-lipschitz},
\begin{align*}
     \E_{\bk \sim [K]}[\accept_{A,b}(t \cdot \underline{R}_{\bk} + (1-t) \cdot \underline{R}_{\bk}' \mid f)]
    &\leq  \E_{\bk \sim [K]}[t \cdot \accept_{A,b}(\underline{R}_{\bk}\mid f) + (1-t) \cdot \accept_{A,b}(\underline{R}_{\bk}' \mid f)]\\
    &= t\cdot\E_{\bk \sim[K]}[\accept_{A,b}(\underline{R}_{\bk} \mid f)] + (1-t) \cdot \E_{\bk \sim[K]}[\accept_{A,b}(\underline{R}'_{\bk} \mid f)].
\end{align*}
Thus, this map is convex. Next,
    \begin{align*}
        \Big|\E_{\bk \sim [K]}[\accept_{A,b}(\underline{\prs}_{\bk} \mid f)]
    - \E_{\bk \sim [K]}[\accept_{A,b}(\underline{\prs}_{\bk}'\mid f)]\Big|
 \leq \E_{\bk \sim [K]}|\accept_{A,b}(\underline{\prs}_{\bk} \mid f)
    - \accept_{A,b}(\underline{\prs}_{\bk}'\mid f)|.
    \end{align*}
    Now, we apply \Cref{lem:accept-lipschitz}, which states that $\accept_{A,b}(\cdot \mid f)$ is $(2/\sqrt{N})$-Lipschitz. Hence, we can upper-bound this by
    \begin{align*}
  \E_{\bk \sim [K]}\frac{2}{\sqrt{N}} \cdot \Vert \underline{\prs}_{\bk} - \underline{\prs}_{\bk}'\Vert_2 
={}& \frac{2}{\sqrt{N}\cdot K} \cdot \sum_{k=1}^K \Vert \underline{\prs}_{k} - \underline{\prs}_{k}'\Vert_2 \\
\leq{}& \frac{2}{\sqrt{N}\cdot K} \cdot \sqrt{K \cdot \sum_{k=1}^K \Vert \underline{\prs}_{k} - \underline{\prs}_{k}'\Vert_2^2 } 
= \frac{2}{\sqrt{KN}} \cdot \Vert \underline{R} - \underline{R}'\Vert_2,
\end{align*}
where the inequality is due to Cauchy-Schwarz.
Thus, this map is $(2/\sqrt{KN})$-Lipschitz.
As for $\distinguish_{A,b}^{}(\cdot \mid f)$,
we recall that it is defined as follows:
\begin{equation*}
    \distinguish_{A,b}^{}(\underline{\prs} \mid f)
    \coloneqq \E_{\bk \sim [K]}[\accept_{A,b}(\underline{\prs}_{\bk} \mid f)]
    - \E_{\bh}[\accept_{A,b}(\bh \mid f)].
\end{equation*}
This is just the map in \Cref{eq:the-map}, offset by a constant.
Hence, it too is convex and $(2/\sqrt{KN})$-Lipschitz.
This completes the proof.
\end{proof}

We have finally reached our goal, which is to show that the $\distinguish_{A,b}^{}(\cdot) = \max_f \distinguish_{A,b}(\cdot \mid f)$ functions are convex and Lipschitz.

\begin{lemma}[The $\distinguish_{A,b}^{}(\cdot)$ functions are convex and Lipschitz]\label{lem:whatever-im-bored}
    Let $b \in \{0, 1\}$.
    Then $\distinguish_{A,b}^{}(\cdot)$ is convex and $(2/\sqrt{KN})$-Lipschitz.
\end{lemma}

\begin{proof}
Consider two bounded functions $\underline{R}, \underline{R}':[K]\times[N]\rightarrow [-1,1]$.
Let $0 \leq t \leq 1$. Then
\begin{equation*}
    \distinguish_{A,b}^{}(t \cdot \underline{R} + (1-t) \cdot \underline{R}')
    = \max_{f:[L] \rightarrow \{\pm 1\}}\{\distinguish_{A,b}^{}(t \cdot \underline{R} + (1-t) \cdot \underline{R}'\mid f)\}
\end{equation*}
By \Cref{lem:noabs-lipschitz}, the function $\distinguish_{A,b}^{}(\cdot \mid f)$ is convex. Hence, this is at most
\begin{align*}
    &\max_{f:[L] \rightarrow \{\pm 1\}}\{t \cdot \distinguish_{A,b}^{}(\underline{R}\mid f) + (1-t) \cdot \distinguish_{A,b}^{}(\underline{R}' \mid f)\}\\
    \leq{}& \max_{f:[L] \rightarrow \{\pm 1\}}\{t \cdot \distinguish_{A,b}^{}(\underline{R}\mid f) + (1-t) \cdot \distinguish_{A,b}^{}(\underline{R}' \mid f)\}\\
    \leq{}& t \cdot \max_{f:[L] \rightarrow \{\pm 1\}}\{\distinguish_{A,b}^{}(\underline{R}\mid f)\} + (1-t) \cdot \max_{f:[L] \rightarrow \{\pm 1\}}\{\distinguish_{A,b}^{}(\underline{R}' \mid f)\}\\
    ={}& t \cdot \distinguish_{A,b}^{}( \underline{R}) + (1-t) \cdot \distinguish_{A,b}^{}( \underline{R}').
\end{align*}
Hence, $\distinguish_{A,b}^{}(\cdot)$ is convex.

Now we show that $\distinguish_{A,b}^{}(\cdot)$
is Lipschitz.
To do so, we will show that for any two bounded functions $\underline{R}, \underline{R}':[K]\times[N]\rightarrow [-1,1]$,
\begin{equation*}
\distinguish_{A,b}^{}(\underline{\prs})
    - \distinguish_{A,b}^{}(\underline{\prs}') \leq \frac{2}{\sqrt{KN}}.
    \end{equation*}
This will show that $\distinguish_{A,b}^{}(\cdot)$
is $(2/\sqrt{KN})$-Lipschitz, as
\begin{equation*}
    |\distinguish_{A,b}^{}(\underline{\prs})
    - \distinguish_{A,b}^{}(\underline{\prs}')|
    = \max\{\distinguish_{A,b}^{}(\underline{\prs})
    - \distinguish_{A,b}^{}(\underline{\prs}'), \distinguish_{A,b}^{}(\underline{\prs}')
    - \distinguish_{A,b}^{}(\underline{\prs})\}
    \leq \frac{2}{\sqrt{KN}}.
\end{equation*}
To begin,
\begin{align*}
    \distinguish_{A,b}^{}(\underline{\prs})
    - \distinguish_{A,b}^{}(\underline{\prs}')
    &= \max_{f:[L] \rightarrow \{\pm 1\}}\{\distinguish_{A,b}^{}(\underline{\prs} \mid f)\} - \max_{f':[L] \rightarrow \{\pm 1\}}\{\distinguish_{A,b}^{}(\underline{\prs}' \mid f')\}.
\end{align*}
Let $f$ be function maximizing the first expression.
Then this is equal to
\begin{align*}
\distinguish_{A,b}^{}(\underline{\prs} \mid f) - \max_{f':[L] \rightarrow \{\pm 1\}}\{\distinguish_{A,b}^{}(\underline{\prs}' \mid f')\}
&\leq 
\distinguish_{A,b}^{}(\underline{\prs} \mid f) - \distinguish_{A,b}^{}(\underline{\prs}' \mid f)\\
&\leq \frac{2}{\sqrt{KN}}\cdot \Vert \underline{R} - \underline{R}'\Vert_2. \tag{by \Cref{lem:noabs-lipschitz}}
\end{align*}
This completes the proof.
\end{proof}

With this in hand, we can finally prove \Cref{lem:weird-advantages}.

\begin{proof}[Proof of \Cref{lem:weird-advantages}]
By \Cref{lem:whatever-im-bored},
$\distinguish_{A,b}^{}(\cdot)$ is convex and has Lipschitz constant $(2/\sqrt{KN})$.
Let $\bprs:[k] \times [N] \rightarrow \{\pm 1\}$ be a uniformly random function family. 
Viewing $\bprs$ as a collection of $KN$ independent $\{\pm 1\}$ random variables, we can apply Talagrand's concentration inequality, which states that there exists an absolute constant $c > 0$ such that
\begin{equation*}
    \Pr_{\bprs}[|\distinguish_{A,b}^{}(\bprs) - \E_{\bprs}[\distinguish_{A,b}^{}(\bprs)]| \geq \epsilon]\leq 2 \cdot \mathrm{exp}\Big(- \frac{c \cdot \epsilon^2}{(2/ \sqrt{KN})^2}\Big)
    = 2 \cdot \mathrm{exp}\Big(- \Big(\frac{c}{4}\Big) \cdot \epsilon^2 K N \Big).
\end{equation*}
Recalling that $\distinguish_{A,b}^{\avg}= \E_{\bprs}[\distinguish_{A,b}^{}(\bprs)]$, this completes the proof.
\end{proof}


\subsection{The adversary's space is bounded without loss of generality}\label{sec:adversary-space}

In this subsection, we will show that if $A^{(\cdot)}$ is an oracle circuit that makes $t$ queries, each of which has size at most $\ell$, then we can assume without loss of generality that $A^{(\cdot)}$ uses at most $t\cdot \ell$ ancilla qubits, in addition to the $n$ qubits in its input register. We prove this by showing that for any such oracle circuit (that potentially uses unbounded space), there is an oracle circuit $B^{(\cdot)}$ that \emph{simulates} $A^{(\cdot)}$ using only $t\cdot \ell$ ancilla qubits.
This will allow us to restrict our attention to adversaries that are space-efficient when proving our one-query lower bounds,
which is necessary given the technical tools we apply.
We begin by defining in what sense $B^{(\cdot)}$ simulates $A^{(\cdot)}$.

\begin{notation}[Query register]
    In this subsection,
    we will assume that every oracle circuit makes an oracle call on a register of exactly $\ell$ qubits.
    We will write $L = 2^{\ell}$ for the dimension of this register,
    and we will write $\queryreg \coloneqq \C^L$
    for the vector space corresponding to this register.
\end{notation}


\begin{definition}[Oracle circuit simulation]
\label{def:equiv-oalg}
    Let $m_B \leq m_A$ be integers. Consider two $t$-query oracle circuits
    \begin{align*}
    A^{(\cdot)} &= (n,m_A = \log_2(\qdim_A),\ell,U_1^A, \ldots, U_{t+1}^A)\text{, and}\\
    B^{(\cdot)} &= (n,m_B = \log_2(\qdim_B),\ell,U_1^B, \ldots, U_{t+1}^B)
    \end{align*}
    with ancilla dimensions $D_A \coloneqq 2^{m_A - \ell}$ and $D_B \coloneqq 2^{m_B - \ell}$, respectively.
    Then $B^{(\cdot)}$ \emph{simulates} $A^{(\cdot)}$ if there exists an isometry $T: \C^{\qdim_B} \rightarrow \C^{\qdim_A}$ such that for all Boolean functions $f : [L] \rightarrow \{\pm 1\}$,
    \begin{align*}
    &{} U_{t+1}^A \cdot (\calO_f \otimes \Id_{D_A}) \cdot U_t^A \cdots (\calO_f \otimes \Id_{D_A}) \cdot U_1^A \cdot (\Id_N \otimes \ket*{0^{m_A - n}}) \\
    ={}& T \cdot U_{t+1}^B \cdot (\calO_f \otimes \Id_{D_B}) \cdot U_t^B \cdots (\calO_f \otimes \Id_{D_B}) \cdot U_1^B \cdot (\Id_N \otimes \ket*{0^{m_B - n}}).
    \end{align*}
\end{definition}

Now we state the main lemma of this section,
namely that an oracle circuit that makes $t$ queries of size $\ell$ can be converted to one of space $n + t\cdot \ell$.
Typical values for these parameters are $t, \ell = \poly(n)$, in which case this results in an oracle circuit of $\poly(n)$ space.

\begin{lemma}[Space reduction for oracle circuits]\label{lemma:main-space-shrink}
    Consider a $t$-query oracle circuit
    \begin{equation*}
        A^{(\cdot)} = (n,m_A,\ell,U_1^A, \ldots, U_{t+1}^A).
    \end{equation*}
    Then $A^{(\cdot)}$ can be simulated by a $t$-query oracle circuit
    \begin{equation*}
        B^{(\cdot)} = (n,m_B,\ell,U_1^B, \ldots, U_{t+1}^B).
    \end{equation*}
    that uses $m_B = (n + t \cdot \ell)$ qubits of space.
\end{lemma}

The key technical ingredient we will use in the proof of this lemma is the following method for compressing an isometry with a large output dimension into an isometry with a small output dimension.

\begin{definition}[Compression of an isometry]
    Let $V: \C^D \rightarrow \queryreg \otimes \C^{S}$ be an isometry.
    Then the \emph{compression of $V$} is the isometry
    $\compress(V): \C^D \rightarrow \queryreg \otimes \C^{D}$ defined as
    \begin{equation*}
        \compress(V) \coloneqq \sum_{z=1}^L \ket{z} \otimes \sqrt{M_z},
    \end{equation*}
    where $M_z\coloneqq V^\dagger \cdot ( \ketbra{z}\otimes \Id_S ) \cdot V.$
\end{definition}
To get intuition for this definition, note that the operators $\{M_z\}$ correspond to the following measurement: first apply the original isometry $V$, and then measure the resulting query register to obtain an outcome $z$. As a result, $\compress(V)$ is the natural isometry that corresponds to the $\{M_z\}$ measurement. We note that $\compress(V)$ \emph{is} indeed an isometry, because
\begin{align*}
    \compress(V)^\dagger \cdot \compress(V)
    &= \Big(\sum_{z=1}^L \bra{z} \otimes \sqrt{M_z}\Big)
        \cdot \Big(\sum_{z=1}^L \ket{z} \otimes \sqrt{M_z}\Big)\\
    &= \sum_{z=1}^L M_z
    = \sum_{z=1}^L V^\dagger \cdot ( \ketbra{z}\otimes \Id_S ) \cdot V
    = V^\dagger \cdot \Id_{\mathrm{query}} \otimes \Id_{S} \cdot V
    = \Id_{B},
\end{align*}
where the last step used the assumption that $V$ is an isometry.
The following technical lemma gives one sense in which $\compress(V)$ does indeed compress $V$, in that whenever $V$ is used to temporarily transition into $\queryreg \otimes \C^{S}$ in order to query an oracle,
we can use $\compress(V)$ to move into $\queryreg \otimes \C^{D}$ instead
with the exact same results.

\begin{lemma}[\textsf{compress} compresses]\label{lem:comp-comp}
    Let $V: \C^D \rightarrow \queryreg \otimes \C^{S}$ be an isometry.
    Then for every function $f:[L]\rightarrow \{\pm 1\}$,
    \begin{equation*}
        \compress(V)^\dagger \cdot \Big( \calO_f \otimes \Id_D \Big) \cdot \compress(V)
        = V^\dagger \cdot \Big( \calO_f \otimes \Id_S \Big) \cdot V.
    \end{equation*}
\end{lemma}
\begin{proof}
The proof is via a straightforward calculation:
    \begin{align*}
    &\compress(V)^\dagger \cdot \Big( \calO_f \otimes \Id_D \Big) \cdot \compress(V) \\
    ={}& \Big(\sum_{z=1}^L \bra{z} \otimes \sqrt{M_z} \Big) \cdot \Big( \sum_{z =1}^L f(z)\cdot \ketbra{z} \otimes \Id_D \Big) \cdot \Big(\sum_{z=1}^L \ket{z} \otimes \sqrt{M_z} \Big)\\
    ={}& \sum_{z=1}^L f(z)\cdot M_z\\
    ={}&  \sum_{z=1}^L f(z) \cdot V^\dagger \cdot (\ketbra{z} \otimes \Id_S) \cdot V \\
    ={}&  V^\dagger \cdot \Big( \sum_{z=1}^L f(z) \cdot \ketbra{z} \otimes \Id_S \Big) \cdot V\\
    ={}&  V^\dagger \cdot \Big( \calO_f \otimes \Id_S \Big) \cdot V.
\end{align*}
That completes the proof.
\end{proof}

Now we use this technical lemma to show that the action of $\compress(V)$ followed by an oracle
is actually \emph{equivalent} to the action of $V$ followed by an oracle,
up to an isometry.

\begin{lemma}[Equivalence of compressed and uncompressed isometry]\label{lem:equivalence-of-compression}
    Let $V: \C^D \rightarrow \queryreg \otimes \C^{S}$ be an isometry.
    Then there exists an isometry $T:\queryreg \otimes \C^D \rightarrow \queryreg \otimes \C^S$ such that for all Boolean functions $f:[L]\rightarrow \{\pm 1\}$,
    \begin{equation*}
        T \cdot (\calO_f \otimes \Id_D) \cdot \compress(V) = (\calO_f \otimes \Id_S) \cdot V.
    \end{equation*}
\end{lemma}

Prior to proving this lemma, we will establish the following linear-algebraic proposition.
We expect that this proposition is well-known,
although we were unable to find a reference for it.

\begin{proposition}[Matching inner products implies an isometry]
\label{prop:inner-prod-isometry}
    Let $d_1 \leq d_2$ be integers.
    Consider two sets of $m$ vectors
    $\ket{x_1},\dots,\ket{x_m} \in \mathbb{C}^{d_1}$
    and $\ket{y_1},\dots,\ket{y_m} \in \mathbb{C}^{d_2}$.
    Suppose that these sets have the same pairwise inner products, meaning that
    \begin{align*}
        \braket{x_i}{x_j} = \braket{y_i}{y_j},
    \end{align*}
    for all $1 \leq i, j \leq m.$
    Then there exists an isometry $T: \mathbb{C}^{d_1} \rightarrow \mathbb{C}^{d_2}$ such that $T \cdot \ket{x_i} = \ket{y_i}$, for all $1 \leq i \leq m$. 
\end{proposition}

\begin{proof}
    Define $X \coloneqq \sum_{i = 1}^m \ketbra{x_i}{i}$ and $Y = \sum_{i=1}^m \ketbra{y_i}{i}$. Because the two sets of vectors have matching inner products,
    \begin{align}
        X^\dagger \cdot X
        = \Big(\sum_{i = 1}^m \ketbra{i}{x_i}\Big) \cdot \Big(\sum_{j = 1}^m \ketbra{x_j}{j}\Big)
        &= \sum_{i,j=1}^m \braket{x_i}{x_j} \cdot \ketbra{i}{j}\label{eq:xx-yy}\\
        &= \sum_{i,j=1}^m \braket{y_i}{y_j} \cdot \ketbra{i}{j}
        = \Big(\sum_{i = 1}^m \ketbra{i}{y_i}\Big) \cdot \Big(\sum_{j = 1}^m \ketbra{y_j}{j}\Big)
        = Y^\dagger \cdot Y.\nonumber
    \end{align}
    Given a complex matrix $A$, we will denote by $A^+$ the \emph{Moore-Penrose pseudo-inverse} of $A$.
    The one fact we will use about the pseudo-inverse, which can be found in \cite[Proposition 4.9.2]{Pet12},
    is that $A^+ \cdot A$ is the projector onto the image of $A^\dagger$.
    Multiplying both sides of \Cref{eq:xx-yy} by $(Y^\dagger)^+$ yields
    \begin{equation*}
        (Y^\dagger)^+ \cdot X^\dagger \cdot X
        = (Y^\dagger)^+ \cdot Y^\dagger \cdot Y.
    \end{equation*}
    From our pseudo-inverse fact,
    $(Y^\dagger)^+ \cdot Y^\dagger$
    is the projector onto the image of $(Y^\dagger)^\dagger = Y$.
    Hence, $(Y^\dagger)^+ \cdot Y^\dagger \cdot Y = Y$.
    \begin{equation*}
        (Y^\dagger)^+ \cdot X^\dagger \cdot X
        = Y.
    \end{equation*}
    Now, let us define $T \coloneqq (Y^\dagger)^+ \cdot X^\dagger$,
    so that $T \cdot X = Y$.
    Note that for all $1 \leq i \leq m$,
    this implies that
    \begin{equation}\label{eq:T-maps-X-to-Y}
        T \cdot \ket{x_i}
        = T \cdot X \cdot \ket{i}
        = Y \cdot \ket{i}
        = \ket{y_i},
    \end{equation}
    as desired.
    Next, write
    \begin{equation*}
        \mathsf{span}_X \coloneqq \mathsf{span}\{\ket{x_1}, \ldots, \ket{x_m}\}
        \quad\text{and}\quad
        \mathsf{span}_Y \coloneqq \mathsf{span}\{\ket{y_1}, \ldots, \ket{y_m}\}.
    \end{equation*}
    Then we claim
    (i) $T$ maps any vector in $\mathsf{span}_X^\perp$ to 0, and
    (ii) $T$ is an isometry from $\mathsf{span}_X$ to $\mathsf{span}_Y$.
    We prove these as follows.
    \begin{enumerate}[label=(\roman*)]
        \item Let $\ket{u} \in \mathsf{span}_X^\perp$. Because $X^\dagger = \sum_{i = 1}^m \ketbra{i}{x_i}$,  we have that $X^\dagger \cdot \ket{u} = 0$.
    Thus, 
    \begin{equation*}
        T \cdot \ket{u} = (Y^\dagger)^+ \cdot X^\dagger \cdot \ket{u} = 0.
    \end{equation*}
        \item To show that $T$ is an isometry mapping $\mathsf{span}_X$ to $\mathsf{span}_Y$,
        it suffices to show that it maps any vector in $\mathsf{span}_X$ to $\mathsf{span}_Y$,
        and that it preserves lengths.
        Let $\ket{v} \in \mathsf{span}_X$.
            Then $\ket{v} = \alpha_1 \cdot \ket{x_1} + \cdots + \alpha_m \cdot \ket{x_m}$ for some complex coefficients $\alpha_1, \ldots, \alpha_m$.
            By \Cref{eq:T-maps-X-to-Y},
        \begin{equation*}
            T \cdot \ket{v}
            = T \cdot\Big(\sum_{i=1}^m \alpha_i \cdot \ket{x_i}\Big)
            = \sum_{i=1}^m \alpha_i \cdot T \cdot \ket{x_i}
            = \sum_{i=1}^m \alpha_i \cdot \ket{y_i},
        \end{equation*}
            which is indeed an element of $\mathsf{span}_Y$.
        Next, the squared length of $\ket{v}$ is
        \begin{align*}
        \braket{v}
        &= \Big(\sum_{i=1}^m \alpha_i^\dagger \cdot \bra{x_i}\Big)\cdot \Big(\sum_{j=1}^m \alpha_j \cdot \ket{x_j}\Big)\\
        &= \sum_{i, j=1}^m \alpha_i^\dagger \alpha_j \cdot \braket{x_i}{x_j}\\
        &= \sum_{i, j=1}^m \alpha_i^\dagger \alpha_j \cdot \braket{y_i}{y_j}\\
        &= \Big(\sum_{i=1}^m \alpha_i^\dagger \cdot \bra{y_i}\Big)\cdot \Big(\sum_{j=1}^m \alpha_j^\dagger \cdot \ket{y_j}\Big)
        = (\bra{v} \cdot T^\dagger) \cdot (T \cdot \ket{v}),
        \end{align*}
        which is the squared length of $T \cdot \ket{v}$.
        This proves the claim.
    \end{enumerate}
    Hence, $T$ is an isometry mapping $\mathsf{span}_X$ to $\mathsf{span}_Y$, acts as 0 outside of $\mathsf{span}_X$,
    and satisfies $T \cdot \ket{x_i} = \ket{y_i}$, for all $1 \leq i \leq m$.
    As a result, it can be extended to an isometry mapping $\mathbb{C}^{d_1}$ to $\mathbb{C}^{d_2}$ which satisfies this property by picking any isometry that maps $\mathsf{span}_X^\perp$ to $\mathsf{span}_Y^\perp$.
    This gives the desired construction.
\end{proof}

Now we prove \Cref{lem:equivalence-of-compression}.

\begin{proof}[Proof of \Cref{lem:equivalence-of-compression}]
    For each $f: [L] \rightarrow \{\pm 1\}$ and $1 \leq x \leq D$, define
    \begin{align*}
    \ket*{\Phi_{f,x}} \coloneqq (\calO_f \otimes \Id_S) \cdot V \cdot\ket{x}, \ \text{and} \ \ket*{\widehat{\Phi}_{f,x}} \coloneqq (\calO_f \otimes \Id_D) \cdot \compress(V)\cdot \ket{x}.
    \end{align*}
    We will prove that there exists an isometry $T:\queryreg \otimes \C^D \rightarrow \queryreg \otimes \C^S$ such that
    \begin{equation*}
    T \cdot \ket*{\widehat{\Phi}_{f,z}} = \ket*{\Phi_{f,x}},
    \end{equation*}
    for all $1 \leq x \leq D$ and Boolean functions $f:[L] \rightarrow \{\pm 1\}$.
    This will in turn imply the desired claim by linearity.
    By~\cref{prop:inner-prod-isometry}, it suffices to show that $\{ \ket*{\Phi_{f,x}}\}_{f,x}$ and $\{ \ket*{\widehat{\Phi}_{f,x}}\}_{f,x}$ have the same pairwise inner products, i.e.
    \begin{align}
    \braket*{\widehat{\Phi}_{f,x}}{\widehat{\Phi}_{g,y}} = \braket*{\Phi_{f,x}}{\Phi_{g,y}},
    \end{align}
     for all $1 \leq x,y \leq D$ and Boolean functions $f,g : [L] \rightarrow \{\pm 1\}$.
    To complete the proof, we verify this by direct calculation for all $x,y,f,g$:
    \begin{align*}
    \braket*{\widehat{\Phi}_{f,x}}{\widehat{\Phi}_{g,y}}
    &= \bra{x} \cdot \compress(V)^\dagger \cdot \Big( \calO_f \otimes \Id_D \Big) \cdot \Big( \calO_g \otimes \Id_D \Big) \cdot \compress(V) \cdot \ket{y} \\
    &= \bra{x} \cdot \compress(V)^\dagger \cdot \Big( \calO_{f \cdot g} \otimes \Id_D \Big) \cdot \compress(V) \cdot \ket{y} \\
    &= \bra{x} \cdot V^\dagger \cdot \Big( \calO_{f \cdot g} \otimes \Id_S \Big) \cdot V \cdot \ket{y} \tag{by \Cref{lem:comp-comp}}\\
    &= \bra{x} \cdot V^\dagger \cdot \Big( \calO_f \otimes \Id_S \Big) \cdot \Big( \calO_g \otimes \Id_S \Big) \cdot V \cdot \ket{y} \\
    &= \braket*{\Phi_{f,x}}{\Phi_{g,y}},
\end{align*}
where $f \cdot g$ is the Boolean function defined as
$(f \cdot g)(x) = f(x) \cdot g(x)$, for all $1 \leq x \leq D$.
This completes the proof.
\end{proof}

With this in hand, we can finally prove the main result of this section, \Cref{lemma:main-space-shrink}.

\begin{proof}[Proof of \Cref{lemma:main-space-shrink}]
    In this proof, we will construct a sequence of isometries $V_1, \ldots, V_t$
    in which for each $1 \leq i \leq t$,
    \begin{equation}
        V_i:\C^N \otimes (\C^L)^{\otimes i - 1} \rightarrow \queryreg \otimes \C^N \otimes (\C^L)^{\otimes i - 1}
        = \C^N \otimes (\C^L)^{\otimes i}. \label{eq:isometry-V}
    \end{equation}
    Given a Boolean function $f:[L]\rightarrow \{\pm 1\}$,
    we will use the shorthand
    \begin{align*}
        \mathsf{Prod}_{U,f, i}&\coloneqq(\calO_f \otimes \Id_{D_A}) \cdot U_i^A \cdots (\calO_f \otimes \Id_{D_A}) \cdot U_2^A \cdot(\calO_f \otimes \Id_{D_A}) \cdot U_1^A \cdot (\Id_N \otimes \ket*{0^{m_A - n}}),\\
        \mathsf{Prod}_{V,f, i}&\coloneqq (\calO_f \otimes \Id_{NL^{i-1}}) \cdot V_i \cdots (\calO_f \otimes \Id_{NL}) \cdot V_2\cdot (\calO_f \otimes \Id_{N}) \cdot V_1 \cdot \Id_N.
    \end{align*}
    Operationally, $\mathsf{Prod}_{U, f, i}$ corresponds to alternating between $i$ unitaries and oracle calls, and similarly for $\mathsf{Prod}_{V,f, i}$.
    
    We will first prove the following statement:
    for each $0 \leq i \leq t$,
    there exists an isometry $T_i:\queryreg \otimes \C^N \otimes (\C^L)^{\otimes i - 1} \rightarrow \C^{\qdim_A}$, such that for every Boolean function $f:[L]\rightarrow \{\pm 1\}$,
    \begin{align}
    T_i \cdot \mathsf{Prod}_{V,f, i} = \mathsf{Prod}_{U,f, i}. \label{eq:isometry-T}
    \end{align}
    At the end, we will derive \cref{lemma:main-space-shrink} from this statement.
    
    The proof is by induction on $t$, the base case being $t = 0$.
    In this case, the statement follows from setting $T_0: \C^N \rightarrow \C^{\qdim_A}$ as $T_0\coloneqq \Id_N \otimes \ket{0^{m_A - n}}$.
    This is because
    \begin{equation*}
        T_0 \cdot \mathsf{Prod}_{V,f, 0} = T_0 \cdot \Id_N = \Id_N \otimes \ket*{0^{m_A - n}} = \mathsf{Prod}_{U,f, 0},
    \end{equation*}
    as desired.
    As for the induction step we suppose it is true for $i \leq t -1$ and prove that it holds for $i + 1$.
    By the induction hypothesis, we have that
    \begin{align}\label{eq:apply-induction}
        \mathsf{Prod}_{U,f, i+1}
        = (\calO_f \otimes \Id_{D_A}) \cdot U_{i+1}^A \cdot \mathsf{Prod}_{U,f, i}
        = (\calO_f \otimes \Id_{D_A}) \cdot U_{i+1}^A \cdot T_i \cdot \mathsf{Prod}_{V,f, i}.
    \end{align}
    Note that $U_{i+1}^A \cdot T_i$ is an isometry mapping $\C^N \otimes (\C^L)^{\otimes i}$ to $\queryreg \otimes \C^{D_A}$.
    Thus, if we set $V_{i+1} \coloneqq \compress(U^A_{i+1} \cdot T_{i})$, then
    \begin{equation*}
        V_{i+1}:\C^N \otimes (\C^L)^{\otimes i} \rightarrow \queryreg \otimes \C^N \otimes (\C^L)^{\otimes i},
    \end{equation*}
    as desired. Applying \Cref{lem:equivalence-of-compression}, there exists an isometry
    \begin{equation*}
        T_{i+1} :\queryreg \otimes C^N \otimes (\C^L)^{\otimes i} \rightarrow \queryreg \otimes \C^{D_A}
    \end{equation*}
    such that for all Boolean functions $f:[L]\rightarrow \{\pm 1\}$,
    \begin{equation*}
        T_{i+1} \cdot (\calO_f \otimes \Id_{NL^i}) \cdot V_{i+1} = (\calO_f \otimes \Id_{D_A}) \cdot U_{i+1}^A \cdot T_i.
    \end{equation*}
    Plugging this into \Cref{eq:apply-induction}, we have that
    \begin{equation*}
        \mathsf{Prod}_{U,f, i+1}
        = T_{i+1} \cdot (\calO_f \otimes \Id_{NL^i}) \cdot V_{i+1} \cdot \mathsf{Prod}_{V,f, i}
        = T_{i+1} \cdot \mathsf{Prod}_{V,f, i+1}.
    \end{equation*}
    Thus, the $(i+1)$ case of the statement is also true, completing the proof by induction.

    It remains to show that the existence of isometries $V_1,\dots,V_t$ and $T_0,\dots,T_t$ satisfying~\cref{eq:isometry-V,eq:isometry-T} implies~\cref{lemma:main-space-shrink}. Recall that our goal is to construct an oracle circuit $B^{(\cdot)} = (n,m_B,\ell,U_1^B,\dots,U_{t+1}^B)$ that uses $m_B = (n + t\cdot \ell)$ qubits of space and \emph{simulates} $A^{(\cdot)}$ in the sense of~\cref{def:equiv-oalg}: namely, there exists an isometry $T: \C^{\qdim_B} \rightarrow \C^{\qdim_A}$ such that for all Boolean functions $f : [L] \rightarrow \{\pm 1\}$,
    \begin{align}
    &{} U_{t+1}^A \cdot (\calO_f \otimes \Id_{D_A}) \cdot U_t^A \cdots (\calO_f \otimes \Id_{D_A}) \cdot U_1^A \cdot (\Id_N \otimes \ket*{0^{m_A - n}}) \nonumber \\
    ={}& T \cdot U_{t+1}^B \cdot (\calO_f \otimes \Id_{D_B}) \cdot U_t^B \cdots (\calO_f \otimes \Id_{D_B}) \cdot U_1^B \cdot (\Id_N \otimes \ket*{0^{m_B - n}}), \label{eq:isometry-guarantee}
    \end{align} 
    where $D_A = 2^{m_A - \ell}$ and $D_B = 2^{m_B - \ell}$. To this end, for $1 \leq i \leq t$, we will extend each isometry $V_i: \C^N \otimes (\C^L)^{\otimes i-1} \rightarrow \C^N \otimes (\C^L)^{\otimes i}$ to a unitary $U_i^B$ acting on $m_B = n+ t\cdot \ell$ qubits as follows. First, for $1 \leq i \leq t$, define $\widetilde{U}_i^{B}$ to be an extension of the isometry $V_i$ to a unitary on $n + \ell \cdot i$ qubits, i.e.,
    \begin{align*}
        \widetilde{U}_i^{B} \cdot (\Id_{N L^{i-1}} \otimes \ket*{0^\ell}) = V_i.
    \end{align*}
    We then extend this to a unitary on $n + t \cdot \ell$ qubits by setting $U_i^{B} \coloneqq \widetilde{U}_i^{B} \otimes \Id_{L^{t-i}}$. Finally, we pick $U_{t+1}^B \coloneqq \Id_{N L^t}$.

    To put everything together,  we need to prove the existence of an isometry $T$ satisfying~\cref{eq:isometry-guarantee}. Plugging in our definitions for $U_i^B$ into~\cref{eq:isometry-T}, there exists an isometry $T_t: \C^{M_B} = \C^N \otimes (\C^L)^{\otimes t} \rightarrow \C^{\qdim_A}$ such that for all Boolean functions $f: [L] \rightarrow \{\pm 1\}$,
    \begin{align}
        &{}(\calO_f \otimes \Id_{D_A}) \cdot U_t^A \cdots \cdot(\calO_f \otimes \Id_{D_A}) \cdot U_1^A \cdot (\Id_N \otimes \ket*{0^{m_A - n}}) \label{eq:isometry-almost-1}\\
        ={}& T_t \cdot (\calO_f \otimes \Id_{NL^{i-1}}) \cdot V_t \cdots (\calO_f \otimes \Id_{N}) \cdot V_1 \cdot \Id_N \nonumber\\
        ={}& T_t \cdot U_{t+1}^B \cdot (\calO_f \otimes \Id_{D_B}) \cdot U_t^B \cdots (\calO_f \otimes \Id_{D_B}) \cdot U_1^B \cdot (\Id_N \otimes \ket*{0^{m_B-n}}). \label{eq:isometry-almost-2}
    \end{align}
    The equation $(\ref{eq:isometry-almost-1}) = (\ref{eq:isometry-almost-2})$ shows that $T_t$ \emph{almost} satisfies the desired properties of the isometry $T: \C^{M_B} \rightarrow \C^{M_A}$ that we want (\cref{eq:isometry-guarantee}), except that in~\cref{eq:isometry-guarantee}, there is an additional $U_{t+1}^A$ unitary applied on the left-hand side. To complete the proof, we set $T = U_{t+1}^A \cdot T_t$.
\end{proof}

\paragraph{Simulating the measurement.} Our notion of what it means for $B^{(\cdot)}$ to \emph{simulate} $A^{(\cdot)}$ only guarantees that there exists an isometry $T$ such that (on any input) running $B^{(\cdot)}$ and then applying $T$ produces the same state as running $A^{(\cdot)}$. However, our aim is to use $B^{(\cdot)}$ in place of $A^{(\cdot)}$ as an adversary in the Oracle State Distinguishing Game; recall that an adversary (\cref{def:adversary}) in this game first applies the oracle circuit on a given input state, and then \emph{measures the first qubit} of the resulting state to produce a guess bit $\bb'$. Thus, what we need is a way to run a low-space oracle circuit $B^{(\cdot)}$ so that when we measure the first qubit of the resulting state, the outcome distribution is the same as if we had run $A^{(\cdot)}$ and measured its first qubit.


Fortunately, we can resolve this issue with standard techniques from quantum information (namely Naimark dilation; see, e.g., page 94 of~\cite{NC10}). First, define the $m_B$-qubit binary-outcome POVM $\{E_0,E_1 = \Id - E_0\}$ where
\begin{align*}
    E_0 \coloneqq T^\dagger \cdot (\ketbra{0} \otimes \Id_2^{\otimes m_A - 1}) \cdot T.
\end{align*}
Now, observe that if we run $B^{(\cdot)}$ and then measure $\{E_0,E_1 = \Id - E_0\}$, the resulting outcome $\bb'$ is distributed exactly the same as it would be if we had instead run $B^{(\cdot)}$, then applied $T$, and measured the first qubit (and the latter is equivalent to running $A^{(\cdot)}$ and measuring the first qubit).

To implement this POVM as a measurement of the \emph{first qubit} of the adversary's state, we will define the isometry $V_{\mathrm{guess}}: \C^{M_B} \rightarrow \C^2 \otimes \C^{M_B}$ as
\begin{align*}
    V_{\mathrm{guess}} \coloneqq \sum_{b \in \{0,1\}} \ket{b} \otimes \sqrt{E_b}. 
\end{align*}
We note that $V_{\mathrm{guess}}$ is in fact an isometry, since
\begin{align*}
    V_{\mathrm{guess}}^\dagger \cdot V_{\mathrm{guess}} = \Big(\sum_{b \in \{0,1\}} \bra{b} \otimes \sqrt{E_b} \Big) \cdot \Big(\sum_{b \in \{0,1\}} \ket{b} \otimes \sqrt{E_b} \Big) = \sum_{b \in \{0,1\}} E_b = \Id_{M_B}.
\end{align*}
Moreover, applying $V_{\mathrm{guess}}$ and measuring the first qubit of the resulting state produces the same distribution as measuring $\{E_0,E_1\}$, since for any state $\ket{\psi} \in \C^{M_B}$ and any $b \in \{0,1\}$,
\begin{align*}
    \bra{\psi} \cdot V_{\mathrm{guess}}^\dagger \cdot \ketbra{b} \cdot V_{\mathrm{guess}} \cdot \ket{\psi} = \bra{\psi} \cdot \sqrt{E_b} \cdot \sqrt{E_b} \cdot \ket{\psi} = \bra{\psi} E_b \ket{\psi}.
\end{align*}
Thus, given any circuit $B^{(\cdot)} = (n,m_B,\ell,U_1^B,\dots,U_{t+1}^B)$ that simulates $A^{(\cdot)}$ in the sense of~\cref{def:equiv-oalg}, we can easily modify $B^{(\cdot)}$ to obtain another low-space $t$-query oracle circuit $C^{(\cdot)} =(n,m_B+1,\ell,U_1^C,\dots,U_{t+1}^C)$ that has the additional guarantee that running $C^{(\cdot)}$ and measuring its first qubit produces a guess from the correct output distribution. 

Concretely, define $U_{\mathrm{guess}}$ to be an $(m_B + 1)$-qubit unitary that extends the isometry $V_{\mathrm{guess}}$ in the sense that $U_{\mathrm{guess}} \cdot (\ket{0} \otimes \Id_{M_B}) = V_{\mathrm{guess}}$. Then define 
\begin{align*}
    U_{t+1}^C \coloneqq U_{\mathrm{guess}} \cdot (\Id_2 \otimes U_{t+1}^B).
\end{align*}
The unitaries corresponding to $1 \leq i \leq t$ are defined as they are in $B^{(\cdot)}$ except that they act on one additional qubit, i.e., $U_i^C \coloneqq \Id_2 \otimes U_i^B$. By the preceding discussion, these definitions guarantee that running $C^{(\cdot)}$ and measuring its first qubit yields the outcome distribution of the original oracle adversary $A^{(\cdot)}$. Thus, we have the following corollary of~\cref{lemma:main-space-shrink}.
\begin{corollary}
\label{cor:small-space-dist}
    Without loss of generality, any $t$-query adversary in the Oracle State Distinguishing Game uses an oracle circuit $(n,m,\ell,U_1,\dots,U_t,U_{t+1})$ that requires $m \leq n + t \cdot \ell + 1$ qubits of space.
\end{corollary}

\subsection{One-query adversary model, final problem setup}\label{sec:one-query-model}
In this section, we give a ``normal form'' for one-query adversaries $A^{(\cdot)}$ with bounded query length. By \cref{cor:small-space-dist}, for any $A^{(\cdot)}$ with query length bounded by $\ell$, we may assume without loss of generality that $A^{(\cdot)}$ uses at most $a \leq \ell + 1$ ancilla qubits, for a total number of  $m= n+a$ qubits. As a result, following \cref{def:oracle-algorithm}, we may assume that $A^{f}$ operates as follows, for some choice of unitaries $U_1, U_2$:
    \begin{enumerate}
        \item Given an $n$ qubit input $\ket{\psi}$, compute the state
        \begin{align*}
            U_2\cdot \calO_f \cdot U_1\cdot \ket{\psi} \ket*{0^{a}}
        \end{align*}
        \item Measure the first qubit of the resulting state in the standard basis. 
    \end{enumerate}
In this special case, we simplify our notation slightly with the following definitions:

\begin{itemize}
    \item[$\circ$] Let $\qdim \coloneqq 2^m$ denote the dimension of the adversary's final Hilbert space.
    \item[$\circ$] Let $V = U_1\cdot(\Id_N \tensor \ket{0^{a}})$ denote the isometry describing $A$'s behavior prior to the query.
    \item[$\circ$] Let $\Pi = U_2^\dagger\cdot (\ketbra{0} \tensor \mathsf{Id}) \cdot U_2$ be the projection describing the adversary's measurement applied to $\mathcal O_f \cdot V \cdot \ket{\psi}$.
\end{itemize}
To summarize, we have modeled the adversary as $A^{(\cdot)} = (\qdim, V, \meas)$,
where $\qdim$ is an integer,
 $V:\C^N \rightarrow \C^{\qdim}$ is an isometry, and
$\meas \in \C^{\qdim \times \qdim}$ is a projection. In this language, the adversary's probability of outputting ``$0$'' on a binary phase state $\ket{\psi_h}$ is given by
\[ p_A(h\mid f) = \bra{\psi_h}\cdot V^\dagger \cdot \oracle_f \cdot \Pi\cdot  \oracle_f \cdot V \cdot \ket{\psi_h}.
\]
Our goal in \cref{sec:lower-bound} will be to prove an upper bound on $\Delta_{A}(\bro)$ (see \cref{def:advantage}) of the form
\[ \distinguish_{A}(\bro) = O\left(\sqrt{\frac{\log \qdim}{K}}\right)
\]
with high probability over $\bro$, which will establish a lower bound of $m=\log(M) = \Omega(K \epsilon^2)$ for adversaries with distinguishing advantage $\epsilon$. 

\newpage

\section{Proof of the one-query lower bound}\label{sec:lower-bound}

In this section, we will consider a single-query adversary $A$ and show that its advantage in the oracle state distinguishing game $\Delta_{A}(\bprs)$ is very small with overwhelming probability over $\bprs$. By \cref{lem:adv-tail}, it suffices to bound the expectation $\E[\Delta_{A}(\bprs)] = \Delta_{A}^{\avg}$ for every $A$.

We will bound this expected value as follows: in \Cref{sec:decoupling},
we will apply a standard decoupling trick to the expression for the adversary's distinguishing advantage.
Next, in \Cref{sec:decouple-relax} we will develop a natural spectral relaxation of this decoupled distinguishing advantage.
Following that, in \Cref{sec:khintchine-proof} we will use a matrix concentration inequality to bound the expectation of the spectral relaxation in terms of a quantity that we call the width of a collection of binary phase states. 
Then, in \Cref{sec:width-bound}, we show how to bound the expected width of a random family of binary phase states.
Finally, in \Cref{sec:the-full-proof-of-khintchine-proof} we will combine these ingredients and complete the proof of the one-query lower bound.

\paragraph{Notation.}
We will first fix some notation to use throughout the section.
By \cref{sec:one-query-model}, we can model the adversary as $A = (\qdim, V, \meas)$,
where $\qdim$ is an integer (a typical value of which is $\qdim = 2^{\poly(n)}$),
 $V:\C^N \rightarrow \C^{\qdim}$ is an isometry, and
$\meas \in \C^{\qdim \times \qdim}$ is a projection.
~Writing $v_{i, x}$ for the $(i, x)$-th entry of $V$,
we can express it as
\begin{equation*}
   V
   = \sum_{i=1}^{\qdim} \sum_{x=1}^N v_{i, x} \cdot \ketbra{i}{x}
   = \sum_{i=1}^{\qdim} \ket{i} \cdot \Big(\sum_{x=1}^N v_{i, x} \bra{x}\Big)
   = \sum_{i=1}^{\qdim} \ketbra{i}{v_i},
\end{equation*}
where $\ket{v_i} \in \C^N$ is the vector
\begin{equation*}
 \ket{v_i} = \sum_{x=1}^N v_{i, x}^\dagger \ket{x}.   
\end{equation*}
Note that
\begin{equation*}
    \sum_{i=1}^{\qdim} \braket{v_i} = \tr(V^\dagger V) =
    \tr(\Id_{N \times N}) = N.
\end{equation*}
This motivates the following definition.
\begin{definition}[Isometry weights]\label{def:isometry-weights}
The \emph{isometry weights} are the numbers
\begin{equation*}
    \weights_{V, i} \coloneqq \tfrac{1}{N} \cdot \braket{v_i},
\end{equation*}
for $1 \leq i \leq \qdim$.
Note that these
sum to one and therefore form a probability distribution.\footnote{It turns out that $\wt_{V,i}$ is the probability that a standard basis measurement of $V\ket{\psi_{\bh}}$, for uniformly random $\bh$, results in an outcome of $i$.}
\end{definition}

\subsection{Decoupling the quadratic form}\label{sec:decoupling}

Our overall goal is to bound the adversary's distinguishing advantage. Writing $\bh:[N] \rightarrow \{\pm 1\}$
for a uniformly random Boolean function, the distinguishing advantage can be written as
\begin{align}
    &\E_{\bprs}[\distinguish_{A}(\bprs)]\nonumber\\
    ={}&  \E_{\bprs} \Big[\underset{f: [\qdim]\rightarrow \{\pm 1\}}{\max} \Big|\E_{\bk \sim [K]} [\accept_{A}(\bprs_{\bk} \mid f)]
    - \E_{\bh} [\accept_{A}(\bh \mid f)] \Big|\Big]\nonumber\\
    ={}& \E_{\bprs} \Big[\max_f \Big|\E_{\bk \sim [K]}[\bra*{\psi_{\bprs_{\bk}}} \cdot V^\dagger \cdot \calO_f \cdot \meas \cdot \calO_f \cdot V \cdot \ket*{\psi_{\bprs_{\bk}}}] - \E_{\bh}[\bra*{\psi_{\bh}} \cdot V^\dagger \cdot \calO_f \cdot \meas \cdot \calO_f \cdot V \cdot \ket*{\psi_{\bh}}]\Big|\Big].\label{eq:TWO-rademachers}
\end{align}
The coefficients of the vector $\ket*{\psi_{\bprs_{\bk}}}$ are independent $\{\pm 1\}$ Rademacher random variables, and indeed there are tools from random matrix theory which allow us to prove concentration bounds on matrices whose entries are linear combinations of Rademachers.
However, the first term in \Cref{eq:TWO-rademachers} is quadratic in the $\ket*{\psi_{\bprs_{\bk}}}$ vector, and so these tools cannot be immediately applied.
What we would like to do is \emph{decouple} the left random vector $\bra*{\psi_{\bprs_{\bk}}}$ from the right random vector $\ket*{\psi_{\bprs_{\bk}}}$ so that this expression becomes a function of two independent random vectors, and is linear in both of them, rather than being quadratic in a single random vector. This motivates the following definition, a natural decoupled analogue of the distinguishing advantage. 

\begin{definition}[The decoupled distinguishing advantage]
Let $\prs, \prs' : [K] \times [N] \rightarrow \{\pm 1\}$
be two function families.
~Let $f: [\qdim]\rightarrow \{\pm 1\}$ be a function, and let $A$ denote a one-query adversary.
Then the corresponding \emph{decoupling distinguishing advantage} is given by
\begin{equation*}
    \distinguish_{\decup}(\prs, \prs' \mid f)
    \coloneqq \Big|\E_{\bk \sim [K]}[\bra*{\psi_{\prs_{\bk}}} \cdot V^\dagger \cdot \calO_f \cdot \meas \cdot \calO_f \cdot V \cdot \ket*{\psi_{\prs'_{\bk}}}] \Big|.
\end{equation*}
\end{definition}

Unlike the normal distinguishing advantage,
the decoupled distinguishing advantage has no natural operational interpretation.
However, it still gives a convenient upper bound to the normal distinguishing advantage, as shown in the following lemma.

\begin{lemma}\label{lem:decouple}
    Let $\bprs, \bprs' : [K] \times [N] \rightarrow \{\pm 1\}$ be two independent and uniformly random function families.
    Then
    \begin{equation*}
        \E_{\bprs}[\distinguish_{A}(\bprs)]
        \leq 4 \cdot \E_{\bR, \bR'}\max_{f} \Big\{\distinguish_{\decup}(\bprs, \bprs' \mid f)\Big\}.
    \end{equation*}
\end{lemma}

Decoupling inequalities are standard in the random matrix theory literature.
Our proof follows an outline similar to other decoupling arguments, for example those in \cite[Lemma 5.2]{Van17} and \cite{Ver11}.

\begin{proof}[Proof of \Cref{lem:decouple}]
Throughout this proof, we will adopt the following shorthand for convenience:
given an oracle $\calO$ acting on $\C^{\qdim}$, we will write
\begin{equation*}
    W_{\calO} \coloneqq V^\dagger \cdot \calO \cdot \meas \cdot \calO \cdot V.
\end{equation*}
For example, if $h:[N]\rightarrow \{\pm 1\}$ is a Boolean function,
and $\prs:[K] \times [N] \rightarrow \{\pm 1\}$ is a function family, then
\begin{align*}
    \accept_{A}(h \mid f)
    &= \bra{\psi_h} \cdot V^\dagger \cdot \calO_f \cdot \meas \cdot \calO_f \cdot V \cdot \ket{\psi_h}\\
    &= \bra{\psi_h} \cdot W_{\calO_f} \cdot \ket{\psi_h}.
\end{align*}

Let $\bprs: [K] \times [N] \rightarrow \{\pm 1\}$ be a uniformly random function family.
Let $\bprs'$ be an independent copy of $\bprs$.
Then for each $1 \leq k \leq K$,
$\bprs'_k$ is distributed as a uniformly random function,
even conditioned on $\bprs$.
As a result, the average distinguishing advantage is given by
\begin{align}
    \E_{\bprs}[\distinguish_{A}(\bprs)]
    &= \E_{\bprs}\Bigg[ \max_f \Big|\E_{\bk \sim [K]} [\accept_{A}(\bprs_{\bk} \mid f)]
    - \E_{\bh} [\accept_{A}(\bh \mid f)] \Big|\Bigg]\nonumber\\
    &= \E_{\bprs} \Bigg[\max_f \Big|\E_{\bk \sim [K]} [\accept_{A}(\bprs_{\bk} \mid f)]
    - \E_{\bprs'}\E_{\bk \sim [K]} [\accept_{A}(\bprs'_{\bk} \mid f)] \Big|\Bigg]\nonumber\\
    &\leq \E_{\bprs, \bprs'} \Bigg[ \max_f \Big|\E_{\bk \sim [K]} [\accept_{A}(\bprs_{\bk} \mid f)]
    - \E_{\bk \sim [K]} [\accept_{A}(\bprs'_{\bk} \mid f)] \Big|\Bigg]\nonumber\\
    &= \E_{\bprs, \bprs'} \Bigg[ \max_f \Big|\E_{\bk \sim [K]} [\accept_{A}(\bprs_{\bk} \mid f)
    - \accept_{A}(\bprs'_{\bk} \mid f)] \Big|\Bigg] \nonumber\\
    &= \E_{\bprs, \bprs'} \Bigg[ \max_f \Big|\E_{\bk \sim [K]} [
        \bra{\psi_{\bprs_{\bk}}} \cdot W_{\calO_f} \cdot \ket{\psi_{\bprs_{\bk}}}
        - \bra*{\psi_{\bprs_{\bk}'}} \cdot W_{\calO_f} \cdot \ket*{\psi_{\bprs_{\bk}'}}] \Big|\Bigg]\nonumber\\
    & = \E_{\bprs, \bprs'} \Bigg[\max_{f} \Big|\Re\Big( \E_{\bk \sim [K]} [
        (\bra*{\psi_{\bprs_{\bk}}} + \bra*{\psi_{\bprs'_{\bk}}}) \cdot W_{\calO_{f}} \cdot (\ket*{\psi_{\bprs_{\bk}}} - \ket*{\psi_{\bprs'_{\bk}}})]\Big) \Big|\Bigg]\nonumber\\
& \leq \E_{\bprs, \bprs'} \Bigg[\max_{f} \Big| \E_{\bk \sim [K]} [
        (\bra*{\psi_{\bprs_{\bk}}} + \bra*{\psi_{\bprs'_{\bk}}}) \cdot W_{\calO_{f}} \cdot (\ket*{\psi_{\bprs_{\bk}}} - \ket*{\psi_{\bprs'_{\bk}}})] \Big|\Bigg].\label{eq:gg-no-re}
\end{align}
For each $1 \leq k \leq K$, consider the two vectors
\begin{equation*}
    \ket*{\psi_{\bprs_{k}}} \pm \ket*{\psi_{\bprs'_{k}}}
    = \frac{1}{\sqrt{N}}\cdot \sum_{x=1}^N (\bprs_k(x) \pm \bprs_k'(x)) \cdot \ket{x}.
\end{equation*}
Note that for each $1 \leq k \leq K$ and $1 \leq x \leq N$,
either $\bprs_k(x) + \bprs_k'(x) \in \{\pm 2\}$
and $\bprs_k(x) - \bprs_k'(x) = 0$, or vice versa.
For each $1 \leq k \leq K$, let $\bS_k^+:[N]\rightarrow \{\pm 1\}$
be a random Boolean function distributed as follows:
if $\bprs_k(x) + \bprs_k'(x) \in \{\pm 2\}$, then $\bS_k^+(x) = \tfrac{1}{2} \cdot (\bprs_k(x) + \bprs_k'(x))$.
Otherwise, when $\bprs_k(x) + \bprs_k'(x) = 0$,
choose $\bS_k^+(x)$ independently and uniformly at random from $\{\pm 1\}$.
Define $\bS_k^-(x)$ similarly. Then the next two equations follow by definition:
\begin{equation*}
    2 \cdot \E[\ket*{\psi_{\bS_k^+}}] = \ket*{\psi_{\bprs_{\bk}}} + \ket*{\psi_{\bprs'_{\bk}}},
    \quad \text{and} \quad 
    2 \cdot \E[\ket*{\psi_{\bS_k^-}}] = \ket*{\psi_{\bprs_{\bk}}} - \ket*{\psi_{\bprs'_{\bk}}}.
\end{equation*}
Plugging this in above,
\begin{align*}
    \eqref{eq:gg-no-re}
    & = 4 \cdot \E_{\bprs, \bprs'} \max_{f} \Big| \E_{\bS^+, \bS^-}\E_{\bk \sim [K]} [
        \bra*{\psi_{\bS^+_{\bk}}} \cdot W_{\calO_{f}} \cdot \ket*{\psi_{\bS^-_{\bk}}}] \Big|\\
    & \leq 4 \cdot \E_{\bprs, \bprs'} \E_{\bS^+, \bS^-}\max_{f} \Big| \E_{\bk \sim [K]} [
        \bra*{\psi_{\bS^+_{\bk}}} \cdot W_{\calO_{f}} \cdot \ket*{\psi_{\bS^-_{\bk}}}] \Big|.
\end{align*}
The expression inside the max only depends on $\bprs$ and $\bprs'$ through $\bS^+$ and $\bS^-$. Hence, this is equal to
\begin{equation*}
    4 \cdot \E_{\bS^+, \bS^-}\max_{f} \Big| \E_{\bk \sim [K]} [
        \bra*{\psi_{\bS^+_{\bk}}} \cdot W_{\calO_{f}} \cdot \ket*{\psi_{\bS^-_{\bk}}}] \Big|
    =  4 \cdot \E_{\bS^+, \bS^-}\max_{f} \Big\{\distinguish_{\decup}(\bS^+, \bS^- \mid f)\Big\}.
\end{equation*}
But it can be checked that $\bS^+$ and $\bS^-$ are just distributed as two independent and uniformly random function families. This completes the proof.
\end{proof}

\subsection{A spectral relaxation for the decoupled distinguishing advantage}\label{sec:decouple-relax}

In this section, we develop a spectral relaxation for the decoupled distinguishing advantage
\begin{equation*}
    \distinguish_{\decup}(R, R' \mid f)
    = \Big|\E_{\bk \sim [K]}[\bra*{\psi_{\prs_{\bk}}} \cdot V^\dagger \cdot \calO_f \cdot \meas \cdot \calO_f \cdot V \cdot \ket*{\psi_{\prs'_{\bk}}}] \Big|.
\end{equation*}
As a precursor to this, we will develop a formula for expressing the vectors $V \cdot \ket*{\psi_{R_{\bk}}}$ and $V \cdot \ket*{\psi_{R'_{\bk}}}$.

Let $h:[N]\rightarrow\{\pm 1\}$ be a Boolean function,
and consider the binary phase state
\begin{equation*}
\ket{\psi_{h}} = \frac{1}{\sqrt{N}} \sum_{x \in [N]} h(x) \ket{x}.
\end{equation*}
The isometry $V$ maps $\ket{\psi_h}$ to
\begin{equation*}
    V \cdot \ket{\psi_h}
    = \Big(\sum_{i=1}^{\qdim} \ketbra{i}{v_i}\Big) \cdot \ket{\psi_h}
    = \sum_{i=1}^{\qdim} \braket{v_i}{\psi_h} \cdot \ket{i}.
\end{equation*}
Thus, the amplitude on the $i$-th basis element is $\braket{v_i}{\psi_h}$.
We would like to estimate the magnitude of this amplitude
for a ``typical'' binary phase state.
This is given by the following proposition.
\begin{proposition}[Typical amplitudes]\label{prop:typical-magnitude}
Let $\bh: [N]\rightarrow \{\pm 1\}$
be a uniformly random Boolean function.
Then
\begin{equation*}
    \E_{\bh} |\braket{v_i}{\psi_{\bh}}|^2 = \weights_{V, i},
\end{equation*}
where $\weights_{V, i}$ is the isometry weight
defined in \Cref{def:isometry-weights}.
\end{proposition}
\begin{proof}
We calculate the expectation as follows:
\begin{align*}
\E_{\bh} |\braket{v_i}{\psi_{\bh}}|^2
&= \E_{\bh} \braket{\psi_{\bh}}{v_i} \cdot \braket{v_i}{\psi_{\bh}}\\
&= \E_{\bh} \Big(\sum_{x =1}^N \tfrac{1}{\sqrt{N}} \bh(x) \cdot v_{i, x}^\dagger\Big)
            \cdot \Big(\sum_{y =1}^N v_{i, y} \cdot \tfrac{1}{\sqrt{N}} \bh(y)\Big)\\
&= \frac{1}{N} \cdot \sum_{x, y = 1}^N v_{i,x}^\dagger v_{i, y} \cdot \E_{\bh}[\bh(x) \bh(y)]\\
&= \frac{1}{N} \cdot \sum_{x = 1}^N |v_{i,x}|^2 = \frac{1}{N} \braket{v_i},
\end{align*}
where the second-to-last equality used the fact that
$\E_{\bh}[\bh(x) \bh(y)] = 1$ if $x = y$ and $0$ otherwise,
because $\bh$ is uniformly random.
The proof concludes by applying the definition of $\weights_{V, i}$.
\end{proof}

In light of this,
it is natural to define the following vector,
which contains the ``typical'' amplitudes
of $V \cdot \ket{\psi_h}$.

\begin{definition}[The weight vector]\label{def:weight-vector}
    The \emph{weight vector}
    is the unit vector given by
    \begin{equation*}
        \ket{\wt_V} = \sum_{i=1}^{\qdim} \sqrt{\weights_{V, i}} \ket{i}.
    \end{equation*}
\end{definition}

We can express $V \cdot \ket{\psi_h}$ in terms of the weight vector as
\begin{equation}\label{eq:rewriting-with-weight-vector}
    V \cdot \ket{\psi_h}
    = \sum_{i=1}^{\qdim} \braket{v_i}{\psi_h} \cdot \ket{i}
    = \sum_{i=1}^{\qdim} \frac{\braket{v_i}{\psi_h}}{\sqrt{\weights_{V, i}}} \cdot \sqrt{\weights_{V, i}} \cdot \ket{i}
    = \Big(\sum_{i=1}^{\qdim} \frac{\braket{v_i}{\psi_h}}{\sqrt{\weights_{v, i}}} \cdot \ketbra{i}\Big) \cdot \ket{\wt_V}.
\end{equation}
This motivates the following definition.

\begin{definition}[The rescaling matrix]\label{def:rescaling-matrix}
Let $h:[N] \rightarrow \{\pm 1\}$ be a Boolean function.
Then the corresponding \emph{rescaling matrix} is the diagonal matrix given by
\begin{equation*}
    \rescale_{V, h} = \sum_{i=1}^{\qdim} D_{V, h, i} \cdot \ketbra{i},
    \text{ where }
    \rescale_{V, h,i} = \frac{\braket{v_i}{\psi_h}}{\sqrt{\weights_{V, i}}}.
\end{equation*}
By construction, we have that
$V \cdot \ket{\psi_h} = \rescale_{V, h} \cdot \ket{\wt_V}$.
\end{definition}

\begin{remark}\label{rem:width-remark}
Loosely speaking,
the size of the rescaling matrix indicates how close the amplitudes of $V \cdot \ket{\psi_h}$ are to their ``typical'' values.
If each diagonal entry of $\rescale_{V, h}$ is close to 1 in magnitude,
then the amplitudes of $V\cdot\ket{\psi_h}$'s are roughly typical;
otherwise, at least one of $V\cdot\ket{\psi_h}$'s amplitudes is atypically large or small.
\end{remark}

We can therefore express the decoupled distinguishing advantage as
\begin{align*}
    \distinguish_{\decup}(\prs, \prs' \mid f)
    &= \Big|\E_{\bk \sim [K]}[\bra*{\psi_{\prs_{\bk}}} \cdot V^\dagger \cdot \calO_f \cdot \meas \cdot \calO_f \cdot V \cdot \ket*{\psi_{\prs'_{\bk}}}] \Big|\\
    &= \Big|\E_{\bk \sim [K]}[\bra*{\wt_V} \cdot D_{V, \prs_{\bk}}^\dagger \cdot \calO_f \cdot \meas \cdot \calO_f \cdot D_{V, \prs_{\bk}'} \cdot \ket{\wt_V}] \Big|.
\end{align*}
Now we observe that $\calO$ and $\rescale_{V, \prs_{\bk}}$ are both diagonal matrices, and hence they both commute (and similarly for $\rescale_{V, \prs_{\bk}'}$).
As a result, this is equal to
\begin{align*}
 \distinguish_{\decup}(\prs, \prs' \mid f)
    &=
    \Big|\E_{\bk \sim [K]}[\bra*{\wt_V} \cdot \calO_f \cdot D_{V, \prs_{\bk}}^\dagger \cdot \meas \cdot D_{V, \prs_{\bk}'} \cdot \calO_f \cdot \ket{\wt_V}] \Big|\\
    &=
    \Big|\bra*{\wt_V} \cdot \calO_f \cdot \E_{\bk \sim [K]}[D_{V, \prs_{\bk}}^\dagger \cdot \meas \cdot D_{V, \prs_{\bk}'}]   \cdot \calO_f \cdot \ket{\wt_V}\Big|.
\end{align*}
Note that for any function $f$, $\calO_{f} \cdot \ket{\wt_V}$ is a unit vector.
We can therefore upper-bound this expression
by relaxing $\calO_f \cdot \ket{\wt_V}$ to be an arbitrary unit vector
maximizing this expression.
This gives the spectral relaxation.

\begin{definition}[Spectral relaxation]
Let $\prs, \prs' : [K] \times [N] \rightarrow \{\pm 1\}$
be two function families, and let $A$ denote an adversary.
~The \emph{spectral relaxation of the decoupled distinguishing advantage} is given by
\begin{equation*}
    \Delta_{\decup}^{\spectral}(\prs, \prs') = \Big\Vert \E_{\bk \sim [K]}[D_{V, \prs_{\bk}}^\dagger \cdot \meas \cdot D_{V, \prs_{\bk}'}]\Big\Vert_{\mathrm{op}}.
\end{equation*}
\end{definition}

From the above discussion, the following lemma is immediate.

\begin{lemma}\label{lem:decouple-relax}
Let $\prs, \prs':[K]\times [N] \rightarrow \{\pm 1\}$ be two function families.
Then
\begin{equation*}
    \max_{f} \Big\{\distinguish_{\decup}(\prs, \prs' \mid f)\Big\} \leq \Delta_{\decup}^{\spectral}(\prs, \prs').
\end{equation*}
\end{lemma}

\subsection{Expectation of the spectral relaxation with one parameter held fixed}\label{sec:khintchine-proof}

The spectral relaxation
is the operator norm of a matrix which is bilinear in both $\prs$ and $\prs'$.
Keeping $\prs$ fixed, we can consider a uniformly random $\bprs':[K] \times [N]\rightarrow \{\pm 1\}$,
and doing so makes this a random matrix whose entries are linear combinations of random $\{\pm 1\}$ variables.
The key technical result we will use to study such matrices is the following, stated in \cite[Theorem 4.1.1]{Tro15}.

\begin{theorem}[Concentration for matrix Rademacher series]\label{thm:khintchine}
    Let $\bx_1, \ldots, \bx_n$ be $n$ independent, uniformly distributed $\{\pm 1\}$ random variables.
    Let $\bZ$ be a $d_1 \times d_2$ complex matrix whose entries are linear combinations of the $\bx_k$'s, i.e.
    \begin{equation*}
        \bZ_{i, j} = c_{i, j,1} \cdot \bx_1 + \cdots + c_{i, j, n} \cdot \bx_n,
    \end{equation*}
    where each $c_{i, j, k}$ is a fixed, complex number.
    Let $v(\bZ)$ be the matrix variance statistic of $\bZ$, i.e.
    \begin{equation*}
        v(\bZ) = \max\{\Vert \E[\bZ \cdot \bZ^\dagger] \Vert_\operator,
                    \Vert \E[\bZ^\dagger \cdot \bZ] \Vert_\operator\}.
    \end{equation*}
    Then
    \begin{equation*}
        \E[\Vert \bZ \Vert_\operator] \leq \sqrt{2 \ln(d_1 + d_2)} \cdot \sqrt{v(\bZ)}.
    \end{equation*}
    Furthermore, for all $t \geq 0$,
    \begin{align*}
    \Pr\left[ \norm{\bZ}_{\operator} \geq t\right] \leq (d_1 + d_2) \cdot \mathrm{exp}\left(-\frac{t^2}{2\cdot v(\bZ)}\right).
    \end{align*}
\end{theorem}

We now use this to upper bound the expectation of the spectral relaxation when one of the parameters is held fixed. 
It states that this expectation can be bounded in terms of a quantity called the \emph{width} of the function family $\prs$. Roughly speaking, the width is a measure of the ``size'' of the diagonal rescaling matrices $\rescale_{V, R_k}$, over all $1 \leq k \leq K$.
As discussed in \Cref{rem:width-remark}, when $\prs$ is a ``typical'' function family, we expect that these rescaling matrices should have small (i.e.\ close to 1) entries on the diagonal,
in which case the width of $\prs$ will be small.
For atypical function families, on the other hand, the width might be large, but we expect such families to be extremely rare.

\begin{lemma}[Expectation of the spectral relaxation with one parameter held fixed]\label{lem:apply-khintchine}
    Let $\prs:[K] \times [N] \rightarrow \{\pm 1\}$ be a fixed function family.
    Define the \emph{width of $\prs$} to be the quantity 
    \begin{equation*}
        \width(R) \coloneqq \max_{1 \leq i \leq \qdim}\Big\{\frac{1}{K} \sum_{k=1}^K\frac{|\braket*{v_i}{\psi_{\prs_{k}}}|^2}{\weights_{V, i}}\Big\}.
    \end{equation*}
    In addition,
    let $\bprs':[K] \times [N] \rightarrow \{\pm 1\}$ be a uniformly random function family.
    Then
    \begin{equation*}
        \E_{\bprs'} [\Delta_{\decup}^{\spectral}(\prs, \bprs')]
        \leq \sqrt{\frac{2 \ln(2\qdim) \cdot \width(R)}{K}}.
    \end{equation*}
\end{lemma}
\begin{proof}
    For the reader's convenience, we will recall the definition of the diagonal rescaling matrix corresponding to a Boolean function $h:[N] \rightarrow \{\pm 1\}$:
    \begin{equation*}
        \rescale_{V, h} = 
     \sum_{i=1}^{\qdim} \frac{\braket{v_i}{\psi_{h}}}{\sqrt{\weights_{V, i}}} \cdot \ketbra{i}.
    \end{equation*}
    Note that each entry of $D_{V, h}$ is a linear combination of the Boolean values $h(1), \ldots, h(N)$.
    In addition, note that
    \begin{align}
        \rescale_{V, h} \cdot \rescale_{V, h}^\dagger
        &= \Big(\sum_{i=1}^{\qdim} \frac{\braket{v_i}{\psi_{h}}}{\sqrt{\weights_{V, i}}} \cdot \ketbra{i}\Big) \cdot \Big(\sum_{i=1}^{\qdim} \frac{\braket{\psi_{h}}{v_i}}{\sqrt{\weights_{V, i}}} \cdot \ketbra{i}\Big)\nonumber\\
        &= \sum_{i=1}^{\qdim} \frac{|\braket{v_i}{\psi_{h}}|^2}{\weights_{V, i}} \cdot \ketbra{i}\nonumber\\
        &= \Big(\sum_{i=1}^{\qdim} \frac{\braket{\psi_h}{v_i}}{\sqrt{\weights_{V, i}}} \cdot \ketbra{i}\Big) \cdot \Big(\sum_{i=1}^{\qdim} \frac{\braket{v_i}{\psi_h}}{\sqrt{\weights_{V, i}}} \cdot \ketbra{i}\Big)
        = \rescale_{V, h}^\dagger \cdot \rescale_{V, h}.\label{eq:double-rescale}
    \end{align}
    Our goal is to compute
    \begin{equation*}
    \E_{\bprs'}[\Delta_{\decup}^{\spectral}(\prs, \bprs')]
    = \E_{\bprs'} \Big\Vert \E_{\bk \sim [K]}[D_{V, \prs_{\bk}}^\dagger \cdot \meas \cdot D_{V, \bprs_{\bk}'}]\Big\Vert_{\mathrm{op}}.
    \end{equation*}
    To this end, define the matrix
    \begin{equation*}
    \bZ
    \coloneqq \E_{\bk \sim [K]}[\rescale_{V, \prs_{\bk}}^\dagger \cdot \meas \cdot \rescale_{V, \bprs_{\bk}'}]
    = \frac{1}{K} \cdot \sum_{k=1}^K \rescale_{V, \prs_{k}}^\dagger \cdot \meas \cdot \rescale_{V, \bprs_{k}'}.
\end{equation*}
For each $1 \leq k \leq K$, $\rescale_{V, \bprs_k'}$ is a matrix whose entries are linear combinations of the $\bR_k'(x)$'s. 
As a result, the entries of $\bZ$ are linear combinations of the $K \cdot N$ many $\{\pm 1\}$-valued random variables in $\bprs'$. Hence, we can apply \Cref{thm:khintchine} to bound $\E_{\bprs'}[\Vert \bZ \Vert_\operator]$. 
To do so, we must first compute the matrix variance statistic of $\bZ$.
To begin,
\begin{align}
    \E_{\bprs'}[\bZ \cdot \bZ^\dagger]
    & =\E_{\bprs'}\Big[\Big(\frac{1}{K} \cdot \sum_{k=1}^K \rescale_{V, \prs_{k}}^\dagger
    \cdot \meas \cdot \rescale_{V, \bprs_{k}'}\Big)\cdot \Big(\frac{1}{K} \cdot \sum_{k'=1}^K \rescale_{V, \bprs_{k'}'}^\dagger \cdot \meas \cdot \rescale_{V, \prs_{k'}}\Big)\Big]\nonumber\\
    & = \frac{1}{K^2} \cdot \sum_{k, k'=1}^K
        \rescale_{V, \prs_k}^\dagger \cdot \meas \cdot \E_{\bprs'}[\rescale_{V, \bprs_k'} \cdot \rescale_{V, \bprs_{k'}'}^\dagger] \cdot \meas \cdot \rescale_{V, \prs_{k'}}.\label{eq:first-step-of-matrix-variance}
\end{align}
Now, if $k \neq k'$, then $\bprs_k'$ and $\bprs_{k'}'$ are distributed independently from each other.
As a result, for any fixed matrix $C$,
\begin{equation}\label{eq:two-different-ks}
\E_{\bprs'}[\rescale_{V, \bprs_k'} \cdot C \cdot \rescale_{V, \bprs_{k'}'}^\dagger]
= \E_{\bprs_k'}[\rescale_{V, \bprs_k'}] \cdot C \cdot \E_{\bprs_{k'}'}[\rescale_{V, \bprs_{k'}'}^\dagger]
= 0,
\end{equation}
because $\rescale_{V, \bprs_k'}$ and $\rescale_{V, \bprs_{k'}'}$ are mean-zero.
(For \Cref{eq:first-step-of-matrix-variance} above we only need the $C = \Id_{\qdim \times \qdim}$ case, but we will apply it below using a different matrix $C$.)
On the other hand, if $k = k'$, then by \Cref{prop:typical-magnitude},
    \begin{align}
        \E_{\bprs'}[\rescale_{V, \bprs_k'} \cdot \rescale_{V, \bprs_k'}^\dagger]
        &= \sum_{i=1}^{\qdim} \E_{\bprs}\Big[\frac{|\braket{v_i}{\psi_{\bprs_k}}|^2}{\weights_{V, i}}\Big] \cdot \ketbra{i}\tag{by \Cref{eq:double-rescale}}\\
        &= \sum_{i=1}^{\qdim}  \ketbra{i}
        = \Id_{\qdim \times \qdim}. \label{eq:average-two-rescales}
    \end{align}
Combining these two facts, we have that
\begin{equation*}
    \eqref{eq:first-step-of-matrix-variance}
    = \frac{1}{K^2} \cdot \sum_{k=1}^K
        \rescale_{V, \prs_k}^\dagger \cdot \meas^2 \cdot \rescale_{V, \prs_{k}}
    \preceq \frac{1}{K^2} \cdot \sum_{k=1}^K
        \rescale_{V, \prs_k}^\dagger \cdot\Id_{\qdim \times \qdim}\cdot \rescale_{V, \prs_{k}}
    = \frac{1}{K^2} \cdot \sum_{k=1}^K
        \rescale_{V, \prs_k}^\dagger \cdot \rescale_{V, \prs_{k}}.
\end{equation*}
Finally, we bound this by
\begin{align}
    \frac{1}{K^2} \cdot \sum_{k=1}^K
        \rescale_{V, \prs_k}^\dagger \cdot \rescale_{V, \prs_{k}}
        &= \frac{1}{K^2} \cdot \sum_{k=1}^K \sum_{i=1}^{\qdim} \frac{|\braket{v_i}{\psi_{\prs_k}}|^2}{\weights_{V, i}} \cdot \ketbra{i}\tag{by \Cref{eq:double-rescale}}\\
        &= \frac{1}{K} \cdot \sum_{i=1}^{\qdim} \Big(\frac{1}{K}\cdot \sum_{k=1}^K \frac{|\braket{v_i}{\psi_{\prs_k}}|^2}{\weights_{V, i}}\Big) \cdot \ketbra{i}\nonumber\\
        &\preceq \frac{1}{K} \cdot \sum_{i=1}^{\qdim} \width(R) \cdot \ketbra{i}
        = \Big(\frac{\width(R)}{K}\Big) \cdot \Id_{\qdim \times \qdim}. \label{eq:here-comes-the-width}
\end{align}
So far, we have shown that
\begin{equation*}
    \Vert \E_{\bprs'}[\bZ \cdot \bZ^\dagger]\Vert_{\operator}
    \leq \Big\Vert \Big(\frac{\width(R)}{K}\Big) \cdot \Id_{\qdim \times \qdim} \Big\Vert_{\operator}
    \leq \frac{\width(R)}{K}.
\end{equation*}
Thus far, we have only computed the first term in the matrix variance statistic of $\bZ$. Now we move on to the second term.
Fortunately, we can reuse many of the steps involved in computing the first term to compute the second term:
\begin{align*}
    \E_{\bR'}[\bZ^\dagger \cdot \bZ]
    & =\E_{\bprs'}\Big[\Big(\frac{1}{K} \cdot \sum_{k=1}^K \rescale_{V, \bprs_{k}'}^\dagger
    \cdot \meas \cdot \rescale_{V, \prs_{k}}\Big)\cdot \Big(\frac{1}{K} \cdot \sum_{k'=1}^K \rescale_{V, \prs_{k'}}^\dagger \cdot \meas \cdot \rescale_{V, \bprs_{k'}'}\Big)\Big]\\
    & = \frac{1}{K^2} \cdot \sum_{k, k'=1}^K\E_{\bprs'}\Big[
        \rescale_{V, \bprs_k'}^\dagger \cdot \meas \cdot \rescale_{V, \prs_k} \cdot \rescale_{V, \prs_{k'}}^\dagger \cdot \meas \cdot \rescale_{V, \bprs_{k'}'}\Big]\\
    & = \frac{1}{K^2} \cdot \sum_{k =1}^K\E_{\bprs'}\Big[
        \rescale_{V, \bprs_k'}^\dagger \cdot \meas \cdot \rescale_{V, \prs_k} \cdot \rescale_{V, \prs_{k}}^\dagger \cdot \meas \cdot \rescale_{V, \bprs_{k}'}\Big]. \tag{by \Cref{eq:double-rescale,eq:two-different-ks}}
\end{align*}
Now, let $\bh:[N]\rightarrow \{\pm 1\}$ be a uniformly random Boolean function.
Then $\bh$ has the same distribution as $\bprs'_k$. for each $1 \leq k \leq K$.
As a result, this is equal to
\begin{align*}
    &\frac{1}{K^2} \cdot \sum_{k =1}^K\E_{\bh}\Big[
        \rescale_{V, \bh}^\dagger \cdot \meas \cdot \rescale_{V, \prs_k} \cdot \rescale_{V, \prs_{k}}^\dagger \cdot \meas \cdot \rescale_{V, \bh}\Big]\\
    ={}&\E_{\bh}\Big[
        \rescale_{V, \bh}^\dagger \cdot \meas \cdot\Big(\frac{1}{K^2} \cdot \sum_{k =1}^K \rescale_{V, \prs_k} \cdot \rescale_{V, \prs_{k}}^\dagger\Big) \cdot \meas \cdot \rescale_{V, \bh}\Big]\\
    \preceq{}&\E_{\bh}\Big[
        \rescale_{V, \bh}^\dagger \cdot \meas \cdot\Big(\Big(\frac{\width(R)}{K}\Big) \cdot \Id_{\qdim \times \qdim}\Big) \cdot \meas \cdot \rescale_{V, \bh}\Big] \tag{by \Cref{eq:here-comes-the-width}}\\
    ={}&\Big(\frac{\width(R)}{K}\Big) \cdot\E_{\bh}\Big[
        \rescale_{V, \bh}^\dagger \cdot \meas^2 \cdot \rescale_{V, \bh}\Big]\\
    \preceq{}&\Big(\frac{\width(R)}{K}\Big) \cdot\E_{\bh}\Big[
        \rescale_{V, \bh}^\dagger \cdot \Id_{\qdim \times \qdim} \cdot \rescale_{V, \bh}\Big]\\
    ={}&\Big(\frac{\width(R)}{K}\Big) \cdot\E_{\bh}\Big[
        \rescale_{V, \bh}^\dagger \cdot \rescale_{V, \bh}\Big]
    ={}\Big(\frac{\width(R)}{K}\Big) \cdot\Id_{\qdim \times \qdim}. \tag{by \Cref{eq:double-rescale,eq:average-two-rescales}}
        \end{align*}
In total, this shows that
\begin{equation*}
    \Vert \E_{\bprs'}[\bZ^\dagger \cdot \bZ]\Vert_{\operator}
    \leq \Big\Vert \Big(\frac{\width(R)}{K}\Big) \cdot \Id_{\qdim \times \qdim} \Big\Vert_{\operator}
    \leq \frac{\width(R)}{K}.
\end{equation*}
As a result, the matrix variance statistic of $\bZ$ is
\begin{equation*}
    v(\bZ) = \max\{\Vert \E[\bZ \cdot \bZ^\dagger] \Vert_\operator,
                    \Vert \E[\bZ^\dagger \cdot \bZ] \Vert_\operator\}
         \leq \frac{\width(R)}{K}.
\end{equation*}
Now we apply \Cref{thm:khintchine}. It states that
\begin{equation*}
    \E_{\bprs'}[\Vert \bZ \Vert_\operator]
    \leq \sqrt{2 \ln(2\qdim)} \cdot \sqrt{v(\bZ)}
    \leq \sqrt{2 \ln(2\qdim)} \cdot \sqrt{\frac{\width(R)}{K}}.
\end{equation*}
This completes the proof.
\end{proof}

\subsection{A bound on the width of a random state family}\label{sec:width-bound}

In the previous section,
we showed that the expectation of the spectral relaxation, when one of the input state families $\bprs':[K]\times [N] \rightarrow \{\pm 1\}$ is randomized, can be bounded by a parameter of the other input family $\prs$ referred to as its \emph{width}. In this section, we show how to bound the expected width of a uniformly random family of binary phase states $\bprs:[K] \times [N] \rightarrow \{\pm 1\}$. For intuition, recall that $\width(\bprs)$ is defined to be the quantity
\begin{equation}\label{eq:width-copied}
    \max_{1 \leq i \leq \qdim}\Big\{\frac{1}{K} \sum_{k=1}^K\frac{|\braket*{v_i}{\psi_{\bprs_{k}}}|^2}{\weights_{V, i}}\Big\}.
\end{equation}
Let us fix a value $1 \leq i \leq \qdim$
and consider the $i$-th average being maximized over.
By \Cref{prop:typical-magnitude}, for each $1 \leq k \leq K$, the $k$-th term in the average has expectation exactly equal to~$1$, and indeed we will show that this term is close to~$1$ with high probability.
As the $i$-th average is an average over $K$ such terms, we expect that it should be extremely close to $1$ with an extremely high probability, a probability so high that we can then union bound over all $1 \leq i \leq \qdim$ and show that $\width(\bprs)$ itself is close to 1 with high probability.
From this, we will be able to conclude that the expectation is close to 1 as well.

To start, let us focus on the $k$-th term in the $i$-th average. It is the absolute value squared of the following quantity:
\begin{equation*}
    \frac{\braket*{v_i}{\psi_{\bprs_{k}}}}{\sqrt{\weights_{V, i}}}
    = \frac{1}{\sqrt{\weights_{V, i}}}\cdot \sum_{x=1}^N \Big(v_{i, x} \cdot \frac{1}{\sqrt{N}} \cdot \bprs_k(x)\Big)
    = \sum_{x=1}^N \Big(\frac{v_{i, x}}{\sqrt{\braket{v_i}}}\Big) \cdot \bprs_k(x).
\end{equation*}
This is just a complex-weighted linear combination of random $\{\pm 1\}$ variables.
In addition, the sum of the squared weights is given by
\begin{equation*}
    \sum_{x=1}^N \Big|\frac{v_{i, x}}{\sqrt{\braket{v_i}}}\Big|^2
    = \frac{1}{\braket{v_i}} \cdot \sum_{x=1}^N |v_{i, x}|^2
    = \frac{1}{\braket{v_i}} \cdot \braket{v_i}
    = 1.
\end{equation*}
We would like to show that weighted sums of this form are highly concentrated. In particular, we will show that they posses a particular concentration property known as being \emph{sub-exponential}.

\begin{definition}[Sub-exponential random variables]
A random variable $\bX$ is \emph{sub-exponential with parameter $\gamma > 0$} if
\begin{equation*}
        \Pr \left[ \abs{\bX} > t \right] \leq 2 \cdot \mathrm{exp} \left(-\frac{t}{\gamma} \right), \ \text{ for all } t \geq 0.
\end{equation*}
\end{definition}

This is shown in the next lemma, whose proof we defer to \Cref{sec:annoying-technical-lemma}.

\begin{lemma}[Each term in the width is sub-exponential]\label{lem:sub-exponential-calculation}
    There exists a constant $\gamma \geq 1$ such that the following is true. 
    Let $\bb_1, \ldots, \bb_m$ be independent and uniform $\pm 1$ random variables,
    and let $a_1, \ldots, a_m$ be complex numbers
    such that $|a_1|^2 + \cdots + |a_m|^2 = 1$.
    Define $\bS = a_1 \cdot \bb_1 + \cdots + a_m \cdot \bb_m$.
    Then the random variable $|\bS|^2 - 1$ is mean-zero and sub-exponential with parameter $\gamma$.
\end{lemma}

Now that we have shown our random variables are well-concentrated,
we would like to that averages of them,
as occur in the formula for the width (\Cref{eq:width-copied} above),
are extremely well-concentrated.
This can be shown using Bernstein's inequality for averages of independent sub-exponential random variables, which is stated in \cite[Corollary 2.8.3]{Ver18}.



\begin{theorem}[Bernstein's inequality]
\label{thm:bernstein-subexp}
    There exists a constant $c > 0$ such that the following is true.
    Let $\bX_1,\dots,\bX_m$ be independent, mean-zero, sub-exponential random variables, each with sub-exponential parameter at most $\gamma$. Then we have
    \begin{equation*}
        \Pr \left[ \abs{\frac{1}{m} \sum_{i=1}^m \bX_i} \geq t \right] \leq 2 \cdot \mathrm{exp} \left( -c \cdot \min\left\{ \frac{t^2}{\gamma^2},\frac{t}{\gamma} \right\} \cdot m\right), \ \text{ for all } t \geq 0.
    \end{equation*}
\end{theorem}

Now we combine these ingredients to show the following tail bound on the width.

\begin{lemma}[Tail bound on the width]
\label{lem:bounded-width}
There exists a constant $c > 0$ such that the following is true.
Let $\bprs:[K] \times [N] \rightarrow \{\pm 1\}$
be a uniformly random function.
Then for all $t \geq 0$,
\begin{equation*}
    \Pr_{\bR}[\width(\bprs) \geq 1 + t] \leq 2\qdim \cdot \mathrm{exp} \left( -c \cdot \min\{ t^2,t \}\cdot K\right).
\end{equation*}
\end{lemma}

\begin{proof}[Proof of \Cref{lem:bounded-width}]
    For each $1 \leq i \leq \qdim$ and $1 \leq k \leq K$,
    let us define the random variables
    \begin{equation*}
        \width_{i, k}(\bprs) \coloneqq \frac{\abs{\braket{v_i}{\psi_{\bprs_k}}}^2}{\weights_{V,i}}
        \quad\text{and}\quad
        \width_{i}(\bprs) \coloneqq \frac{1}{K} \cdot \sum_{k=1}^K\width_{i,k}(\bprs).
    \end{equation*}
    \Cref{lem:sub-exponential-calculation} states that there is a constant $\gamma \geq 1$ such that $(\width_{i, k}(\bprs) -1)$ is sub-exponential with parameter $\gamma$, for all $1 \leq i \leq \qdim$ and $1 \leq k \leq K$.
    Now, fix a value $1 \leq i \leq \qdim$.
    Since each $\width_{i, k}(\bprs)$ only depends on $\bprs_k$,
    the random variables $(\width_{i, k}(\bprs) -1)$ are independent across all $1 \leq k \leq K$.
    As a result, Bernstein's inequality states that
    there exists a constant $c > 0$ such that for all $t \geq 0$,
    \begin{align*}
        \Pr_{\bprs}[|\width_i(\bprs) - 1| \geq t]
        &=\Pr_{\bprs} \Big[ \Big|\frac{1}{K} \cdot \sum_{k=1}^K (\width_{i, k}(\bprs)-1)\Big|\geq t \Big]\\
        &\leq 2 \cdot \mathrm{exp} \left( -c \cdot \min\left\{ \frac{t^2}{\gamma^2},\frac{t}{\gamma} \right\} \cdot K\right)\\
        &\leq 2 \cdot \mathrm{exp} \left( -c \cdot \min\left\{ \frac{t^2}{\gamma^2},\frac{t}{\gamma^2} \right\} \cdot K\right) \tag{because $\gamma \geq 1$}\\
        &= 2 \cdot \mathrm{exp} \left( -\Big(\frac{c}{\gamma^2}\Big) \cdot \min\{ t^2, t \} \cdot K\right).
    \end{align*}
    Now, since the width is defined as $\width(\bprs) = \max_{1 \leq i \leq \qdim} \{\width_i(\bprs)\}$, we have that
    \begin{align*}
        \Pr_{\bprs}[\width(\bprs) \geq 1 + t]
        &= \Pr_{\bprs}[\exists_{1 \leq i \leq \qdim} \{\width_i(\bprs) \geq 1 + t\}]\\
        &\leq \sum_{i=1}^{\qdim} \Pr_{\bprs}[\width_i(\bprs) \geq 1 + t] \tag{by the union bound}\\
        &\leq \sum_{i=1}^{\qdim} \Pr_{\bprs}[|\width_i(\bprs) - 1| \geq t] \\
        &\leq \sum_{i=1}^{\qdim} 2 \cdot \mathrm{exp} \left( -\Big(\frac{c}{\gamma^2}\Big) \cdot \min\{ t^2, t \} \cdot K\right) \\
        & = 2\qdim \cdot \mathrm{exp} \left( -\Big(\frac{c}{\gamma^2}\Big) \cdot \min\{ t^2, t \} \cdot K\right).
    \end{align*}
    This completes the proof, by taking the constant ``$c$'' in the lemma statement to be $c/\gamma^2$.
\end{proof}

Finally, we use our tail bound to derive an expectation bound on the width.
Our proof will allow us to prove a bound of $1 + o(1)$ for a wide range of parameters $\qdim$ and $K$, as our initial intuition suggested.
However, to get a bound which applies to the widest relevant range of parameters, we will prove a slightly weaker $O(1)$ bound, which is still sufficient for our applications.

\begin{lemma} [Expectation bound on the width]
\label{lem:width-expectation}
There exists a constant $C \geq 1$ such that the following is true.
Let $\bprs:[K] \times [N] \rightarrow \{\pm 1\}$
be a uniformly random function family.
Suppose that $\qdim \leq e^K$. Then
    \begin{align*}
        \E_{\bR}\left[ \width(\bprs) \right] \leq C.
    \end{align*}
\end{lemma}

\begin{proof}
    Fix some $\alpha \geq 1$, to be determined later. Then
    \begin{align*}
        \E_{\bR}\left[ \width(\bprs) \right] &= \int_{0}^\infty \Pr[\width(\bprs) \geq t]dt\\
        &= \int_{0}^{1 + \alpha} \Pr[\width(\bprs) \geq t]dt + \int_{1 + \alpha}^\infty \Pr[\width(\bprs) \geq t]dt\\
        &\leq 1 + \alpha + \int_{1+\alpha}^\infty \Pr[\width(\bprs) \geq t]dt\\
        &= 1 + \alpha + \int_{\alpha}^\infty \Pr[\width(\bprs) \geq 1+t]dt\\
        &\leq 1 + \alpha + \int_{\alpha}^\infty 2\qdim \cdot \mathrm{exp} \left( -cK \cdot \min\{ t^2,t \right) \} dt\tag{by \Cref{lem:bounded-width}}\\
        &= 1 + \alpha + \int_{\alpha}^\infty 2\qdim \cdot \mathrm{exp} \left( -cK \cdot t\right) dt. \tag{because $\alpha \geq 1$}
    \end{align*}
    We can compute the integral exactly:
    \begin{equation*}
        2\qdim \cdot \int_{\alpha}^\infty \mathrm{exp} \left( -c K \cdot t \right) dt = -\frac{2\qdim}{cK} \cdot  \mathrm{exp}(-cK \cdot t) \bigg\rvert_{\alpha}^{\infty} =  \frac{2\qdim}{cK} \cdot  \mathrm{exp}(-cK \cdot \alpha).
    \end{equation*}
    In total, this gives us a bound of
    \begin{equation*}
        \E_{\bR}\left[ \width(\bprs) \right]
        \leq 1 + \alpha + \frac{2\qdim}{cK} \cdot  \mathrm{exp}(-cK \cdot \alpha)
        \leq 1 + \alpha + \frac{2\qdim}{c} \cdot  \mathrm{exp}(-cK \cdot \alpha).
    \end{equation*}
    Now we select $\alpha$ to be $\alpha = \max\{1, c^{-1}\}$. Then we get a bound of
    \begin{equation*}
        1 + \max\{1, c^{-1}\} + \frac{2\qdim}{c} \cdot  \mathrm{exp}(-cK \cdot \max\{1, c^{-1}\})
        \leq 1 + \max\{1, c^{-1}\} + \frac{2\qdim}{c} \cdot  \mathrm{exp}(-K).
    \end{equation*}
    When $\qdim \leq e^K$, this is at most
    \begin{equation*}
        1 + \max\{1, c^{-1}\} + \frac{2}{c},
    \end{equation*}
    which is a constant. Picking this for the ``$C$'' in the lemma statement completes the proof.
\end{proof}
\ignore{
OLD

\begin{proof}
    Fix some $0 < \delta < 1$, to be determined later. Then
    \begin{align*}
        \E_{\bR}\left[ \width(\bprs) \right] &= \int_{0}^\infty \Pr[\width(\bprs) \geq t]dt\\
        &= \int_{0}^{1 + \delta} \Pr[\width(\bprs) \geq t]dt + \int_{1 + \delta}^\infty \Pr[\width(\bprs) \geq t]dt\\
        &\leq 1 + \delta + \int_{1+\delta}^\infty \Pr[\width(\bprs) \geq t]dt\\
        &= 1 + \delta + \int_{\delta}^\infty \Pr[\width(\bprs) \geq 1+t]dt\\
        &\leq 1 + \delta + \int_{\delta}^\infty 2\qdim \cdot \mathrm{exp} \left( -cK \cdot \min\{ t^2,t \right) \} dt\tag{by \Cref{lem:bounded-width}}\\
        &= 1 + \delta + \int_{\delta}^1 2\qdim \cdot \mathrm{exp} \left( -c K \cdot t^2\right) dt + \int_{1}^\infty 2\qdim \cdot \mathrm{exp} \left( -cK \cdot t\right) dt
    \end{align*}
    Let's compute both of these integrals separately. The second integral is
    \begin{equation*}
        2\qdim \cdot \int_{1}^\infty \mathrm{exp} \left( -c K \cdot t \right) dt = -\frac{2\qdim}{cK} \cdot  \mathrm{exp}(-c \cdot t \cdot K) \bigg\rvert_1^{\infty} =  \frac{2\qdim}{cK} \cdot  \mathrm{exp}(-cK).
    \end{equation*}
    The first integral can be bounded as
    \[ \int_{\delta}^1 2\qdim \cdot \mathrm{exp} \left( -c K \cdot t^2\right) dt \leq 2\qdim\mathrm{exp} \left( -c K \cdot \delta^2\right)
    \]
    since $\mathrm{exp} \left( -c K \cdot t^2\right)$ is decreasing on $[\delta,1]$. In total, this gives us a bound of
    \begin{equation*}
        \E_{\bR}\left[ \width(\bprs) \right]
        \leq 1 + \delta + 2\qdim\mathrm{exp} \left( -c K \cdot \delta^2\right) + \frac{2\qdim}{cK} \cdot  \mathrm{exp}(-cK).
    \end{equation*}
    Now we select $\delta$ to be $\delta = K^{-1/3}$. (Indeed, picking $\delta = K^{\varepsilon - 1/2}$ for any constant $0 < \varepsilon < 1/2$ will suffice.) Then we get a bound of
    \begin{align*}
        & 1 + \frac{1}{K^{1/3}} + \frac{2\qdim}{\mathrm{exp}(c\cdot K^{1/3})} + \frac{2\qdim}{cK} \cdot  \mathrm{exp}(-cK) = 1 + o(1),
    \end{align*}
    where here we are using our choice of parameters $K \geq n^{\omega(1)}$ and $\qdim = 2^{\poly(n)}$.
\end{proof}

XXXXXXXXXXXX

To start, let us define individual random variables
corresponding to each term in the width.

\begin{definition}[Individual widths]
    Let $\prs:[K] \times [N] \rightarrow \{\pm 1\}$ be a  function family. For each $1 \leq i \leq \qdim$ and $1 \leq k \leq K$ we define the corresponding \emph{individual width} as
    \begin{equation*}
        \width_{i, k}(\prs) = \frac{\abs{\braket{v_i}{\psi_{\prs_k}}}^2}{\weights_{V,i}}.
    \end{equation*}
\end{definition}

The expression $\braket{v_i}{\psi_{\prs_k}}$ in the individual width is a linear combination of the Boolean values $\prs_k(1), \ldots, \prs_k(N)$.
To show concentration for this expression, we use the following tail bound, a version of Hoeffding's inequality for complex random variables.

\begin{theorem}[Sub-Gaussian concentration for sums of complex random variables]\label{thm:complex-concentration}
Let $\bb_1, \ldots, \bb_m$ be independent and uniform $\pm 1$ random variables,
and let $a_1, \ldots, a_m$ be complex numbers.
Then $\bS = a_1 \cdot \bb_1 + \cdots + a_m \cdot \bb_m$ satisfies
\begin{equation*}
\Pr[|\bS| \geq t] \leq 2\cdot \mathrm{exp} \left(-\frac{t^2}{\sum_{i \in [m]} |a_i|^2}\right).
\end{equation*}
\end{theorem}

Curiously, the only proof we know for the above bound is via the \emph{matrix concentration} tail bound stated in~\cref{thm:khintchine}.

\begin{proof}[Proof of~\Cref{thm:complex-concentration}]
    We invoke~\cref{thm:khintchine} by treating each $\bb_i$ as a $1 \times 1$-dimensional complex matrix. In particular, we define the $1 \times 1$-dimensional matrix $\bS = a_1 \cdot \bb_1 + \cdots + a_m \cdot \bb_m$ and compute its ``matrix'' variance parameter
    \begin{align*}
        v(\bS) = \max\{\norm*{ \E[\bS \cdot \bS^\dagger] }_\operator, \norm*{ \E[\bS^\dagger \cdot \bS] }_\operator\} = \E \Big[ \sum_{i,j \in [m]} a_i^\dagger a_j \bb_i \bb_j \Big] = \sum_{i \in [m]} \abs{a_i}^2.
    \end{align*}
    Then the tail bound of~\cref{thm:khintchine} implies that for all $t \geq 0$,
    \begin{align*}
    \Pr\left[ \abs{\bS} \geq t \right] = \Pr\left[ \norm{\bS}_{\operator} \geq t\right] \leq 2 \cdot \mathrm{exp}\left(-\frac{t^2}{\sum_{i \in [m]} \abs{a_i}^2}\right).
    \end{align*}
\end{proof}


We will use \Cref{thm:complex-concentration} to show that the individual widths posses a concentration property known as being \emph{sub-exponential}.

\begin{definition}[Sub-exponential random variables]
A random variable $\bX$ is \emph{sub-exponential with parameter $\gamma$} if
\begin{equation*}
        \Pr \left[ \abs{\bX} > t \right] \leq 2 \cdot \mathrm{exp} \left(-\frac{t}{\gamma} \right), \ \text{ for all } t \geq 0.
\end{equation*}
\end{definition}

\begin{lemma}[Individual widths are sub-exponential]
    Let $\bprs:[K] \times [N] \rightarrow \{\pm 1\}$
be a uniformly random function family.
Then for each $1 \leq i \leq \qdim$ and $1 \leq k \leq K$,
$(\width_{i, k}(\bprs) - 1)$ is a mean-zero, sub-exponential random variable with parameter 1.
\end{lemma}
\begin{proof}
    By \Cref{prop:typical-magnitude},
    \begin{equation*}
        \E_{\bprs}[\width_{i, k}(\bprs) - 1]
        = \E_{\bprs}\Big[\frac{\abs{\braket{v_i}{\psi_{\bprs_k}}}^2}{\weights_{V,i}}-1\Big]
        = \Big(\frac{\weights_{V,i}}{\weights_{V,i}}-1\Big)
        = 0.
    \end{equation*}
    This proves the mean-zero claim.
    As for being sub-exponential:
    \begin{align*}
        \Pr\left[ \frac{\abs{\braket{v_i}{\psi_{\bR_k}}}^2}{\weights_{V,i}} -1 \geq t\right] &= \Pr\left[ \abs{\braket{v_i}{\psi_{\bR_k}}} \geq \sqrt{(t+1) \cdot \weights_{V,i}} \right]\\
        &= \Pr\left[ \abs{\frac{1}{\sqrt{N}} \sum_{x \in [N]} v_{i,x}^\dagger \bR_k(x) } \geq \frac{1}{\sqrt{N}}\sqrt{(t+1)\cdot \sum_{x \in [N]} \abs{v_{i,x}}^2} \right]\\
        &= \Pr\left[ \abs{ \sum_{x \in [N]} v_{i,x}^\dagger \bR_k(x) } \geq \sqrt{(t+1)\cdot \sum_{x \in [N]} \abs{v_{i,x}}^2} \right]\\
        &\leq 2 \cdot \mathrm{exp} \left( -\frac{(t+1)\cdot \sum_{x \in [N]} \abs{v_{i,x}}^2}{\sum_{x \in [N]} \abs{v_{i,x}}^2} \right) = 2 \cdot \mathrm{exp} \left( -t-1\right) \leq 2 \cdot \mathrm{exp} \left( -t\right).
    \end{align*}
    where the inequality is Hoeffding's inequality for complex random variables (\cref{thm:complex-concentration}).
\end{proof}

This argument is formalized in the following tail bound on the width, which we will later convert into a bound on the expectation.

\begin{lemma}[Tail bound on the width]
\label{lem:bounded-width}
There exists a constant $c > 0$ such that the following is true.
Let $\bprs:[K] \times [N] \rightarrow \{\pm 1\}$
be a uniformly random function family.
Then for all $t \geq 0$,
\begin{equation*}
    \Pr_{\bR}[\width(\bprs) \geq 1 + t] \leq 2\qdim \cdot \mathrm{exp} \left( -c \cdot \min\{ t^2,t \}\cdot K\right).
\end{equation*}
\end{lemma}


The other technical tool we'll need is a concentration inequality for averages of sub-exponential random variables. Recall that a random variable $\bX$ is sub-exponential with parameter $\gamma$ if
\begin{align*}
        \Pr \left[ \abs{\bX} > t \right] \leq 2 \cdot \mathrm{exp} \left(-\frac{t}{\gamma} \right) \ \text{ for all }, t \geq 0.
\end{align*}


\begin{proof}[Proof of \Cref{lem:bounded-width}]
    Recall that $\width(\bprs)$ is defined as
    \begin{equation}
        \width(\bprs) = \max_{1 \leq i \leq \qdim}\Big\{\frac{1}{K} \sum_{k \in [K]} \frac{|\braket*{v_i}{\psi_{\bR_{k}}}|^2}{\weights_{V, i}} \Big\}.\label{eq:recall-width}
    \end{equation}
    To prove a tail bound on $\width(\bprs)$, we'll first show that for any fixed $1 \leq i \leq \qdim$ and $1 \leq k \leq K$, the random variable
    \begin{align*}
        \frac{\abs{\braket{v_i}{\psi_{\bR_k}}}^2}{\weights_{V,i}} -1
    \end{align*}
    is a mean-zero, sub-exponential random variable with parameter $1$. The mean-zero claim follows from \Cref{prop:typical-magnitude}, which states that
    \begin{align*}
        \weights_{V,i} = \E_{\bR_k} \abs{\braket{v_i}{\psi_{\bR_k}}}^2.
    \end{align*}
    We can bound the tails of this random variable as follows:
    \begin{align*}
        \Pr\left[ \frac{\abs{\braket{v_i}{\psi_{\bR_k}}}^2}{\weights_{V,i}} -1 \geq t\right] &= \Pr\left[ \abs{\braket{v_i}{\psi_{\bR_k}}} \geq \sqrt{(t+1) \cdot \weights_{V,i}} \right]\\
        &= \Pr\left[ \abs{\frac{1}{\sqrt{N}} \sum_{x \in [N]} v_{i,x}^\dagger \bR_k(x) } \geq \frac{1}{\sqrt{N}}\sqrt{(t+1)\cdot \sum_{x \in [N]} \abs{v_{i,x}}^2} \right]\\
        &= \Pr\left[ \abs{ \sum_{x \in [N]} v_{i,x}^\dagger \bR_k(x) } \geq \sqrt{(t+1)\cdot \sum_{x \in [N]} \abs{v_{i,x}}^2} \right]\\
        &\leq 2 \cdot \mathrm{exp} \left( -\frac{(t+1)\cdot \sum_{x \in [N]} \abs{v_{i,x}}^2}{\sum_{x \in [N]} \abs{v_{i,x}}^2} \right) = 2 \cdot \mathrm{exp} \left( -t-1\right) \leq 2 \cdot \mathrm{exp} \left( -t\right).
    \end{align*}
    where the inequality is Hoeffding's inequality for complex random variables (\cref{thm:complex-concentration}).

    For any fixed $i$, this random variable is independent across all $k \in [K]$. Thus, by~\cref{thm:bernstein-subexp}, there exists a constant $c > 0$ such that for all $t \geq 0$,
    \begin{align*}
        \Pr \left[ \abs{\frac{1}{K}\sum_{k \in [K]}  \Bigg(\frac{\abs{\braket{v_i}{\psi_{\bR_k}}}^2}{\weights_{V,i}} -1\Bigg)} \geq t \right] \leq 2 \cdot \mathrm{exp} \left( -c \cdot \min\left( t^2,t \right) \cdot K\right),
    \end{align*}
    which in particular implies that for all $t \geq 0$,
    \begin{align*}
        \Pr \left[ \frac{1}{K}\sum_{k \in [K]}  \frac{\abs{\braket{v_i}{\psi_{\bR_k}}}^2}{\weights_{V,i}} \geq 1 + t \right] \leq 2 \cdot \mathrm{exp} \left( -c \cdot \min\left( t^2,t \right) \cdot K\right).
    \end{align*}
    Next, by a union bound over all $i$, we have for all $t \geq 0$,
    \begin{align*}
        \Pr[\width(\bprs) \geq 1 + t] &= \Pr \left[ \max_{1 \leq i \leq \qdim}\Big\{\frac{1}{K} \sum_{k \in [K]} \frac{|\braket*{v_i}{\psi_{\bR_{k}}}|^2}{\weights_{V, i}} \Big\}\geq 1 + t \right] \\
        &\leq 2\qdim \cdot \mathrm{exp} \left( -c \cdot \min\left( t^2,t \right) \cdot K\right).
    \end{align*}

    Thus, each term inside the max in~\cref{eq:recall-width} is the expectation of $K$ i.i.d. sub-exponential random variables each with parameter 
    
\end{proof}
}

\subsection{The one-query lower bound}\label{sec:the-full-proof-of-khintchine-proof}

Now we complete the proof of the one-query lower bound. We begin by proving a bound on the expected value of the distinguishing advantage.

\begin{theorem}[Expectation bound for the distinguishing advantage]\label{thm:khintchine-bound}
There exists a constant $C > 0$ such that the following is true.
Let $\bprs:[K] \times [N] \rightarrow \{\pm 1\}$
be a uniformly random function family. Then
\begin{equation*}
    \E_{\bprs}[\distinguish_{A}(\bprs)] \leq C \cdot\sqrt{ \frac{\ln(\qdim)}{K} }.
\end{equation*}
\end{theorem}

\begin{proof}[Proof of \Cref{thm:khintchine-bound}]
    We will consider two regimes of parameters: $\qdim \leq e^K$ and $\qdim > e^K$.
    Let us first consider the case of $\qdim \leq e^K$.
    Let $\bprs, \bprs':[K]\times[N]\rightarrow \{\pm 1\}$ be two independent and uniformly random function families.
    Then
    \begin{align*}
    \E_{\bprs}[\distinguish_{A}(\bprs)]
        &\leq 4 \cdot \E_{\bR, \bR'}\Big[\max_{f} \Big\{\distinguish_{\decup}(\bprs, \bprs' \mid f)\Big\}\Big] \tag{by \Cref{lem:decouple}}\\
    &\leq 4 \cdot \E_{\bprs, \bprs'} [\Delta_{\decup}^{\spectral}(\bprs, \bprs')] \tag{by \Cref{lem:decouple-relax}}\\
    & \leq 4\cdot \E_{\bprs} \Big[ \sqrt{\frac{2 \ln(2\qdim) \cdot \width(\bprs)}{K}} \Big] \tag{by \Cref{lem:apply-khintchine}}\\
    & \leq 4\cdot\sqrt{ \E_{\bprs} \Big[ \frac{2 \ln(2\qdim) \cdot \width(\bprs)}{K} \Big]} \tag{by Jensen's inequality}\\
    & \leq 4\cdot\sqrt{ \frac{2 \ln(2\qdim) \cdot C}{K} } \tag{by \Cref{lem:width-expectation}, for some constant $C\geq 1$}\\
    & \leq 4\cdot\sqrt{ \frac{2 \cdot (2\ln(\qdim)) \cdot C}{K} } \tag{because $\qdim \geq 2$} \\
    &= 8\sqrt{C} \cdot\sqrt{ \frac{\ln(\qdim)}{K} }.
    \end{align*}
    Picking the ``$C$'' in the theorem statement to be $8 \sqrt{C}$,
    this completes the $\qdim \leq e^K$ case.
    As for the $\qdim > e^K$ case, we note that because the distinguishing advantage is a difference of two probabilities, it is always at most $1$. Hence,
    \begin{equation*}
        \E_{\bprs}[\distinguish_{A}(\bprs)] \leq 1
        \leq 8\sqrt{C} \leq 8\sqrt{C} \cdot\sqrt{ \frac{\ln(\qdim)}{K}}.
    \end{equation*}
    The first inequality is because $C \geq 1$,
    and the second inequality is because $\qdim > e^K$.
    This completes the $\qdim > e^K$ case,
    and therefore completes the proof. 
\end{proof}

Combining this with \cref{lem:adv-tail},
we have our main technical result.

\begin{theorem}[Main theorem]\label{thm:main-at-end-of-proof}
There exists a constants $C_1, C_2 > 0$ such that the following is true.
Let $\bprs:[K] \times [N] \rightarrow \{\pm 1\}$
be a uniformly random function family. Then
\begin{equation*}
    \Pr_{\bprs}\Big[\distinguish_{A}(\bprs) \geq C_1 \cdot\sqrt{ \frac{\ln(\qdim)}{K} } + \epsilon\Big] \leq 4\cdot\exp(-C_2 \cdot\epsilon^2 KN).
\end{equation*}
\end{theorem}

In particular, this implies \cref{thm:main-distinguishing-intro}.

\subsection{Technical lemma: sub-exponential random variables}\label{sec:annoying-technical-lemma}

Now we prove \Cref{lem:sub-exponential-calculation}. For convenience, we restate it here.

\begin{lemma}[\Cref{lem:sub-exponential-calculation} restated]\label{lem:sub-exponential-calculation-restated}
    There exists a constant $\gamma \geq 1$ such that the following is true. 
    Let $\bb_1, \ldots, \bb_m$ be independent and uniform $\pm 1$ random variables,
    and let $a_1, \ldots, a_m$ be complex numbers
    such that $|a_1|^2 + \cdots + |a_m|^2 = 1$.
    Define $\bS = a_1 \cdot \bb_1 + \cdots + a_m \cdot \bb_m$.
    Then the random variable $|\bS|^2 - 1$ is mean-zero and sub-exponential with parameter $\gamma$.
\end{lemma}

To compute the mean of $|\bS|^2 - 1$, we will use the following proposition.

\begin{proposition}\label{prop:s^2-mean}
    Let $\bb_1, \ldots, \bb_m$ be independent and uniform $\pm 1$ random variables,
    and let $a_1, \ldots, a_m$ be complex numbers.
    Then $\bS = a_1 \cdot \bb_1 + \cdots + a_m \cdot \bb_m$ satisfies
    \begin{equation*}
        \E[|\bS|^2] = \sum_{i=1}^m |a_i|^2.
    \end{equation*}
    \end{proposition}
\begin{proof}
    We calculate
    \begin{align*}
        \E[|\bS|^2]
        = \E[\bS^\dagger \cdot \bS]
        &= \E[\Big(\sum_{i=1}^m a_i \cdot \bb_i\Big)^\dagger \cdot \Big(\sum_{j=1}^m a_j \cdot \bb_i\Big)]\\
        &= \E \Big[ \sum_{i,j=1}^m a_i^\dagger a_j \cdot \bb_i \bb_j \Big]
        =  \sum_{i,j=1}^m a_i^\dagger a_j \cdot \E[\bb_i \bb_j]
        = \sum_{i =1}^m \abs{a_i}^2,
    \end{align*}
    where the final step used $\E[\bb_i \bb_j] = 1$ if $i = j$ and $0$ if $i \neq j$. This completes the proof.
\end{proof}

To show concentration for $|\bS|^2 - 1$, we use the following tail bound, a version of Hoeffding's inequality for complex-weighted random sums.

\begin{theorem}[Sub-Gaussian concentration for sums of complex random variables]\label{thm:complex-concentration}
Let $\bb_1, \ldots, \bb_m$ be independent and uniform $\pm 1$ random variables,
and let $a_1, \ldots, a_m$ be complex numbers.
Then $\bS = a_1 \cdot \bb_1 + \cdots + a_m \cdot \bb_m$ satisfies
\begin{equation*}
\Pr[|\bS| \geq t] \leq 2\cdot \mathrm{exp} \left(-\frac{t^2}{2\cdot \sum_{i =1}^m |a_i|^2}\right).
\end{equation*}
\end{theorem}

As it turns out, this can be proved as a (very) special case of the \emph{matrix concentration} tail bound stated in~\cref{thm:khintchine}.

\begin{proof}[Proof of~\Cref{thm:complex-concentration}]
    We invoke~\cref{thm:khintchine} by treating $\bS$ as a $1 \times 1$ complex-valued matrix. In particular,  define the $1 \times 1$ matrix
    \begin{equation*}
        \widehat{\bS}
        \coloneqq \begin{pmatrix}
            \bS
        \end{pmatrix}
        = \begin{pmatrix}
            a_1 \cdot \bb_1 + \cdots + a_m \cdot \bb_m
        \end{pmatrix}.
    \end{equation*}
    Then its ``matrix'' variance parameter is
    \begin{equation*}
        v(\widehat{\bS})
        = \max\{\norm*{ \E[\widehat{\bS} \cdot \widehat{\bS}^\dagger] }_\operator, \norm*{ \E[\widehat{\bS}^\dagger \cdot \widehat{\bS}] }_\operator\}
         = \norm*{\E\begin{pmatrix}|\bS|^2 \end{pmatrix}}_{\operator}\\
        = \E[|\bS|^2]
        = \sum_{i =1}^m \abs{a_i}^2,
    \end{equation*}
    where the final step used \Cref{prop:s^2-mean}.
    Then the tail bound of~\cref{thm:khintchine} implies that for all $t \geq 0$,
    \begin{align*}
    \Pr\left[ \abs{\bS} \geq t \right] = \Pr\left[ \norm*{\widehat{\bS}}_{\operator} \geq t\right] \leq 2 \cdot \mathrm{exp}\left(-\frac{t^2}{2\cdot\sum_{i \in [m]} \abs{a_i}^2}\right).
    \end{align*}
    This completes the proof.
\end{proof}

The following is an immediate corollary of \Cref{thm:complex-concentration}.

\begin{corollary}\label{cor:s^2-is-subexponential}
Let $\bS$ be as in \Cref{lem:sub-exponential-calculation-restated}. Then $|\bS|^2$  is a sub-exponential random variable with parameter $\gamma = 2$.
\end{corollary}
\begin{proof}
    By \Cref{thm:complex-concentration},
    \begin{equation*}
        \Pr[|\bS| \geq t]
        \leq 2\cdot \mathrm{exp} \left(-\frac{t^2}{2\cdot \sum_{i =1}^m |a_i|^2}\right)
        = 2\cdot \mathrm{exp}(-t^2/2).
    \end{equation*}
    Hence,
    \begin{equation*}
        \Pr[|\bS|^2 \geq t]
        \leq 2\cdot \mathrm{exp}(-t/2).
    \end{equation*}
    This means that $|\bS|^2$ is a sub-exponential random variable with parameter $\gamma = 2$.
\end{proof}

    Now we want to show that $|\bS|^2 - 1$ is also sub-exponential,
    taking advantage of the fact that $\E [|\bS|^2] = 1$.
    To do so, we will use standard facts about sub-exponential random variables from \cite[Section 2.7]{Ver18}.
    In particular, we will rely on an alternative method of parameterizing sub-exponential random variables in terms of their \emph{moment generation functions (MGFs)}.

    \begin{definition}[Sub-exponential norm]
        Given a real random variable $\bX$, the MGF of $|\bX|$ is \emph{bounded at point $\kappa > 0$} if
    \begin{equation*}
        \E[\exp(|\bX|/\kappa)] \leq 2.
    \end{equation*}
    The smallest $\kappa$ for which this equation is holds is given by the \emph{sub-exponential norm of $\bX$}, denoted $\Vert \bX \Vert_{\psi_1}$, and is defined formally as follows:
    \begin{equation*}
        \Vert \bX \Vert_{\psi_1} = \inf \{ t > 0 : \E[\exp(|\bX|/\kappa)] \leq 2\}.
    \end{equation*}
    \end{definition}

     We require two facts about this method of parameterizing sub-exponential random variables.
     The first is stated in \cite[Proposition 2.7.1]{Ver18} and the second is stated in \cite[Exercise 2.7.10]{Ver18}.

     \begin{proposition}[Approximate equivalence of the two parameterizations]\label{prop:both-params-same}
    There is an absolute constant $C_1 > 0$ such that the following is true. If the MGF of $|\bX|$ is bounded at point $\kappa$, then $\bX$ is sub-exponential with parameter $\gamma$, for some $\gamma \leq C_1 \cdot \kappa$. Likewise, if $\bX$ is sub-exponential with parameter $\gamma$, then the MGF of $|\bX|$ is bounded at point $\kappa$, for some $\kappa \leq C_1 \cdot \gamma$.
    \end{proposition}

    \begin{proposition}[Centering]\label{prop:centering} There is an absolute constant $C_2 > 0$ such that the following is true. If $\bX$ is a sub-exponential random variable, then so is $\bX - \E[\bX]$, and it satisfies
        \begin{equation*}
            \Vert \bX - \E[\bX] \Vert_{\psi_1} 
            \leq C_2 \cdot \Vert \bX \Vert_{\psi_1}.
        \end{equation*}
    \end{proposition}

Now we prove \Cref{lem:sub-exponential-calculation-restated}.

\begin{proof}[Proof of \Cref{lem:sub-exponential-calculation-restated}]
    First, \Cref{prop:s^2-mean} states that
    \begin{equation*}
        \E[|\bS|^2] = \sum_{i=1}^m |a_i|^2 = 1.
    \end{equation*}
    Hence, $|\bS|^2 - 1$ is mean-zero.
    Next,
    \Cref{cor:s^2-is-subexponential} states that $|\bS|^2$ is a sub-exponential random variable with parameter $\gamma_1 = 2$.
    \Cref{prop:both-params-same} then implies that
    the MGF of $||\bS|^2|$ is bounded at point $\kappa_1$, for some \begin{equation*}
        \kappa_1 \leq C_1 \cdot \gamma_1 = 2\cdot C_1.
    \end{equation*}
    By definition of the sub-exponential norm, this immediately implies that $\Vert |\bS|^2 \Vert_{\psi_1} \leq 2\cdot C_1$.
    \Cref{prop:centering} then implies that
    \begin{equation*}
        \Vert |\bS|^2 - 1 \Vert_{\psi_1} 
        = \Vert |\bS|^2 - \E[|\bS|^2] \Vert_{\psi_1} 
        \leq C_2 \cdot \Vert |\bS|^2 \Vert_{\psi_1}
        \leq C_2 \cdot 2\cdot C_1.
    \end{equation*}
    Now $|\bS|^2 - 1$ is a non-constant random variable, and in particular it is nonzero with finite probability.
    In addition, it only obtains a discrete set of values.
    Hence, the infimum over $\{t > 0\}$ in the definition of the sub-exponential norm $\Vert |\bS|^2 - 1 \Vert_{\psi_1}$ is achieved at a nonzero minimizing value $\kappa_2 > 0$; in other words, if we set
    \begin{equation*}
        \kappa_2 = \Vert |\bS|^2 - 1 \Vert_{\psi_1}
        \leq 2\cdot C_1 \cdot C_2,
    \end{equation*}
    then the MGF of $||\bS|^2 - 1|$ is bounded at point $\kappa_2$.
    Applying \Cref{prop:both-params-same} again,
    this implies that $|\bS|^2 - 1$ is sub-exponential with parameter $\gamma_2$, for some
    \begin{equation*}
        \gamma_2 \leq C_1 \cdot \kappa_2 \leq 2\cdot C_1^2 \cdot C_2.
    \end{equation*}
    Now, we note that if a random variable $\bX$ is sub-exponential with parameter $a > 0$, then it is also sub-exponential with parameter $b$, for any $b \geq a$.
    This is because for all $t > 0$,
    \begin{equation*}
        \Pr \left[ \abs{\bX} > t \right]
        \leq 2 \cdot \mathrm{exp} \left(-\frac{t}{a} \right)
        \leq 2 \cdot \mathrm{exp} \left(-\frac{t}{b} \right).
    \end{equation*}
    Hence, because $|\bS|^2 - 1$ is sub-exponential for parameter $\gamma_2 \leq 2\cdot C_1^2 \cdot C_2$,
    it is also sub-exponential for parameter $\gamma = \max\{1, 2\cdot C_1^2 \cdot C_2\}$.
    This is a constant which is greater than or equal to 1, which completes the proof.
\end{proof}

\section{Pseudorandom states relative to a random oracle}\label{sec:prs}
In this section, we use \cref{thm:main-at-end-of-proof} to derive \cref{thm:main-prs-intro}, our lower bound for breaking pseudorandom state families. We begin with a definition of (single-copy) pseudorandom states in the plain model, for reference.

\begin{definition}[Pseudorandom state family]\label{def:prs}
Let $n:\N\rightarrow \N$ be a function
and $\{\ket{\psi_{\secp, k}}\}_{k \in \{0,1\}^\secp}$ be a family of $n(\secp)$-qubit quantum states for each $\secp \in \mathbb N$.
Then the state family ensemble
\begin{equation*}
    \{\{\ket{\psi_{\secp, k}}\}_{k\in\{0,1\}^\secp}\}_{\secp \in \N}
\end{equation*}
is a \emph{pseudorandom state (PRS) family} if it has the following properties.

\begin{itemize}
    \item[$\circ$] \textbf{Efficient constructability:}
    there is a polynomial-time quantum algorithm that on input $(1^\secp, k)$, for $k\in \{0,1\}^\secp$, outputs $\ket{\psi_{\lambda,k}}$.
    \item[$\circ$] \textbf{Stretch:} $n(\secp) \geq \secp + 1$, for all $\secp$. 
    \item[$\circ$] \textbf{Pseudorandomness:} for all algorithms $A$ described by polynomial-size quantum circuit families, we have that
    \[ \Big|\Pr_{\bk \sim \{0, 1\}^{\secp}}\Big[ A(\ket{\psi_{\secp, \bk}})\text{ outputs } ``0"\Big] - \Pr_{\ket{\bpsi}}\Big[ A(\ket{\bpsi}) \text{ outputs } ``0"\Big]\Big| = \negl(\secp),
    \]
    where $\ket{\bpsi}$ is drawn from the Haar distribution on $n(\secp)$-qubit states. 
\end{itemize}

\end{definition}

\paragraph{Our Oracle and Adversary Model.} In this paper, we consider pseudorandom state families defined relative to an oracle $\ro: \{0,1\}^*\times \{0,1\}^* \rightarrow \{\pm 1\}$. In that case, the efficient constructability property requires that there is a quantum polynomial-time oracle algorithm that on input $(1^\secp, k)$, for $k\in \{0,1\}^\secp$, outputs $\ket{\psi_{\lambda,k}}$, given oracle access to $\ro$.

In addition, the pseudorandomness property should require that the PRS family be secure against all algorithms $A^f$, where $A^{(\cdot)}$ is an oracle algorithm described by a polynomial-size oracle circuit family and $f=f_{\prs}$ is an arbitrary $\prs$-dependent oracle;
equivalently, $A^{(\cdot)}$ is computable by a family of quantum circuits output by a polynomial-time Turing machine with the help of polynomial-size non-uniform advice.

The main result of this section (\cref{thm:main-prs-body} below) proves that relative to a random oracle, there are PRS families secure against all one-query attacks. Explicitly, the adversary model we consider is as follows:

\begin{itemize}
    \item[$\circ$] For a given function $\ro: \{0,1\}^*\times \{0,1\}^*\rightarrow \{\pm 1\}$, the adversary is described by an $\ro$-dependent Turing machine and $R$-dependent collection of advice strings $(z_{\secp})_{\secp \in \mathbb N}$.
    \item[$\circ$] On input $z_\secp$, the Turing machine outputs the description of a one-query oracle circuit $A^{(\cdot)} \coloneqq A^{(\cdot)}_{\ro, z_\secp}$. 
    \item[$\circ$] On input the state $\ket{\psi}$, the adversary executes the oracle circuit $A^{f_{\ro}}(\ket{\psi})$ for a function $f_{\ro}: \{0,1\}^*\rightarrow \{\pm 1\}$ that may depend on $\ro$. 
\end{itemize}

\begin{theorem}[\Cref{thm:main-prs-intro} formalized]\label{thm:main-prs-body}
    Let $n(\secp)$ be any efficiently computable polynomial function in $\secp$ such that $n(\secp) \geq \secp + 1$ for all $\secp$. Then with probability~$1$ over the choice of a random oracle $\bprs: \{0,1\}^* \times \{0,1\}^* \rightarrow \{\pm 1\}$, the following is true relative to $\bprs$.
    There exists a PRS family consisting of $n(\secp)$-qubit quantum states that is secure against all polynomial-time quantum algorithms $A^{f}$ that have polynomial-size non-uniform classical advice and make one query to an arbitrary Boolean function $f: \{0,1\}^* \rightarrow \{\pm 1\}$.
\end{theorem}

\begin{proof}
For each $\lambda \in \N$, we define the function family
$
    \bprs^\lambda:\{0, 1\}^\lambda \times \{0, 1\}^{n(\lambda)} \rightarrow \{\pm 1\}
$
by setting
\begin{equation*}
    \bprs^\lambda_k(x) \coloneqq \bprs(k, x),
\end{equation*}
for each $k \in \{0, 1\}^\lambda$ and $x \in \{0, 1\}^{n(\lambda)}$.
Then the candidate PRS family is the state family ensemble which contains the family of $n(\lambda)$-qubit quantum states
\begin{equation*}
    \{\ket*{\psi_{\bprs^{\lambda}_k}}\}_{k \in \{0, 1\}^\lambda},
\end{equation*}
for each security parameter $\lambda \in \N$.
By construction, the state $\ket*{\psi_{\bprs^{\lambda}_k}}$ can be generated in time $\poly(\lambda)$ given a single oracle call to $\bprs$. Thus, all that remains is to establish security. 

Security \emph{nearly} follows from \cref{thm:main-at-end-of-proof}, except that the order of quantifiers is wrong: in \cref{thm:main-at-end-of-proof}, the oracle circuit $A^{(\cdot)}$ is not allowed to depend on $\bprs$, although the function $f$ it queries is. However, in this setting, $A^{(\cdot)}$ \emph{is} allowed to depend on $\bprs$. We handle this by a standard quantifier-switching argument using the Borel-Cantelli lemma~\cite{BG81,IR89}, which applies even in the case of $A$ with bounded non-uniformity.

The argument is as follows. We abuse notation and let $A(\cdot)$ denote a polynomial-time Turing machine that on input $z_\secp$ outputs a one-query oracle circuit $A^{(\cdot)}_{z_\secp}(\cdot)$. The adversary runs $A^{f}_{z_\secp}$ on input state $\ket{\psi}$ using an arbitrary $\bprs$-dependent oracle $f = f_{\bprs}$.
Here, $z = \{z_\secp\}_\secp$ is a collection of advice strings in which $z_\secp$ has length $\poly(\lambda)$.
Because $A(\cdot)$ runs in polynomial time, the query length of $A^{f}_{z_\secp}$ is bounded by some $p(\secp) = \poly(\lambda)$. As a result, by \cref{thm:main-at-end-of-proof} (setting $\epsilon = \frac 1 {\sqrt K}$), we know that for every security parameter $\lambda \in \N$, 
\begin{equation*}
\Pr_{\bR^\lambda}\Big[ \Delta_{A_{z_\secp}}(\bR^\secp) \geq \frac{(1+ C_1\cdot\sqrt{p(\secp)})}{\sqrt{K}} \Big]
\leq 4\cdot \exp(-C_2\cdot N)
= 4\cdot \mathrm{exp}(-C_2\cdot 2^{n(\lambda)})
\leq 4\cdot \mathrm{exp}(-C_2\cdot 2^{\lambda}),
\end{equation*}
where the last inequality uses the fact that $n(\secp)\geq \secp + 1$. We may then union bound over the $2^{p(\secp)}$ possible advice strings $z_\secp$ and conclude that 
\[ \Pr_{\bR^\lambda}\Big[ \exists z: \Delta_{A_{z_\secp}}(\bR^\secp) \geq \frac{(1+C_1\cdot\sqrt{p(\secp)})}{\sqrt{K}} \Big] \leq 2^{p(\lambda)} \cdot 4\cdot \mathrm{exp}(-C_2\cdot 2^{\lambda})
\leq 4\cdot \mathrm{exp}(-c\cdot 2^{\lambda}),
\]
for a universal constant $c > 0$ and all sufficiently large $\secp$. 

Let $\calE_\secp$ denote the above event. Then, we know that the summation
\begin{equation*}
    \sum_{\secp\in \mathbb N} \underset{\bro: \{0,1\}^*\times \{0,1\}^* \rightarrow \{\pm 1\}}\Pr[\calE_\secp] < \infty
\end{equation*}
converges to a real number. Therefore, by the Borel-Cantelli lemma, 
\[ \Pr_{\bprs: \{0,1\}^*\times \{0,1\}^* \rightarrow \{\pm 1\}}\Big[ \calE_\secp \text{ occurs for infinitely many }\secp \Big] = 0.
\]
Therefore, for all sufficiently large $\lambda \in \N$,
\begin{equation}\label{eq:sad-fact}
    \Delta_{A_{z_\secp}}(\bR^\secp) \leq \frac{\poly(\secp)}{\sqrt{K}},
\end{equation}
no matter what advice $z = \{z_\secp\}_\secp$ the algorithm is given.
Finally, we observe that the probability space above is uncountable. Therefore, we may union bound over all countably many polynomial-time Turing machines $A(\cdot)$ and conclude that \Cref{eq:sad-fact} holds for all $A(\cdot)$ and all sufficiently large $\lambda \in \N$.
This shows that the PRS family satisfies the claimed pseudorandomness property, concluding the proof.
\end{proof}

\bibliographystyle{alpha}
\bibliography{biblio}

\newcommand{\etalchar}[1]{$^{#1}$}
\begin{thebibliography}{BCKM21}

\bibitem[Aar16]{Aar16}
Scott Aaronson.
\newblock The complexity of quantum states and transformations: from quantum
  money to black holes.
\newblock Technical report, arXiv:1607.05256, 2016.

\bibitem[Aar21]{Aar21}
Scott Aaronson.
\newblock Open problems related to quantum query complexity, comment \#36,
  2021.
\newblock \url{https://scottaaronson.blog/?p=5837}.

\bibitem[AK07]{AK07}
Scott Aaronson and Greg Kuperberg.
\newblock Quantum versus classical proofs and advice.
\newblock In {\em Proceedings of the 22nd Annual IEEE Conference on
  Computational Complexity}, pages 115--128, 2007.

\bibitem[AQY22]{AQY22}
Prabhanjan Ananth, Luowen Qian, and Henry Yuen.
\newblock Cryptography from pseudorandom quantum states.
\newblock In {\em Proceedings of the 42nd Annual International Cryptology
  Conference}, pages 208--236, 2022.

\bibitem[BCKM21]{BCKM21}
James Bartusek, Andrea Coladangelo, Dakshita Khurana, and Fermi Ma.
\newblock One-way functions imply secure computation in a quantum world.
\newblock In {\em Proceedings of the 41st Annual International Cryptology
  Conference}, pages 467--496, 2021.

\bibitem[BCQ23]{BCQ23}
Zvika Brakerski, Ran Canetti, and Luowen Qian.
\newblock On the computational hardness needed for quantum cryptography.
\newblock In {\em Proceedings of the 14th Innovations in Theoretical Computer
  Science}, pages 24:1--24:21, 2023.

\bibitem[BEM{\etalchar{+}}23]{BEM+23}
John Bostanci, Yuval Efron, Tony Metger, Alexander Poremba, Luowen Qian, and
  Henry Yuen.
\newblock Unitary complexity and the {U}hlmann transformation problem.
\newblock {\em Technical report, arXiv:2306.13073}, 2023.

\bibitem[BG81]{BG81}
Charles Bennett and John Gill.
\newblock Relative to a random oracle $a$, $\mathrm{P}^a \neq \mathrm{NP}^a
  \neq \mathrm{co-NP}^a$ with probability 1.
\newblock {\em SIAM Journal on Computing}, 10(1):96--113, 1981.

\bibitem[BG22]{BG22}
Anne Broadbent and Alex Grilo.
\newblock {QMA}-hardness of consistency of local density matrices with
  applications to quantum zero-knowledge.
\newblock {\em SIAM Journal on Computing}, 51(4):1400--1450, 2022.

\bibitem[BV97]{BV97}
Ethan Bernstein and Umesh Vazirani.
\newblock Quantum complexity theory.
\newblock {\em SIAM journal on computing}, 26(5):1411--1473, 1997.

\bibitem[CGLQ20]{CGLQ20}
Kai-Min Chung, Siyao Guo, Qipeng Liu, and Luowen Qian.
\newblock Tight quantum time-space tradeoffs for function inversion.
\newblock In {\em Proceedings of the 61st Annual IEEE Symposium on Foundations
  of Computer Science}, pages 673--684, 2020.

\bibitem[CLQ20]{CLQ20}
Kai-Min Chung, Tai-Ning Liao, and Luowen Qian.
\newblock Lower bounds for function inversion with quantum advice.
\newblock In {\em Proceedings of the 1st Conference on Information-Theoretic
  Cryptography}, 2020.

\bibitem[DGLM23]{DGLM23}
Hugo Delavenne, Fran{\c{c}}ois~Le Gall, Yupan Liu, and Masayuki Miyamoto.
\newblock Quantum {M}erlin-{A}rthur proof systems for synthesizing quantum
  states.
\newblock Technical report, arXiv: 2303.01877, 2023.

\bibitem[GJMZ23]{GJMZ23}
Sam Gunn, Nathan Ju, Fermi Ma, and Mark Zhandry.
\newblock Commitments to quantum states.
\newblock In {\em Proceedings of the 55th Annual ACM Symposium on Theory of
  Computing}, pages 1579--1588, 2023.

\bibitem[GLSV21]{GLSV21}
Alex Grilo, Huijia Lin, Fang Song, and Vinod Vaikuntanathan.
\newblock Oblivious transfer is in {M}ini{QC}rypt.
\newblock In {\em Proceedings of the 40th Annual International Cryptology
  Conference}, pages 531--561, 2021.

\bibitem[HXY19]{HXY19}
Minki Hhan, Keita Xagawa, and Takashi Yamakawa.
\newblock Quantum random oracle model with auxiliary input.
\newblock In {\em Proceedings of the 25th Annual International Conference on
  the Theory and Application of Cryptology and Information Security}, pages
  584--614, 2019.

\bibitem[INN{\etalchar{+}}22]{INN+22}
Sandy Irani, Anand Natarajan, Chinmay Nirkhe, Sujit Rao, and Henry Yuen.
\newblock Quantum search-to-decision reductions and the state synthesis
  problem.
\newblock In {\em Proceedings of the 37th Annual IEEE Conference on
  Computational Complexity}, pages 1--19, 2022.

\bibitem[IR89]{IR89}
Russell Impagliazzo and Steven Rudich.
\newblock Limits on the provable consequences of one-way permutations.
\newblock In {\em Proceedings of the 19th Annual ACM Symposium on Theory of
  Computing}, pages 44--61, 1989.

\bibitem[JLS18]{JLS18}
Zhengfeng Ji, Yi-Kai Liu, and Fang Song.
\newblock Pseudorandom quantum states.
\newblock In {\em Proceedings of the 38th Annual International Cryptology
  Conference}, pages 126--152, 2018.

\bibitem[KQST23]{KQST23}
William Kretschmer, Luowen Qian, Makrand Sinha, and Avishay Tal.
\newblock Quantum cryptography in {A}lgorithmica.
\newblock In {\em Proceedings of the 55th Annual ACM Symposium on Theory of
  Computing}, 2023.

\bibitem[Kre21]{Kre21}
William Kretschmer.
\newblock Quantum pseudorandomness and classical complexity.
\newblock In {\em Proceedings of the 16th Conference on the Theory of Quantum
  Computation, Communication and Cryptography}, 2021.

\bibitem[Kre23]{Kre23}
William Kretschmer.
\newblock Does quantum cryptography imply classical lower bounds?
\newblock Talk at the Simons Institute, 2023.

\bibitem[Liu23]{Liu23}
Qipeng Liu.
\newblock Non-uniformity and quantum advice in the quantum random oracle model.
\newblock In {\em Proceedings of the 42nd Annual International Cryptology
  Conference}, pages 117--143, 2023.

\bibitem[LMR14]{LMR14}
Seth Lloyd, Masoud Mohseni, and Patrick Rebentrost.
\newblock Quantum principal component analysis.
\newblock {\em Nature Physics}, 10(9):631--633, 2014.

\bibitem[MY22]{MY22}
Tomoyuki Morimae and Takashi Yamakawa.
\newblock Quantum commitments and signatures without one-way functions.
\newblock In {\em Proceedings of the 42nd Annual International Cryptology
  Conference}, pages 269--295, 2022.

\bibitem[MY23]{MY23}
Tony Metger and Henry Yuen.
\newblock state{QIP}= state{PSPACE}.
\newblock In {\em Proceedings of the 64th Annual IEEE Symposium on Foundations
  of Computer Science}, 2023.

\bibitem[NC10]{NC10}
Michael Nielsen and Isaac Chuang.
\newblock {\em Quantum computation and quantum information}.
\newblock Cambridge university press, 2010.

\bibitem[Pet12]{Pet12}
Peter Petersen.
\newblock {\em Linear algebra}.
\newblock Springer, 2012.

\bibitem[Ros22]{Ros22}
Gregory Rosenthal.
\newblock Query and depth upper bounds for quantum unitaries via {G}rover
  search.
\newblock Technical report, arXiv:2111.07992, 2022.

\bibitem[Ros23a]{Ros23}
Gregory Rosenthal.
\newblock Efficient quantum state synthesis with one query.
\newblock Technical report, arXiv:2306.01723, 2023.

\bibitem[Ros23b]{Ros23b}
Gregory Rosenthal.
\newblock {\em Quantum State and Unitary Complexity}.
\newblock PhD thesis, University of Toronto, 2023.

\bibitem[RY22]{RY22}
Gregory Rosenthal and Henry Yuen.
\newblock Interactive proofs for synthesizing quantum states and unitaries.
\newblock In {\em Proceedings of the 13th Innovations in Theoretical Computer
  Science}, 2022.

\bibitem[Sha49]{Sha49}
Claude Shannon.
\newblock The synthesis of two-terminal switching circuits.
\newblock {\em The Bell System Technical Journal}, 28(1):59--98, 1949.

\bibitem[Tro12]{Tro12}
Joel Tropp.
\newblock User-friendly tail bounds for sums of random matrices.
\newblock {\em Foundations of computational mathematics}, 12:389--434, 2012.

\bibitem[Tro15]{Tro15}
Joel Tropp.
\newblock {\em An introduction to matrix concentration inequalities}, volume~8.
\newblock 2015.

\bibitem[Ver11]{Ver11}
Roman Vershynin.
\newblock A simple decoupling inequality in probability theory.
\newblock Found at
  \url{https://www.math.uci.edu/~rvershyn/papers/decoupling-simple.pdf}, 2011.

\bibitem[Ver18]{Ver18}
Roman Vershynin.
\newblock {\em High-dimensional probability: an introduction with applications
  in data science}.
\newblock Cambridge University Press, 2018.

\bibitem[vH17]{Van17}
Ramon van Handel.
\newblock Structured random matrices.
\newblock {\em Convexity and concentration}, pages 107--156, 2017.

\bibitem[Wat18]{Wat18}
John Watrous.
\newblock {\em The theory of quantum information}.
\newblock Cambridge University Press, 2018.

\bibitem[Yan22]{Yan22}
Jun Yan.
\newblock General properties of quantum bit commitments.
\newblock In {\em Proceedings of the 28th Annual International Conference on
  the Theory and Application of Cryptology and Information Security}, pages
  628--657, 2022.

\bibitem[Yue22a]{Yue22}
Henry Yuen.
\newblock Lecture {6} from {COMS E6998}:\ {F}rontiers of quantum complexity and
  cryptography.
\newblock Found at
  \url{https://www.henryyuen.net/spring2022/lec6-statesynthesis.pdf} and
  \url{https://www.henryyuen.net/spring2022/lec6-unitarysynthesis.pdf}, 2022.

\bibitem[Yue22b]{Yue22b}
Henry Yuen.
\newblock Lecture {7} from {COMS E6998}:\ {F}rontiers of quantum complexity and
  cryptography.
\newblock Found at
  \url{https://www.henryyuen.net/spring2022/lec7-quantumprograms.pdf}, 2022.

\end{thebibliography}

\appendix

\section{A matrix Chernoff proof of the one-query lower bound}\label{sec:appendix-matrix-chernoff}

In this section, we will give an alternative proof of the one-query lower bound using a variant of the matrix Chernoff bound known as the matrix Hoeffding bound.
Although this proof strategy ultimately results in worse bounds than the proof presented in \Cref{sec:lower-bound}, we have included it because we believe its techniques will be more familiar to a computer science audience.
This section is organized as follows: in \Cref{sec:spectral-relaxation} we will develop a natural spectral relaxation of the distinguishing advantage $\distinguish_{A}(\prs)$.
Next, we will perform a slight modification to this spectral relaxation in \Cref{sec:truncate} to handle function families $\prs$ which produce outlier values.
Finally, in \Cref{sec:finally-the-proof} we will use these ingredients to complete the proof of the one-query lower bound.

We will largely follow the same notation as the proof in \Cref{sec:lower-bound}, which can be found in \cref{sec:notation} as well as in \Cref{sec:decouple-relax}. For convenience, we repeat several important definitions and results here.

\begin{definition}[Isometry weights, \Cref{def:isometry-weights} restated]\label{def:isometry-weights-restated}
The \emph{isometry weights} are the numbers
\begin{equation*}
    \weights_{V, i} \coloneqq \tfrac{1}{N} \cdot \braket{v_i},
\end{equation*}
for $1 \leq i \leq \qdim$.
Note that these
sum to one and therefore form a probability distribution.
\end{definition}

\begin{proposition}[Typical amplitudes, \Cref{prop:typical-magnitude} restated]\label{prop:typical-magnitude-restated}
Let $\bh: [N]\rightarrow \{\pm 1\}$
be a uniformly random Boolean function.
Then
\begin{equation*}
    \E_{\bh} |\braket{v_i}{\psi_{\bh}}|^2 = \weights_{V, i}.
\end{equation*}
\end{proposition}

\begin{definition}[The weight vector, \Cref{def:weight-vector} restated]
    The \emph{weight vector}
    is the unit vector given by
    \begin{equation*}
        \ket{\wt_V} = \sum_{i=1}^{\qdim} \sqrt{\weights_{V, i}} \ket{i}.
    \end{equation*}
\end{definition}

\begin{definition}[The rescaling matrix, \Cref{def:rescaling-matrix} restated]
Let $h:[N] \rightarrow \{\pm 1\}$ be a Boolean function.
Then the corresponding \emph{rescaling matrix} is the diagonal matrix given by
\begin{equation*}
    \rescale_{V, h} = \sum_{i=1}^{\qdim} D_{V, h, i} \cdot \ketbra{i},
    \text{ where }
    \rescale_{V, h,i} = \frac{\braket{v_i}{\psi_h}}{\sqrt{\weights_{V, i}}}.
\end{equation*}
By construction, we have that
$V \cdot \ket{\psi_h} = \rescale_{V, h} \cdot \ket{\wt_V}$.
\end{definition}

\begin{theorem}[Sub-Gaussian concentration for sums of complex random variables, \Cref{thm:complex-concentration} restated]\label{thm:complex-concentration-restated}
Let $\bb_1, \ldots, \bb_m$ be independent and uniform $\pm 1$ random variables,
and let $a_1, \ldots, a_m$ be complex numbers.
Then $\bS = a_1 \cdot \bb_1 + \cdots + a_m \cdot \bb_m$ satisfies
\begin{equation*}
\Pr[|\bS| \geq t] \leq 2\cdot \mathrm{exp} \left(-\frac{t^2}{2\cdot \sum_{i =1}^m |a_i|^2}\right).
\end{equation*}
\end{theorem}

Now we move to the proof.

\subsection{A spectral relaxation for the distinguishing advantage}\label{sec:spectral-relaxation}

Let $h:[N]\rightarrow\{\pm 1\}$ be a Boolean function,
and consider the adversary's execution on the binary phase state
$\ket{\psi_{h}}$.
First, it applies the isometry, resulting in the state
\begin{equation*}
    V \cdot \ket{\psi_h}
    =  D_{V, h} \cdot \ket{\wt_V}.
\end{equation*}
Next, it applies an oracle $\calO_f$.
This will produce the state
\begin{equation*}
    \calO_f \cdot D_{V, h} \cdot \ket{\wt_V}.
\end{equation*}
Now we observe that $\calO_f$ and $\rescale_{V, h}$ are both diagonal matrices, and hence they both commute.
As a result,
\begin{equation*}
    \calO_f \cdot \rescale_{V, h} \cdot \ket{\wt_V}
    = \rescale_{V, h} \cdot \calO_f \cdot \ket{\wt_V}.
\end{equation*}
Note that $\calO_f \cdot \ket{\wt_V}$ is always a unit vector,
and that it is independent of $h$.

Finally, the adversary performs the measurement $\{\meas, I-\meas\}$ and accepts if it observes the first outcome. We can therefore calculate the acceptance probability of the adversary $A$ with oracle access to a function $f: [\qdim]\rightarrow \{\pm 1\}$ as:
\begin{align*}
        \accept_{A}(h \mid f) &= \bra{\psi_h} V^\dagger \cdot \oracle_f \cdot \meas \cdot \oracle_f \cdot V \ket{\psi_h}\\
        &=\bra{\wt_{V}}  \oracle_f \cdot \rescale_{V, h}^\dagger \cdot \meas \cdot \rescale_{V, h}  \cdot \oracle_f  \ket{\wt_{V}}.
\end{align*}

Let $\prs:[K]\times [N] \rightarrow \{\pm 1\}$ be a function family. By the above calculation and the definition of the distinguishing advantage $\distinguish_A(\prs)$ allows us to conclude that

\begin{align*}
    \distinguish_{A}(\prs)
    &= \underset{f: [\qdim]\rightarrow \{\pm 1\}}{\max}\Big|\E_{\bk \sim [K]} [\accept_{A}(\prs_{\bk} \mid f)]
    - \E_{\bh} [\accept_{A}(\bh \mid f)] \Big|\\
    &= \max_f \Big|\E_{\bk \sim [K]} \bra{\wt_{V}} \oracle_f  \cdot \rescale_{V, \prs_{\bk}}^\dagger \cdot \meas \cdot \rescale_{V,\prs_{\bk}}  \cdot \oracle_f  \ket{\wt_{V}}
    - \E_{\bh} \bra{\wt_{V}}  \oracle_f \cdot \rescale_{V, \bh}^\dagger \cdot \meas \cdot \rescale_{V, \bh}  \cdot \oracle_f  \ket{\wt_{V}}\Big|\\
    &= \max_f \Big|\bra{\wt_{V}} \oracle_f \cdot \Big(\E_{\bk \sim [K]}  \rescale_{V, \prs_{\bk}}^\dagger \cdot \meas \cdot R_{V, \prs_{\bk}}
    - \E_{\bh}  \rescale_{V, \bh}^\dagger \cdot \meas \cdot \rescale_{V, \bh}\Big) \cdot \oracle_f \ket{\wt_{V}}\Big|.
\end{align*}
Recall that $\oracle_f \cdot \ket{\wt_{V}}$ is a unit vector
which depends on $V$ and $f$.
We can therefore upper-bound this expression
by relaxing $\oracle_f \cdot \ket{\wt_{V}}$ to be an arbitrary unit vector
maximizing this expression.
This gives the spectral relaxation.

\begin{definition}[Spectral relaxation]
Let $\prs:[K]\times [N] \rightarrow \{\pm 1\}$ be a function family.
The \emph{spectral relaxation} of the distinguishing probability on $\prs$ is given by
\begin{equation*}
    \Delta_{A}^{\spectral}(\prs) = \Big\Vert \E_{\bk \sim [K]}  \rescale_{V, \prs_{\bk}}^\dagger \cdot \meas \cdot \rescale_{V, \prs_{\bk}}
    - \E_{\bh}  \rescale_{V, \bh}^\dagger \cdot \meas \cdot \rescale_{V, \bh}\Big\Vert_{\mathrm{op}}.
\end{equation*}
\end{definition}

From the above discussion, the following lemma is immediate.

\begin{lemma}
Let $\prs:[K]\times [N] \rightarrow \{\pm 1\}$ be a function family.
Then $\Delta_{A}(\prs) \leq \Delta_{A}^{\spectral}(\prs)$.
\end{lemma}

In the worst case, this relaxation can be quite poor. The following is an example in which the relaxation is equal to $\sqrt{N}-1$,
even though the distinguishing value $\Delta_{A}(\prs)$ can never be more than one.
\begin{example}[A large relaxation value]\label{ex:large-relaxation}
For this example,
we will view the space $\C^N$
as corresponding to $n$ qubits,
so that the standard basis contains the vector
$\ket{x}$ for each $x \in \{0, 1\}^n$.
With this viewpoint, a binary phase state is specified by a Boolean function $h:\{0, 1\}^n \rightarrow \{\pm 1\}$ and is given by
\begin{equation*}
    \ket{\psi_h} = \frac{1}{\sqrt{N}} \sum_{x \in \{0, 1\}^n} h(x) \ket{x}.
\end{equation*}
Suppose the adversary's strategy does not expand the Hilbert space (so that $\qdim = N$).
In addition, suppose that the isometry $V$ is just the $n$-qubit Hadamard transform $V = H^{\otimes n}$,
and that the measurement $\meas$ is just the $n$-qubit identity matrix $\meas = I_{N \times N}$.
In this case, the rows of $V$ are just the binary phase states $\ket{\psi_{\chi_\alpha}}$, where $\chi_\alpha:\{0, 1\}^n \rightarrow \{\pm 1\}$ is the Boolean function $\chi_\alpha(x) = (-1)^{\langle \alpha, x\rangle}$;
in other words,
\begin{equation*}
    V = \sum_{\alpha \in \{0, 1\}^n} \ketbra{\alpha}{\psi_{\chi_{\alpha}}}.
\end{equation*}
As a result, the weight $\weights_{V, \alpha} = 1/N$ for all $\alpha \in \{0, 1\}^n$, and so the rescaling matrix is given by
\begin{equation*}
    R_{V, h} = \sum_{\alpha \in \{0, 1\}^n} \sqrt{N} \cdot \braket{\psi_{\chi_{\alpha}}}{\psi_h} \cdot \ketbra{\alpha}.
\end{equation*}
Now we compute the two terms in the spectral relaxation.
The second term is independent of the function family $\prs$, and so we compute it first:
\begin{equation*}
    \E_{\bh}  \rescale_{V, \bh}^\dagger \cdot \meas \cdot \rescale_{V, \bh}
    = \E_{\bh}  \rescale_{V, \bh}^\dagger \cdot \rescale_{V, \bh}
    = \E_{\bh}  \sum_{\alpha \in \{0, 1\}^n} N \cdot |\braket{\psi_{\chi_{\alpha}}}{\psi_{\bh}}|^2 \cdot \ketbra{\alpha}
    = I_{N \times N},
\end{equation*}
where the last equality used the fact that 
$\E_{\bh} \braket{\psi_{\chi_{\alpha}}}{\psi_{\bh}}|^2 = \weights_{V, \alpha} = N$ due to \Cref{prop:typical-magnitude-restated}.
As for the first term, consider a worst-case function family $\prs$ in which every $\prs_k$ is equal to the same parity $\prs_k = \chi_\alpha$, for some $\alpha \in \{0, 1\}^n$.
Then for every $1 \leq k \leq K$,
the rescaling matrix is given by
$\rescale_{V, \prs_k} = \sqrt{N} \cdot \ketbra{\alpha}{\alpha}$.
As a result,
\begin{equation*}
    \E_{\bk \sim [K]}  \rescale_{V, \prs_{\bk}}^\dagger \cdot \meas \cdot \rescale_{V, \prs_{\bk}}
    - \E_{\bh}  \rescale_{V, \bh}^\dagger \cdot \meas \cdot \rescale_{V, \bh} = \sqrt{N} \cdot \ketbra{\alpha}{\alpha} - I_{N \times N}.
\end{equation*}
The operator norm of this matrix is $\sqrt{N}-1$,
and so $\Delta_{A}^{\spectral}(\prs) = \sqrt{N}-1$.
\end{example}

The reason that this example has such a large relaxation value
is that the rescaling matrices $\rescale_{V, \prs_k}$ all have an extremely large diagonal entry and therefore an extremely large operator norm.
We would like to rule out examples like this
by only considering function families $\prs$ in which the rescaling matrices have operator norms which are not too much larger than 1. This motivates the following definition.

\begin{definition}[$B$-bounded function families]
    A function $h:[N]\rightarrow \{\pm 1\}$
    is \emph{$B$-bounded} if $|\rescale_{V, h, i}| \leq B$
    for all $1 \leq i \leq \qdim$.
    In addition, 
    a function family $\prs:[K]\times [N]\rightarrow \{\pm 1\}$
    is \emph{$B$-bounded} if $\prs_k$ is $B$-bounded for all $1 \leq k \leq K$.
\end{definition}

\Cref{ex:large-relaxation}
showed that there exist worst-case function families $\prs$ which are not $B$-bounded for small values of $B$.
However, the next lemma shows that an average-case function family will in fact be $B$-bounded with extremely high probability.

\begin{lemma}[Random function families are bounded]\label{lem:bounded-prs}
    Let $\bprs : [K]\times[N] \rightarrow \{\pm 1\}$
    be a uniformly random function family. Then
    \begin{equation*}
        \Pr_{\bprs}[\text{$\bprs$ is not $B$-bounded}] \leq 4 K\qdim \cdot e^{-B^2/4}.
    \end{equation*}
\end{lemma}

This lemma follows as a simple corollary of the following lemma
by applying it to each function $\bprs_k$ separately
and then union bounding over all $1 \leq k \leq K$.

\begin{lemma}[Random functions are bounded]\label{lem:bounded-function}
    Let $\bh : [N] \rightarrow \{\pm 1\}$
    be a uniformly random Boolean function. Then
    \begin{equation*}
        \Pr_{\bh}[\text{$\bh$ is not $B$-bounded}] \leq 2 \qdim \cdot e^{-B^2/2}.
    \end{equation*}
\end{lemma}

The key technical tool in the proof of this lemma is \Cref{thm:complex-concentration-restated}

\begin{proof}[Proof of \Cref{lem:bounded-prs}]
Fix a $1 \leq i \leq \qdim$, and let us consider $\rescale_{V, \bh, i}$. By definition,
\begin{equation*}
\rescale_{V, \bh, i}
= \frac{1}{\sqrt{\weights_{V, i}}} \cdot \braket{v_i}{\psi_{\bh}}
= \frac{1}{\sqrt{\weights_{V, i}}} \cdot \sum_{x=1}^N v_{i, x} \cdot \tfrac{1}{\sqrt{N}} \bh(x)
= \sum_{x=1}^N a_x \cdot \bh(x).
\end{equation*}
where
\begin{equation*}
    a_x = \frac{1}{\sqrt{\weights_{V, i} \cdot N}} \cdot  v_{i, x}.
\end{equation*}
Note that
\begin{equation*}
    \sum_{x=1}^N |a_x|^2
    = \sum_{x=1}^N \frac{1}{\weights_{v,i} \cdot N} \cdot |v_{i, x}|^2
    = \frac{1}{\weights_{v,i} \cdot N}\cdot \braket{v_i}{v_i}
    = \frac{1}{\weights_{v,i}} \cdot \weights_{v, i} = 1.
\end{equation*}
As a result, \Cref{thm:complex-concentration-restated}  says that
\begin{equation*}
    \Pr[|\rescale_{V, \bh, i}| \geq B] \leq 2\cdot \mathrm{exp} \left(-\frac{B^2}{2\cdot \sum_{x=1}^N |a_x|^2}\right)
    = 2\cdot \mathrm{exp} \left(-\frac{B^2}{2}\right).
\end{equation*}
Union bounding over all $1 \leq i \leq \qdim$, we have that
\begin{equation*}
    \Pr_{\bh}[\text{$\bh$ is not $B$-bounded}]
    \leq \sum_{i=1}^{\qdim} \Pr[|\rescale_{V, \bh, i}| \geq B]
    \leq \sum_{i=1}^{\qdim} 2\cdot e^{-B^2/2}
    = 2 \qdim \cdot e^{-B^2/2}.
\end{equation*}
This completes the proof.
\end{proof}

\subsection{Truncating the spectral relaxation}\label{sec:truncate}

Although \Cref{lem:bounded-prs} shows that the overwhelming majority of function families $\prs$ are $B$-bounded,
it will still be convenient to modify the spectral relaxation
slightly so that the rare bad events do not lead to extremely large values,
as in \Cref{ex:large-relaxation}.
We will handle this by truncation.
\begin{definition}[$B$-truncation]
Let $\trunc_B: \C \rightarrow \C$ be the function which, on input $t \in \C$, acts as follows:
\begin{equation*}
\trunc_B(t)
= \left\{
\begin{array}{cl}
t & \text{if $|t| \leq B$}\\
B \cdot t/|t| & \text{if $|t| > B$.}
\end{array}\right.
\end{equation*}
\end{definition}
By design, $|\trunc_B(t)| \leq B$ for all $t \in \C$.
Now we use this to define a truncated version of the rescaling matrix.

\begin{definition}[$B$-truncated rescaling matrix]
Let $h:[N] \rightarrow \{\pm 1\}$ be a Boolean function.
Then the \emph{$B$-truncated rescaling matrix} is the diagonal matrix given by
\begin{equation*}
    \rescale_{V, h, B} = \sum_{i=1}^{\qdim} \trunc_B(D_{V, h, i}) \cdot \ketbra{i}.
\end{equation*}
\end{definition}

With this in hand, we can define truncated analogues
of the distinguishing advantage and the spectral relaxation.

\begin{definition}[$B$-truncated advantage and spectral relaxation]
Let $\prs:[K]\times [N] \rightarrow \{\pm 1\}$ be a function family.
Let $h:[N]\rightarrow \{\pm 1\}$ be a Boolean function.
Given an oracle $\calO$, the \emph{$B$-truncated acceptance probability} is defined as
\begin{equation*}
    \accept_{A, B}(h \mid f) = \bra{\wt_{V}}\cdot \oracle_f \cdot\rescale_{V, h, B}^\dagger \cdot \meas \cdot \rescale_{V, h, B} \cdot \oracle_f \cdot \ket{\wt_{V}}
\end{equation*}
In addition, the \emph{$B$-truncated distinguishing advantage} and \emph{$B$-truncated spectral relaxation} are defined as follows.
\begin{align*}
\distinguish_{A, B}(\prs)&= \max_f 
\Big|\E_{\bk \sim [K]} \accept_{A, B}(\prs_{\bk} \mid f) - \E_{\bh} \accept_{A, B}(\bh \mid f)\Big|\\
&=\max_f \Big|\bra{\wt_{V}}\cdot \oracle_f \cdot \Big(\E_{\bk \sim [K]}  \rescale_{V, \prs_{\bk}, B}^\dagger \cdot \meas \cdot \rescale_{V, \prs_{\bk}, B}
    - \E_{\bh}  \rescale_{V, \bh, B}^\dagger \cdot \meas \cdot \rescale_{V, \bh, B}\Big) \cdot \oracle_f \cdot \ket{\wt_{V}}\Big|,\\
\distinguish_{A, B}^{\spectral}(\prs) &= \Big\Vert \E_{\bk \sim [K]}  \rescale_{V, \prs_{\bk}, B}^\dagger \cdot \meas \cdot \rescale_{V, \prs_{\bk}, B}
    - \E_{\bh}  \rescale_{V, \bh, B}^\dagger \cdot \meas \cdot \rescale_{V, \bh, B}\Big\Vert_{\mathrm{op}}.
\end{align*}
Note that the $B$-truncated spectral relaxation remains a spectral relaxation of the $B$-truncated distinguishing advantage, in that $\distinguish_{A, B}(\prs) \leq \distinguish_{A, B}^{\spectral}(\prs)$.
\end{definition}

As we have seen,
a random function family is $B$-bounded with overwhelming probability.
This suggests that the $B$-truncated analogue of the distinguishing advantage should not be too far from the regular distinguishing advantage.
This is shown in the next lemma.

\begin{lemma}[Truncating doesn't change the distinguishing advantage by much]\label{lem:no-change-from-trunc}
Let $\prs:[K] \times [N] \rightarrow \{\pm 1\}$
be a $B$-bounded function family. Then
\begin{equation*}
    \Delta_{A}(\prs) \leq \distinguish_{A, B}(\prs) + 4\qdim \cdot e^{-B^2/2}.
\end{equation*}
\end{lemma}

Before proving this, we will need to establish the following technical lemma.

\begin{lemma}\label{lem:bounded-probabilities}
For any function family $\prs:[K]\times[N] \rightarrow \{\pm 1\}$, any function $h:[N]\rightarrow \{\pm 1\}$, and any function $f$,
\begin{equation*}
    |\accept_{A}(h \mid f)
     - \accept_{A, B}(h \mid f)| \leq 2.
\end{equation*}
\end{lemma}
\begin{proof}
    The first step of the proof is a simple triangle inequality:
    \begin{align*}
        |\accept_{A}(h \mid f)
     - \accept_{A, B}(h \mid f)| 
     &\leq |\accept_{A}(h \mid f)|
     + |\accept_{A, B}(h \mid f)| \\
     &\leq 1
     + |\accept_{A, B}(h \mid f)| \\
     &= 1 + |\bra{\wt_{V}}\cdot \oracle_f\cdot \rescale_{V, h,B}^\dagger \cdot \meas \cdot \rescale_{V, h,B} \cdot \oracle_f \cdot \ket{\wt_{V}}
    |,
    \end{align*}
    where the second inequality is because $\accept_{A}(h \mid f)$ is an acceptance probability and therefore at most 1.
    As for the second term, we note that
    \begin{equation*}
        \rescale_{V, h, B} \cdot \oracle_f \cdot \ket{\wt_{V}}
        = \oracle_f \cdot \rescale_{V, h, B} \cdot \ket{\wt_{V}}
    \end{equation*} 
    because $\rescale_{V, h, B}$ and $\oracle_f$ are diagonal matrices. Expanding $\rescale_{V, h, B} \cdot \ket{\wt_{V}}$,
\begin{align*}
\rescale_{V, h, B} \cdot \ket{\wt_{V}}
& = \rescale_{V, h, B} \cdot \Big(\sum_{i=1}^{\qdim} \sqrt{\weights_{V, i}} \ket{i}\Big)\\
& = \sum_{i=1}^{\qdim} \trunc_B(D_{V, h, i}) \cdot \sqrt{\weights_{V, i}} \ket{i}\\
& = \sum_{i=1}^{\qdim} \trunc_{B \cdot \sqrt{\weights_{V, i}}}(D_{V, h, i} \cdot \sqrt{\weights_{V, i}}) \cdot \ket{i}\\
& = \sum_{i=1}^{\qdim} \trunc_{B \cdot \sqrt{\weights_{V, i}}}(\braket{v_i}{\psi_h}) \cdot \ket{i}.
\end{align*}
We note that
\begin{equation*}
    |\trunc_{B \cdot \sqrt{\weights_{V, i}}}(\braket{v_i}{\psi_h})|
    \leq |\braket{v_i}{\psi_h}|,
\end{equation*}
which is the amplitude on $\ket{i}$ in the state $V \cdot \ket{\psi_h}$ due to \Cref{eq:rewriting-with-weight-vector}.
As a result, $\rescale_{V, h, B} \cdot \ket{\wt_{V}}$ is a subnormalized vector, and therefore so is $\oracle_f \cdot \rescale_{V, h, B} \cdot \ket{\wt_{V}}$ because $\oracle_f$ is a unitary matrix.
Putting everything together, this tells us that
\begin{equation*}
    |\bra{\wt_{V}}\cdot \oracle_f \cdot \rescale_{V, h,B}^\dagger \cdot \meas \cdot \rescale_{V, h,B} \cdot \oracle_f \cdot \ket{\wt_{V}}
    |\leq 1,
\end{equation*}
because $0 \preceq \meas \preceq\Id$.
Thus, the sum of the two terms is at most $2$.
\end{proof}

Now we use this to prove \Cref{lem:no-change-from-trunc}.

\begin{proof}[Proof of \Cref{lem:no-change-from-trunc}]
By definition,
\begin{align*}
\distinguish_{A}(\prs)
    &= \max_f \Big|\E_{\bk \sim [K]} \accept_{A}(\prs_{\bk} \mid f)
    - \E_{\bh} \accept_{A}(\bh \mid f)\Big|\\
    &= \max_f \Big|\E_{\bk \sim [K]} \accept_{A, B}(\prs_{\bk} \mid f)
    - \E_{\bh} \accept_{A}(\bh \mid f)\Big|,
\end{align*}
where the second equality holds because $\prs$ is $B$-bounded, so $\accept_{A}(\prs_{\bk} \mid f) = \accept_{A, B}(\prs_{\bk} \mid f)$ for every $1 \leq k \leq K$.
By the triangle inequality, this can be upper-bounded by
\begin{align*}
    &\max_f \Big|\E_{\bk \sim [K]} \accept_{A, B}(\prs_{\bk} \mid f)
    - \E_{\bh} \accept_{A, B}(\bh \mid f)\Big|
    + \max_f \Big| \E_{\bh} \accept_{A, B}(\bh \mid f) - \E_{\bh} \accept_{A}(\bh \mid f)\Big|\\
    ={}& \distinguish_{A, B}(\prs)
    + \max_f \Big| \E_{\bh} \accept_{A, B}(\bh \mid f) - \E_{\bh} \accept_{A}(\bh \mid f)\Big|\\
    \leq{}& \distinguish_{A, B}(\prs)
    + \max_f \E_{\bh}\Big| \accept_{A, B}(\bh \mid f) -  \accept_{A}(\bh \mid f)\Big|.
\end{align*}
Let us first focus on the second term. 
If $\bh$ is $B$-bounded, then $\accept_{A, B}(\bh \mid f) =  \accept_{A}(\bh \mid f)$,
and the term inside the expectation is zero.
Otherwise the term inside the expectation is at most 2, by \Cref{lem:bounded-probabilities}.
As a result, for every function $f$, the expectation is at most
\begin{equation*}
    2 \cdot \Pr_{\bh}[\text{$\bh$ is not $B$-bounded}] \leq 4 \qdim \cdot e^{-B^2/2},
\end{equation*}
by \Cref{lem:bounded-function}.
Putting everything together,
\begin{equation*}
    \distinguish_{A}(\prs)
    \leq \distinguish_{A, B}(\prs) + 4 \qdim \cdot e^{-B^2/2}.
\end{equation*}
This completes the proof.
\end{proof}

\subsection{The one-query lower bound}\label{sec:finally-the-proof}

Now we complete the proof of the one-query lower bound.
The quantitative bound we prove is as follows.

\begin{theorem}\label{thm:main-bound}
Let $\bprs:[K] \times [N] \rightarrow \{\pm 1\}$
be a uniformly random function family. Then
\begin{equation*}
    \Pr_{\bprs}[\Delta_{A}(\bprs) \geq 1/K^{1/4} + 4\qdim \cdot e^{-K^{1/8}/2}]
        \leq 6K\qdim \cdot e^{-K^{1/8}/32}.
\end{equation*}
\end{theorem}

We note that although this bound is quantitatively weaker than \Cref{thm:main-at-end-of-proof}, it still gives a strong lower bound.
For our typical settings of $K$ and $\qdim$, it states that $\Delta_{A}(\bprs)$ is roughly bounded by $1/K^{1/4}$ with all but a negligible probability.
The key technical result we use to prove this is the following variant of the matrix Chernoff bound, stated in \cite[Theorem 1.3]{Tro12}.

\begin{theorem}[Matrix Hoeffding]\label{thm:matrix-hoeffding}
Let $\{\bZ_k\}$ be a set of independent, Hermitian, random matrices with dimension $D$.
Let $\{C_k\}$ be a set of fixed Hermitian matrices.
Assume that for all $k$,
$\E\bZ_k = 0$ and $\bZ_k^2 \preceq C_k^2$.
Set
\begin{equation*}
\sigma^2 = \Big \Vert \sum_k C_k^2 \Big \Vert.
\end{equation*}
Then
\begin{equation*}
\Pr\Big[\lambda_{\max} \Big(\sum_k \bZ_k\Big) \geq t\Big] \leq D\cdot e^{-t^2/8\sigma^2}.
\end{equation*}
\end{theorem}

By applying matrix Hoeffding
to both $(\sum_k \bZ_k)$ and $-(\sum_k \bZ_k)$,
we can derive the following concentration bound
for the operator norm.

\begin{corollary}[Matrix Hoeffding for operator norm]\label{cor:hoeffding-operator-norm}
    Under the same assumptions as \Cref{thm:matrix-hoeffding},
    we have that
\begin{equation*}
\Pr\Big[\Big\Vert \sum_k \bZ_k\Big\Vert_{\mathrm{op}} \geq t\Big] \leq 2D\cdot e^{-t^2/8\sigma^2}.
\end{equation*}
\end{corollary}

\begin{proof}[Proof of \Cref{thm:main-bound}]
    Let $B \geq 0$ be a nonnegative real number to be determined later.
    By \Cref{lem:no-change-from-trunc}, we have that
    \begin{equation*}
    \Delta_{A}(\prs) \leq \distinguish_{A, B}(\prs) + 4\qdim \cdot e^{-B^2/2}.
    \end{equation*}
    for any $B$-bounded function family $\prs$.
    This means that if $\Delta_{A}(\prs) \geq \epsilon + 4\qdim \cdot e^{-B^2/2}$ for some number $\epsilon$ to be determined later, it must either be the case that $\distinguish_{A, B}(\prs) \geq \epsilon$
    or that $\prs$ is not $B$-bounded.
    Hence, by the union bound,
    if $\bprs:[K] \times [N] \rightarrow \{\pm 1\}$ is a random function family,
    \begin{align*}
        \Pr_{\bprs}[\Delta_{A}(\bprs) \geq \epsilon + 4\qdim \cdot e^{-B^2/2}]
         &\leq \Pr_{\bprs}[\distinguish_{A, B}(\bprs) \geq \epsilon]
        + \Pr_{\bprs}[\text{$\bprs$ is not $B$-bounded}]\\
        &\leq \Pr_{\bprs}[\Delta^{\spectral}_{A, B}(\bprs) \geq \epsilon]
        + 4 K\qdim \cdot e^{-B^2/2},
    \end{align*}
    where the second inequality is due to \Cref{lem:bounded-prs}
    and the fact that $\distinguish_{A, B}(\bprs) \leq \distinguish_{A, B}^{\spectral}(\bprs)$.
    We will now focus on bounding the first term.
    By definition of the $B$-bounded spectral relaxation,
    \begin{equation*}
    \distinguish_{A, B}^{\spectral}(\bprs)
    = \Big\Vert\E_{\bk \sim [K]}  \rescale_{V, \bprs_{\bk},B}^\dagger \cdot \meas \cdot \rescale_{V, \bprs_{\bk},B}
    - \E_{\bh}  \rescale_{V, \bh,B}^\dagger \cdot \meas \cdot \rescale_{V, \bh,B}\Big\Vert_{\mathrm{op}}.
    \end{equation*}
    To analyze this, we note that for each $1 \leq k \leq K$, $\bprs_k$ is distributed as a uniformly random function,
and so $\bprs_k$ has the same distribution as $\bh$. Hence, if we keep $k$ fixed and randomize over $\bprs$,
\begin{equation*}
\E_{\bprs}
\rescale_{V,\bprs_{k},B}^\dagger \cdot \meas \cdot \rescale_{V,\bprs_{k},B}
 =\E_{\bh}\rescale_{V, \bh,B}^\dagger \cdot \meas \cdot \rescale_{V, \bh,B}.
\end{equation*}
This means that the random matrix
\begin{equation*}
X_{\bprs_k} \coloneqq 
\rescale_{V,\bprs_{k},B}^\dagger \cdot \meas \cdot \rescale_{V, \bprs_{k},B}
- \E_{\bh}\rescale_{V, \bh,B}^\dagger \cdot \meas \cdot \rescale_{V, \bh,B}
\end{equation*}
has $\E_{\bprs} [X_{\bprs_k}] = 0$.
In terms of these matrices, our goal is to bound
\begin{equation*}
\Vert \E_{\bk \sim [K]}\rescale_{V,\bprs_{\bk},B}^\dagger \cdot \meas \cdot \rescale_{V, \bprs_{\bk},B}
- \E_{\bh}\rescale_{V, \bh,B}^\dagger \cdot \meas \cdot \rescale_{V, \bh,B} \Vert
= \Vert \E_{\bk} X_{\bprs_{\bk}} \Vert.
\end{equation*}
Note the following properties of $X_{\bprs_k}$:
\begin{enumerate}
\item For each $k$, $X_{\bprs_k}$ only depends on $\bprs_k$. Hence, the random variables $X_{\bprs_k}$, over all $1 \leq k \leq K$, are independent and identically distributed.
\item $X_{\bprs_k}$ is an $\qdim \times \qdim$ matrix.
\item To bound the operator norm of $X_{\bprs_k}$, we begin with the bound
	\begin{align*}
	\Vert X_{\bprs_k} \Vert &= \Vert \rescale_{V,\bprs_{k},B}^\dagger \cdot \meas \cdot \rescale_{V, \bprs_{k},B}
- \E_{\bh}\rescale_{V, \bh,B}^\dagger \cdot \meas \cdot \rescale_{V, \bh,B} \Vert\\
	&\leq \Vert \rescale_{V,\bprs_{k},B}^\dagger \cdot \meas \cdot \rescale_{V, \bprs_{k},B}\Vert + \E_{\bh} \Vert\rescale_{V, \bh,B}^\dagger \cdot \meas \cdot \rescale_{V, \bh,B} \Vert.
	\end{align*}
	Now, let us bound the operator norm of $\rescale_{V,\bprs_{k},B}^\dagger \cdot \meas \cdot \rescale_{V, \bprs_{k},B}$.
	Let $\ket{v}$ be any unit vector. Then because $\rescale_{V, \bprs_{k},B}$ is a diagonal matrix whose diagonal entries have magnitude at most~$B$,
	$\rescale_{V, \bprs_{k},B}\cdot \ket{v}$ has norm at most $B$. Hence,
	\begin{equation*}
	\bra{v} \cdot \rescale_{V,\bprs_{k},B}^\dagger \cdot \meas \cdot \rescale_{V, \bprs_{k},B} \cdot \ket{v} \leq B^2.
	\end{equation*}
	As a result, $\rescale_{V,\bprs_{k},B}^\dagger \cdot \meas \cdot \rescale_{V, \bprs_{k},B}$ has spectral norm at most $B^2$;
	a similar argument will show that $\rescale_{V, \bh,B}^\dagger \cdot \meas \cdot \rescale_{V, \bh,B}$ has spectral norm at most $B^2$ as well.
	Thus, the spectral norm of $X_{\bprs_k}$ is at most $2B^2$,
	and so,  $X_{\bprs_k}^2 \leq 4B^4\cdot \Id$ always.
\end{enumerate}

Now we are in a good place to apply matrix Hoeffding.
In our setting, the matrices $X_{\bprs_1}, \ldots, X_{\bprs_K}$ are independent, Hermitian, and have dimension $\qdim$.
Furthermore, we know that $X_{\bprs_k}^2 \leq 4B^4\cdot \Id$ always.
Hence, our value of $\sigma^2$ is
\begin{equation*}
\sigma^2 = \Big \Vert \sum_{k=1}^K 4B^4\Id\Big \Vert = 4KB^4.
\end{equation*}
Now, our goal is to bound
\begin{align*}
\Pr_{\bprs}[\distinguish_{A,B}^{\spectral}(\bprs) \geq \epsilon]
&=\Pr_{\bprs}\Big[\Big\Vert \E_{\bk \in [K]} X_{\bprs_{\bk}} \Big\Vert_{\mathrm{op}} \geq \eps\Big]\\
&= \Pr_{\bprs} \Big[\Big\Vert \frac{1}{K} \sum_{k=1}^K X_{\bprs_k} \Big\Vert_{\mathrm{op}} \geq \eps\Big]
= \Pr_{\bprs} \Big[\Big\Vert  \sum_{k=1}^K X_{\bprs_k} \Big\Vert_{\mathrm{op}} \geq \eps K\Big].
\end{align*}
This we can apply \Cref{cor:hoeffding-operator-norm} to, which tells us that
\begin{equation*}
\Pr_{\bprs}[\distinguish_{A,B}^{\spectral}(\bprs) \geq \epsilon]
\leq 2 \qdim \cdot e^{- \eps^2 K^2 / 8 (4 K B^4)}
= 2 \qdim \cdot e^{-\eps^2 K / (32B^4)}.
\end{equation*}
Putting everything together, we have
\begin{equation*}
    \Pr_{\bprs}[\Delta_{A}(\bprs) \geq \epsilon + 4\qdim \cdot e^{-B^2/2}]
        \leq 2 \qdim \cdot e^{-\eps^2 K / (32B^4)}
        + 4 K\qdim \cdot e^{-B^2/2}.
\end{equation*}
Now we select our constants to be $\epsilon = 1/K^{1/4}$ and $B = K^{1/16}$.
Then this states that
\begin{equation*}
    \Pr_{\bprs}[\Delta_{A}(\bprs) \geq 1/K^{1/4} + 4\qdim \cdot e^{-K^{1/8}/2}]
        \leq 2 \qdim \cdot e^{-K^{1/4} / 32}
        + 4 K\qdim \cdot e^{-K^{1/8}/2}
        \leq 6K\qdim \cdot e^{-K^{1/8}/32}.
\end{equation*}
This completes the proof.
\end{proof}

\providecommand{\Id}{\mathsf{Id}}

\section{On the power of counting arguments}\label{sec:appendix-counting}

In this section, we will consider the power of \emph{counting arguments}
to show lower bounds for the Oracle State Distinguishing Game.
Counting arguments apply to the case when the adversary cannot compute too many unitaries, which we formalize as follows.

\begin{definition}[Small oracle circuits]
    An oracle circuit $A^{(\cdot)}$ is \emph{$S$-small} if the number of distinct unitaries $A^f$, ranging over all oracles $f$, is at most $S$.
\end{definition}

Although most oracle circuits are not ``small'' enough to be useful,
there are a few interesting families of small oracle circuits,
which we state below.
As always, we will write $N \coloneqq 2^n$
for the size of the oracle circuit's input register,
and $M \coloneqq 2^m$
for the total size of the oracle circuit's registers.

\begin{example}[Aaronson-Kuperberg adversaries]
    As described in \Cref{sec:related-work}, Aaronson and Kuperberg~\cite{AK07} considered oracle circuits $A^{(\cdot)}$
    in which for every oracle $f$,
    $A^f$ exactly computes some unitary transformation on its first $n$ qubits.
    They showed that any such oracle circuit is $4^{N}$-small~\cite[Theorem 6.7]{AK07}.
\end{example}

\begin{example}[Adversaries with no ancillas but multiple oracles]\label{ex:no-ancilla}
    Consider an oracle circuit with $n$ input qubits and no ancilla qubits which makes $t$ queries.
    For this example only, we will depart from our usual notation and allow the queries to be made to $t$ potentially different functions $f_1, \ldots, f_t$.
    Then there are at most $2^{N}$ choices for each function $f_i$,
    and so this oracle circuit is $(2^{N})^t = 2^{N t}$-small.
    If $t = \poly(n)$, then this is $2^{N \cdot \poly(n)}$-small.
\end{example}

\begin{example}[Small-ancilla adversaries]\label{ex:small-ancilla}
    Let $A^{(\cdot)}$ be an oracle circuit which makes multiple queries to a single function $f$ and uses $n$ input qubits and $a$ ancilla qubits, for a total of $m = n + a$ qubits.
    Then $A^{(\cdot)}$ is $2^{M}$-small.
\end{example}

Note that the bound in \Cref{ex:small-ancilla} subsumes \Cref{ex:no-ancilla}.
This is because by \Cref{rem:multi-query},
an adversary which makes $t$ queries to different functions $f_1, \ldots, f_t$ can be simulated by an oracle circuit $A^{(\cdot)}$ which uses $\lceil \log_2 t \rceil$ additional ancilla qubits and queries a single function~$f$.

Now we state our main bound, which rules out adversaries for the Oracle State Distinguishing Game which are ``small''.

\begin{theorem}[Counting bound]\label{thm:appendix-small-family}
    There is a universal constant $c>0$ such that the following is true. 
    Consider the Oracle State Distinguishing Game played with a uniformly random function family $\bprs:[K] \times [N] \rightarrow \{\pm 1\}$. 
    Let $A^{(\cdot)}$ be an adversary which is $S$-small,
    for $S = \exp(c \cdot \eps^2 K N)$.
    Then
    \[ \underset{\bR}{\Pr}\Big[ \distinguish_{A}(\bR) \geq \epsilon \Big] \leq 2\cdot \exp(-c\cdot \epsilon^2 KN).
    \]
\end{theorem}

In the context of our examples,
this rules out Aaronson-Kuperberg adversaries,
so long as $K = \Omega(1/\epsilon^2)$.
This also rules out small-ancilla adversaries.
For example, if adversary uses $a = \tfrac{1}{2} \cdot \log_2(K)$ ancilla qubits,
then its total size is $M = N \cdot \sqrt{K}$,
and so it is $2^{N \sqrt{K}}$-small,
which is small enough (assuming reasonable settings of parameters)
for \Cref{thm:appendix-small-family} to apply.
On the other hand, \Cref{thm:appendix-small-family} cannot rule out general adversaries which use $a = \log_2(K)$ ancilla qubits or more.
For example, a natural adversary might intend to perform the query $(k, x) \mapsto R(k, x)$ in superposition, and to do so it needs $\log_2(K) + n$ qubits,
putting it in the range where \Cref{thm:appendix-small-family} no longer applies.
This shows the limitation of this style of counting argument:
it becomes ineffective once the adversary has even a small number of ancilla qubits.

Now we prove \Cref{thm:appendix-small-family}.
We will do so using a standard ``concentration and union bound'' approach:
we prove a tail inequality on the probability that $A^f$ results in a good attack for a fixed $f$, and the we union bound over all $f$.

\begin{lemma}[Success probability of no-query adversaries]\label{lem:main-appendix-counting}
There is a universal constant $c>0$ such that the following is true.
Let $\bR: [K]\times [N]\rightarrow \{\pm 1\}$ be a uniformly random function family.
Let $A^{(\cdot)}$ be an adversary that does not make any queries.
Then
    \[ \underset{\bR}{\Pr}\Big[ \distinguish_{A}(\bR) \geq \epsilon \Big] \leq 2\cdot \exp(-c\cdot \epsilon^2 KN).
    \]
\end{lemma} 
\begin{proof}
    We prove this using tools from \Cref{sec:game-concentration}.
    Since $A^{()}$ makes no queries, let us fix an arbitrary $f$
    and note that $\distinguish_{A}(R) = \distinguish_{A}(R \mid f)$ for all function families $R$. Then
    \begin{equation}\label{eq:blah}
        \distinguish_{A}(\prs \mid f)
    = \Big|\E_{\bk \sim [K]}[\accept_{A}(\prs_{\bk} \mid f)]
    - \E_{\bh}[\accept_{A}(\bh \mid f)]\Big|
    = \Big|\E_{\bk \sim [K]}[\accept_{A,0}(\prs_{\bk} \mid f)]
    - \E_{\bh}[\accept_{A,0}(\bh \mid f)]\Big|,
    \end{equation}
    where $\accept_{A,0}(\cdot \mid f)$ is the extension of $\accept_{A}(\cdot \mid f)$ to bounded functions $\underline{R}:[K]\times[N] \rightarrow [-1, 1]$
    from \Cref{def:bounded-inputs}.
    But the function $\underline{R} \mapsto \E_{\bk \sim [K]}[\accept_{A,0}(\underline{\prs}_{\bk} \mid f)]$
    is convex and $(2/\sqrt{KN})$-Lipschitz by \Cref{lem:noabs-lipschitz}. Hence, if $\bprs:[K] \times [N] \rightarrow \{\pm 1\}$ is a uniformly random function family,
    Talagrand's inequality (\Cref{thm:talagrand}) implies that
    \begin{equation*}
        \Pr_{\bprs}\Big[\Big|\E_{\bk \sim [K]}[\accept_{A,0}(\bprs_{\bk} \mid f)]
    - \E_{\bh}[\accept_{A,0}(\bh \mid f)]\Big|\geq \epsilon\Big] \leq 2 \cdot \mathrm{exp} \left( -\frac{c \cdot \epsilon^2 KN}{4} \right)
    \end{equation*}
    for some absolute constant $c > 0$.
    But by \Cref{eq:blah} the left-hand side is $\Pr_{\bprs}[\distinguish_{A}(\bR) \geq \epsilon]$, and so this completes the proof,
    with the ``$c$'' in the lemma statement being equal to $c/4$.
\end{proof}

Deriving \Cref{thm:appendix-small-family} from \Cref{lem:main-appendix-counting} is relatively straightforward.

\begin{proof}[Proof of \Cref{thm:appendix-small-family}]
Let $A^{(\cdot)}$ be an $S$-small adversary, for a value of $S$ to be determined later.
Then there exist $S$ functions $f_1, \ldots, f_S$
such that the set of unitaries $A^{f_1}, \ldots, A^{f_S}$
contains every unitary computable by $A^{(\cdot)}$.
Fix a $1 \leq i \leq S$.
Then by hard-coding the function $f_i$ into $A^{(\cdot)}$,
we can view $A^{f_i}$ as an oracle circuit that does not make any queries.
Thus, \Cref{lem:main-appendix-counting} says that
\begin{equation*}
    \Pr_{\bprs}[\distinguish_{A^{f_i}}(\bR) \geq \epsilon] \leq 2\cdot \exp(-c\cdot \epsilon^2 KN).
\end{equation*}
As a result, we can upper-bound the maximum distinguishing probability by
\begin{align*}
   \Pr_{\bprs}[\distinguish_{A}(\bR) \geq \epsilon]
    &= \Pr_{\bprs}\Big[\max_{f}\{\distinguish_{A}(\bR \mid f)\} \geq \epsilon\Big]\\
    &= \Pr_{\bprs}\Big[\max_{i}\{\distinguish_{A}(\bR \mid f_i)\} \geq \epsilon\Big] \\
    &\leq \sum_{i=1}^S \Pr_{\bprs}[\distinguish_{A}(\bR \mid f_i) \geq \epsilon]\tag{by the union bound}\\
    &= \sum_{i=1}^S \Pr_{\bprs}[\distinguish_{A^{f_i}}(\bR) \geq \epsilon]\\
    &\leq \sum_{i=1}^S 2\cdot \exp(-c\cdot \epsilon^2 KN)
    = S \cdot 2\cdot \exp(-c\cdot \epsilon^2 KN).
\end{align*}
Now, let us choose $S$ to be $S = \exp(c/2 \cdot \epsilon^2 K N)$.
Then this upper bound on the maximum distinguishing probability equals $S \cdot 2\cdot \exp(-c/2\cdot \epsilon^2 KN)$.
This concludes the proof, with the constant ``$c$'' in the statement of the proof equal to $c/2$.
\end{proof}

\section{A one-query attack with advantage $\Omega(1/\sqrt{K})$}\label{sec:appendix-attack}

In this section, we give a one-query adversary for the Oracle State Distinguishing Game achieving advantage $\Omega(1/\sqrt{K})$ using only one ancilla qubit. This demonstrates that the dependence on $K$ in our main theorem (\cref{thm:main-at-end-of-proof}) is tight.
The adversary is given as follows.

\begin{definition}[Hadamard adversary]
    On input an $n$-qubit state $\ket{\psi}$, the \emph{Hadamard adversary} $A_{\mathrm{Had}}^{(\cdot)}$ acts as follows.
    \begin{enumerate}
        \item Apply the $n$-qubit Hadamard $H^{\otimes n}$ to $\ket{\psi}$.
        \item Measure $H^{\otimes n} \cdot \ket{\psi}$ in the standard basis. Let $\by \in \{0, 1\}^n$ be the measurement outcome.
        \item Query a bit flip oracle $f:\{0, 1\}^n \rightarrow \{0, 1\}$ on $\by$.
        Let $\bb' = f(\by) \in \{0, 1\}$ be the result.
        \item Output $\bb'$.
    \end{enumerate}
\end{definition}

The Hadamard's adversary's one ancilla qubit is used to store the outcome of the query to the bit flip oracle.
Note that this can be simulated by an adversary which makes a query to a phase oracle instead,
as discussed following the statement of \Cref{def:phase}.
We also remark that because the Hadamard adversary applies an $n$-qubit Hadamard, it will be more convenient to think of the adversary's state space as consisting of $n$-qubits, rather than being a single space of overall dimension $N \coloneqq 2^n$.
As a result, using the correspondence between $\{0, 1\}^n$ and $[N]$ mentioned in \Cref{not:size-versus-dim}, we will prefer to format our function families as $R:[K] \times \{0, 1\}^n \rightarrow \{\pm 1\}$, with the $k$-th binary phase state being
\begin{equation*}
    \ket{\psi_{R_k}} = \frac{1}{\sqrt{N}} \cdot \sum_{x \in \{0, 1\}^n} R_k(x) \cdot \ket{x}.
\end{equation*}
Our main goal is to prove the following bounds on the Hadamard adversary's distinguishing probability.

\begin{theorem}[Distinguishing advantage of the Hadamard adversary]\label{thm:disting-of-hadamard}
    There exists a constant $c > 0$ such that the following is true.
    Let $K, N \geq c$, and 
    let $\bprs:[K] \times \{0, 1\}^n \rightarrow \{\pm 1\}$ be a uniformly random function family.
    Then
    \begin{equation*}
        \E_{\bprs}[\distinguish_{A_{\mathrm{Had}}}(\bprs)] \geq \Omega\Big(\frac{1}{\sqrt{K}}\Big).
    \end{equation*}
\end{theorem}

When the Oracle State Distinguishing Game is played with some function family $R:[K] \times [N] \rightarrow \{\pm 1\}$, with probability $\frac{1}{2}$ the Hadamard adversary is given the state $\ket{\psi_{\prs_{\bk}}}$ for $\bk$ chosen uniformly at random.
We will write $\calM_R$ for the probability distribution on the measurement outcome $\by$ in this case.
In other words,
\begin{equation*}
    \calM_R(y) \coloneqq
    \E_{\bk \sim [K]} |\bra{y} \cdot H^{\otimes n} \cdot \ket{\psi_{R_{\bk}}}|^2.
\end{equation*}
With the remaining $\tfrac{1}{2}$ probability,
the Hadamard adversary is given a uniformly random phase state;
equivalently, it is given the maximally mixed state $\Id_N/N$.
In this case, the measurement outcome $\by$ is distributed as a uniformly random string in $\{0, 1\}^n$.
We will write $\calU_N$ for this uniform probability distribution, i.e.\ $\calU_N(y) := 1/N$.

The Hadamard adversary measures a $\by$ which is sampled either from $\calM_R$ or $\calU_N$, and it feeds $\by$ into the function $f$, which can be thought of as a statistical test to distinguish these two distributions. The following lemma characterizes the Hadamard adversary's distinguishing advantage in terms of the \emph{total variation distance} $\mathrm{d}_{\mathrm{TV}}(\cdot, \cdot)$ between these two distributions.

\begin{lemma}[Distinguishing advantage equals TV distance]\label{lem:equals-tv}
    \begin{equation*}
        \distinguish_{A_{\mathrm{Had}}}(R) = \mathrm{d}_{\mathrm{TV}}(\calM_R, \calU_N)
        = \frac 1 2  \cdot \sum_{y\in \{0, 1\}^n} \left| \calM_R(y) - \calU_N(y)\right|.
    \end{equation*}
\end{lemma}
\begin{proof}
    Let $f:\{0, 1\}^n \rightarrow \{0, 1\}$ be a Boolean function.
    By definition,
    \begin{align*}
    \distinguish_{A_{\mathrm{Had}}}(\prs \mid f)
    &= \Big|\E_{\bk \sim [K]}[\accept_{A_{\mathrm{Had}}}(\prs_{\bk} \mid f)]
    - \E_{\bh}[\accept_{A_{\mathrm{Had}}}(\bh \mid f)]\Big|\\
    &= \Big|\E_{\bk \sim [K]} \Pr[\text{$A_{\mathrm{Had}}^f$ outputs ``$0$'' on $\ket{\psi_{\prs_{\bk}}}$}] - \E_{\bh} \Pr[\text{$A_{\mathrm{Had}}^f$ outputs ``$0$'' on $\ket{\psi_{\bh}}$}]\Big|\\
    & = \Big| \sum_{y : f(y) = 0} \calM_R(y) - \sum_{y : f(y) = 0}\calU_N(y)\Big|.
\end{align*}
The maximum distinguishing advantage is then computed by optimizing this expression over all $f$, but that is exactly the definition of the total variation distance.
\end{proof}

Our goal is to calculate the expectation $\E_{\bprs}[\distinguish_{A_{\mathrm{Had}}}(\bprs)]$.
The following lemma gives an alternative expression for this expectation in terms of a new random variable.

\begin{lemma}\label{lem:new-var}
    Given a function family $\prs:[K] \times \{0, 1\}^n \rightarrow \{\pm 1\}$, define the quantity
    \[ X_{\prs} \coloneqq  \frac 1 N\cdot \underset{\bk \sim [K]}{\E} \Big( \sum_{x \in \{0, 1\}^n} \ro_{\bk}(x) \Big)^2.
\]
Then for a uniformly random function family $\bprs:[K] \times [N] \rightarrow \{\pm 1\}$,
\begin{equation*}
    \E_{\bprs}[\distinguish_{A_{\mathrm{Had}}}(\bprs)] = \frac{1}{2} \cdot \E_{\bprs}|X_{\bprs} - 1|.
\end{equation*}
\end{lemma}
\begin{proof}
    By \Cref{lem:equals-tv},
\begin{align}
    \E_{\bprs}[\distinguish_{A_{\mathrm{Had}}}(\bprs)]
    = \frac 1 2  \cdot \sum_{y\in \{0, 1\}^n} \E_{\bprs} \left| \calM_{\bprs}(y) - \calU_N(y)\right|
    &= \frac 1 2  \cdot \sum_{y\in \{0, 1\}^n} \E_{\bprs} \Big| \calM_{\bprs}(y) - \frac{1}{N}\Big|\nonumber\\
    &
    = \frac 1 2  \cdot \frac{1}{N} \cdot \sum_{y\in \{0, 1\}^n} \E_{\bprs} \Big| N \cdot  \calM_{\bprs}(y) -1\Big|.\label{eq:brought-out-an-N}
\end{align}
Now, fix a $y \in \{0, 1\}^n$.
Consider the random variable
\begin{equation*}
    N \cdot \calM_{\bprs}(y)
    =
    N \cdot \E_{\bk \sim [K]}\left|\bra{y} \cdot H^{\tensor n} \cdot \ket{\psi_{\bro_{\bk}}} \right|^2
    = \frac 1 {N}\cdot \E_{\bk \sim [K]}\Big( \sum_{x\in \{0, 1\}^n} \bro_{\bk}(x) \cdot (-1)^{x\cdot y} \Big)^2.
\end{equation*}
For each value of $\bk$, the corresponding term is distributed as the square of the sum of $N$ independent and uniformly random $\{\pm 1\}$ numbers,
and the terms are independent across different values of $\bk$.
Hence, this random variable is distributed identically to $X_{\bprs}$.
In addition, these $K$ random variables are independent.
As a result, by linearity of expectation,
\begin{equation*}
    \eqref{eq:brought-out-an-N}
    = \frac 1 2  \cdot \frac{1}{N} \cdot \sum_{y\in \{0, 1\}^n} 
    \E_{\bprs} | X_{\bprs} -1|
    = \frac 1 2  \cdot 
    \E_{\bprs} | X_{\bprs} -1|.
\end{equation*}
This completes the proof.
\end{proof}

Now we study the distribution of the random variable $X_{\bprs}$.
To begin, we compute its mean and variance.

\begin{lemma}\label{lem:mean-var}
    Let $\bprs:[K] \times \{0, 1\}^n \rightarrow \{\pm 1\}$
    be a uniformly random function family.
    Then the random variable $X_{\bprs}$ has expectation 1 and variance $2/K \cdot (N-1)/N$.
\end{lemma}

\begin{proof}
    First, we compute the mean:
\begin{align*}
    \underset{\bro}{\E}[X_{\bprs}] &= \frac 1 N \cdot \underset{\bro, \bk}{\E}\Bigg[\Big(\sum_{x\in \{0, 1\}^n} \bro_{\bk}(x)\Big)^2\Bigg] = \frac 1 N\cdot  \underset{\bro, \bk}{\E}\Bigg[\sum_{x,y\in \{0, 1\}^n} \bro_{\bk}(x) \cdot \bro(\bk,y)\Bigg]\\
    &= \frac{1}{N} \cdot \E_{\bk \sim [K]} \Bigg[  \sum_{x,y \in \{0, 1\}^n} \E_{\bR} \Big[\bR_{\bk}(x) \cdot \bR_{\bk}(y) \Big]\Bigg] = \frac{1}{N} \cdot \E_{\bk \sim [K]} [ N ] = 1.
\end{align*}
Next, we compute the variance. To begin, note that for each $R$,
\begin{equation*}
    X_{R} - \E_{\bprs}[X_{\bprs}]
    = X_R - 1
    = \frac 1 N\cdot \underset{\bk \sim [K]}{\E} \Big( \sum_{x \in \{0, 1\}^n} \ro_{\bk}(x) \Big)^2 - 1
    = \frac{1}{N} \cdot \underset{\bk \sim [K]}{\E} \Big(\sum_{x \neq y} \ro_{\bk}(x) \cdot \ro_{\bk}(y)\Big).
\end{equation*}
Thus, the variance is given by
\begin{align}
\E_{\bprs}[(X_{\bprs} - \E_{\bprs}[X_{\bprs}])^2]
&= \E_{\bprs}\Big[\Big(\frac{1}{N} \cdot \underset{\bk \sim [K]}{\E} \Big(\sum_{x \neq y} \bro_{\bk}(x) \cdot \bro_{\bk}(y)\Big)\Big)^2\Big]\nonumber\\
&= \frac{1}{N^2} \cdot \E_{\bprs}\E_{\bk, \bk' \sim[K]}\Big[\sum_{x \neq y} \sum_{z \neq w} \bprs_{\bk}(x) \cdot \bprs_{\bk}(y) \cdot \bprs_{\bk'}(z) \cdot \bprs_{\bk'}(w)\Big]\nonumber\\
&= \frac{1}{N^2} \cdot \sum_{x \neq y} \sum_{z \neq w} \E_{\bk, \bk' \sim[K]}\E_{\bprs}\Big[ \bprs_{\bk}(x) \cdot \bprs_{\bk}(y) \cdot \bprs_{\bk'}(z) \cdot \bprs_{\bk'}(w)\Big].\label{eq:whatevers}
\end{align}
The expectation over $\bR$ is zero if $\bk \neq \bk'$.
On the other hand, if $\bk = \bk'$, then the expectation is 1 if $\{x, y\} = \{z, w\}$ and $0$ otherwise. As a result,
\begin{equation*}
    \eqref{eq:whatevers}
    = \frac{1}{N^2} \cdot \sum_{x \neq y} \sum_{z \neq w} \frac{1}{K} \cdot \bone[\{x, y\} = \{z, w\}]
    = \frac{1}{N^2} \cdot\frac{1}{K} \cdot \sum_{x \neq y} \sum_{z \neq w}  \bone[\{x, y\} = \{z, w\}]
    = \frac{1}{N^2} \cdot\frac{1}{K} \cdot 2 N(N-1).
\end{equation*}
This completes the proof.
\end{proof}

From \Cref{lem:mean-var},
we roughly expect that $X_{\bprs}$ tends to be around the values
$1 \pm \sqrt{2/K}$.
If this were true, then the expectation $\E_{\bprs}|X_{\bprs} -1|$ we are trying to compute would be roughly $\sqrt{2/K}$, and we would be done.
However, it could be that the $X_{\bprs}$'s variance being roughly $2/K$ could be due to some small probability events where $X_{\bprs}$ is very far from 1, whereas with high probability $X_{\bprs}$ is much closer to 1 than $\sqrt{2/K}$.
To rule this out, we prove the following concentration bound for $X_{\bprs}$.

\begin{lemma}\label{lem:concentration}
    There exists a constant $c > 0$ such that the following is true.
    Let $\bprs:[K] \times \{0, 1\}^n \rightarrow \{\pm 1\}$
    be a uniformly random function family.
    Then for all $t >0$,
\begin{align*}
    \underset{\bro}{\Pr}\left[ |X_{\bprs} - 1| \geq t \right] \leq 2\cdot \exp(-c \cdot K \cdot \min\{t^2, t\}).
\end{align*}
\end{lemma}
\begin{proof}
    For each $1 \leq k \leq K$, define
    \begin{equation*}
        X_{\bprs, k} \coloneqq \frac{1}{N} \cdot \Big(\sum_{x \in \{0, 1\}^n} \bprs_k(x)\Big)^2
        =\Big(\sum_{x \in \{0, 1\}^n}\frac{1}{\sqrt{N}} \cdot  \bprs_k(x)\Big)^2,
    \end{equation*}
    Then \Cref{lem:sub-exponential-calculation} implies that there exists a constant $\gamma \geq 1$ such that for each $1 \leq k \leq K$, the random variable $|X_{\bprs,k}|^2-1$ is sub-exponential with parameter $\gamma$.
    Since $X_{\bprs} = \E_{\bk \sim [K]} [X_{\bprs, \bk}]$,
    Bernstein's inequality (\Cref{thm:bernstein-subexp})
    states that
    \begin{equation*}
        \underset{\bro}{\Pr}\left[ |X_{\bprs} - 1| \geq t \right]
        \leq 2\cdot \exp(-c \cdot \min\Big\{\frac{t^2}{\gamma^2}, \frac{t}{\gamma}\Big\} \cdot K)
        \leq 2\cdot \exp(-c \cdot \min\Big\{\frac{t^2}{\gamma^2}, \frac{t}{\gamma^2}\Big\} \cdot K),
    \end{equation*}
    because $\gamma \geq 1$.
    This completes the proof, by setting the ``$c$'' in the lemma statement to $c/\gamma^2$.
\end{proof}

Now we prove our bound on the expected distinguishing advantage of the Hadamard tester.

\begin{proof}[Proof of \Cref{thm:disting-of-hadamard}]
By \Cref{lem:new-var}, it suffices to show that $\E_{\bprs}|X_{\bprs} - 1| \geq \Omega(1/\sqrt{K})$.
We will show this by deriving the following weak anti-concentration result: there exists a constant $\epsilon > 0$ such that 
\begin{align}\label{eq:weak-anti}
    \underset{\bro}{\Pr}\left[ |X_{\bprs} - 1| \geq  \frac \epsilon {\sqrt{K}} \right] \geq \epsilon.
\end{align}
Assuming this is true, then our main result can be shown as follows:
\begin{align*}
    \underset{\bro}{\E}|X_{\bprs} - 1| \geq \frac{\epsilon}{\sqrt{K}} \cdot \Pr\Big[|X_{\bprs} - 1| \geq \frac \epsilon {\sqrt{K}}\Big] \geq \frac{\epsilon^2}{\sqrt{K}}.
\end{align*}
Now we prove \Cref{eq:weak-anti}.
The proof is by contradiction:
for sake of contradiction, let us assume that it is false.
Then one can obtain an upper bound for the variance of $X_{\bprs}$ as follows:
\begin{align}
    & \underset{\bro}{\E}\Big[|X_{\bprs}-1|^2\Big]\nonumber\\={}& \int_0^\infty \Pr\Big[|X_{\bprs} - 1|^2 \geq t\Big] dt \nonumber\\
    ={}& \int_0^{\frac {\epsilon^2} {K}} \Pr\Big[|X_{\bprs} - 1|^2 \geq t\Big] dt + \int_{\frac {\epsilon^2}{K}}^{\frac {1}{\epsilon \cdot K}} \Pr\Big[|X_{\bprs} - 1|^2 \geq t\Big] dt + \int_{\frac{1}{\epsilon \cdot K}}^\infty \Pr\Big[|X_{\bprs} - 1|^2 \geq t\Big] dt.\label{eq:asgawg}
     \end{align}
We can upper-bound the first term by $\epsilon^2/K$ since probabilities are always at most one.
As for the second term, since we are assuming that \Cref{eq:weak-anti} is false, we have that
\begin{equation*}
    \int_{\frac {\epsilon^2}{K}}^{\frac {1}{\epsilon \cdot K}} \Pr\Big[|X_{\bprs} - 1|^2 \geq t\Big] dt
    \leq \int_{\frac {\epsilon^2}{K}}^{\frac {1}{\epsilon \cdot K}} \Pr\Big[|X_{\bprs} - 1|^2 \geq \frac{\epsilon^2}{K}\Big] dt
    <\int_{\frac {\epsilon^2}{K}}^{\frac {1}{\epsilon \cdot K}} \epsilon\cdot dt
    < \epsilon \cdot \frac{1}{\epsilon \cdot K} = \frac{1}{K}.
\end{equation*}
Finally, we can bound the third term using \Cref{lem:concentration}.
In total, we have that
     \begin{align*}
     \eqref{eq:asgawg}
     <{}& \frac {\epsilon^2} { K} +  \frac {1}{K} + \int_{\frac {1} {\epsilon \cdot K}}^\infty 2\cdot \exp(-c\cdot K \cdot \min(t, \sqrt{t})) dt\\
    \leq{}& \frac {\epsilon^2} { K} + \frac 1 {K}  + \int_{\frac {1} {\epsilon \cdot K}}^\infty 2\cdot \exp(-c\cdot K \cdot t) dt + \int_{1}^\infty 2\cdot \exp(-c\cdot K \cdot t) \cdot 2t \cdot dt\\
    ={}& \frac {\epsilon^2} { K} + \frac 1 {K}  +  \frac {2}{c\cdot K} \cdot \exp(-\frac c \epsilon) + 4\cdot \Big(\frac 1 {c\cdot K} + \frac 1 {(c\cdot K)^2}\Big) \cdot \exp(-c\cdot K).
\end{align*}
For $\epsilon$ a sufficiently small constant and $K$ a sufficiently large constant, this is at most $\tfrac{3}{2} \cdot \frac{1}{K}$.
But we already calculated that the variance of $X_{\bprs}$ is $\tfrac{N-1}{N} \cdot \frac{2}{K}$, in \Cref{lem:mean-var}.
Hence, we have a contradiction for sufficiently large $N$, completing the proof.
\end{proof}

\end{document}